\documentclass{jpp}

\usepackage{graphicx}% Include figure files
\usepackage{dcolumn}% Align table columns on decimal point
\usepackage{bm}% bold math
\usepackage{xcolor}
\usepackage{breqn}
\usepackage{hyperref}% add hypertext capabilities
\begin{document}

\title{The Equations of Reduced Magnetohydrodynamics in Dipole Coordinates}% Force line breaks with \\

\author{Ian M. DesJardin\aff{1,2}\corresp{\email{ian.m.desjardin@nasa.gov}}, Christopher Bard\aff{2}, Natalia Y. Buzulukova\aff{3,2}, John C. Dorelli\aff{2}}

\affiliation{\aff{1}Physics Department, The Catholic University of America
  \aff{2} Geospace Physics Laboratory
  NASA Goddard Spaceflight Laboratory
  \aff{3} Department of Astronomy,
  University of Maryland, College Park
}%

\date{\today}% It is always \today, today,

\maketitle

\begin{abstract}
The equations of reduced magnetohydrodynamics (RMHD) isolate the Alfv\'enic energy transfer and turbulence dynamics from MHD in a computationally tractable way. In this study, we derive the equations of RMHD for a low $\beta$ plasma in a dipole coordinate system aligned with an inhomogeneous background that is general enough to represent the Earth's magnetosphere using a multiscale analysis and the Kreiss theorem. The most important effect beyond homogeneous RMHD is the propensity for background Alfv\'en speed gradients to drive nonlinear wave effects. This work has immediate application to the magnetosphere-ionosphere coupling problem in the Earth's magnetosphere, where the Alfv\'en speed varies by an order of magnitude over a short distance. We estimate the turbulent energy transfer driven by this. In Earth's magneotsphere, a substantial portion of energy ($>20\%$) could be transferred in this way along auroral field lines. A revised field aligned current density mapping technique to account for this energy loss mechanism is proposed to replace using $j_\parallel/B_0$ being constant along the field line. We also predict the existence of a ``saturation current,'' where there is a maximum $j_\parallel$ that exist before driving a nonlinear cascade. At Earth, this value is $\sim 5\; \mu \text{A}/\text{m}^2$ at the ionosphere, which corresponds to the maximum observed value by ground magnetometers. Furthermore, this study makes use of a novel methodology, a computer algebra system, in performing algebraic work under complicated coordinate systems. We hope this approach serves as a template for performing reproducible and verifiable multiscale perturbation analysis under arbitrarily complex geometries. 
\end{abstract}

\section{\label{sec:introduction}Introduction}
Magnetohydrodynamic turbulence, Alfv\'enic wave propagation, and Alfv\'enic energy transfer are often studied by isolating the slower convective timescale dynamics into a set of equations known as the reduced MHD (RMHD) equations. RMHD is advantageous due to its analytical simplicity and computational efficiency. To derive RMHD, one takes the ideal MHD equations with a strong background magnetic field and solves for perturbations under the assumption of long wavelengths parallel to the field and short wavelengths perpendicular to it \citep{kadomtsev_nonlinear_1973,strauss_nonlinear_1976,strauss_reduced_1997}. Under these circumstances, the fast and slow mode are advected by the shear Alfv\'enic dynamics which dominate the system. RMHD (and kinetic extensions) have been used to study the development of turbulence and nonlinear structures in astrophysical (outside the solar system) and heliophysical (inside the solar system) plasmas. This includes solar turbulence \citep{perez_direct_2013}, planetary magnetospheres \citep{watanabe_feedback_2010, watanabe_generation_2016, sakaki_convective_2024, sakaki_nonlinear_2026}, and in other situations concerning MHD turbulence \citep{schekochihin_mhd_2022}.

Although originally proposed by \citet{kadomtsev_nonlinear_1973, strauss_nonlinear_1976}, one of the most powerful derivations of RMHD was done in \citet{zank_equations_1992} and subsequent papers. They used the Kreiss theorem and a multi-time multi-scale analysis to show that isolation of the turbulence dynamics can be achieved by the assumption that long wavelength solutions \textit{only} depend on the magnetic field aligned component and that there is no short time dependence, although under a homogeneous isotropic slab geometry. An outcome of this perturbation analysis is a system of constraints that must be self-consistently satisfied by the RMHD equations. Their methodology provides a logical way to derive RMHD along with the hallmark behaviors of it, namely time independent pressure balance and perpendicular incompressibility. Although many studies present the final result, they do not emphasize the constituent assumptions. Our study follows this approach to derive necessary conditions in the background geometry for RMHD to apply to the magnetosphere in dipole coordinates.

Turbulence theories derived from RMHD, often with kinetic extensions, in Cartesian geometries with a homogeneous background have been used to explain observations in the near-Earth low-beta plasmas (e.g. \citet{hallinan_smallscale_1970,stasiewicz_origin_1984,lynch_multiple-point_1999,stasiewicz_identification_2000,milanese_dynamic_2020,chaston_auroral_2021,di_mare_new_2024}) and modeled small scale structure of auroral arcs driven by Alfv\'en waves \citep{seyler_mathematical_1990,chaston_smallscale_2010}. Other prominent heliophysical phenomena, such as auroral arc formation, dynamics of auroral forms and particle acceleration in auroral region are all tightly connected to the turbulence process and Alfv\'enic dynamics \citep{kataoka_small-scale_2021,schroeder_laboratory_2021,tian_evidence_2026}. The role of turbulence is evident by the well-known rippling and folding behavior of auroral arcs during intense geomagnetic storms \citep{kataoka_turbulent_2011,ivarsen_turbulence_2024}. RMHD, together with additional physics of parallel electric field included (see, e.g., Eq. 1 from \citet{kataoka_small-scale_2021}), provides a unified framework to understand these processes, as well as the transfer of power by Alfv\'enic waves, cross-scale coupling of the auroral forms, and the development of turbulence in the auroral zone. When applied to the Earth's magnetosphere, this class of problems is sometimes called ``magnetosphere-ionosphere coupling.'' Many previous studies' wave propagation models in this region, e.g. \citet{knudsen_alfven_1992, lysak_generalized_1993, thompson_electron_1996, lysak_magnetosphere-ionosphere_2004, watt_inertial_2006, woodroffe_ultra-low_2012, desjardin_steepening_2026}, neglect nonlinear advection terms that can create a turbulent cascade even though turbulence is a prominent feature of geomagnetic substorms \citep{chaston_turbulent_2008, hull_alfvenic_2016, miles_alfvenic_2018, chaston_auroral_2021}. Turbulence associated with Alfv\'enic field aligned currents has been observed as low as the E-layer of the ionosphere by incoherent scatter radars \citep{ivarsen_turbulence_2024-1}.

Despite both experimental and theoretical understanding of the role of turbulence in MI coupling, current state-of-the-art global magnetospheric models largely ignore it. They are unable to directly model the interface between these regions due to the numerical stiffness of converging field lines and an increasing Alfv\'en speed \citep{yu_including_2010}. Instead, magnetosphere codes operate with an inner boundary at several (2-3) characteristic radii (e.g. Earth's radius for Earth's magnetosphere) from the planet's center. Field aligned currents from this inner boundary are projected onto the ionosphere upper boundary assuming conservation of $j_\parallel/B$, and quasi-static Poisson-like equation is used to calculate ionospheric potential form field-aligned current and ionospheric conductivity \citep{atkinson_auroral_1970,merkin_effects_2010,zhang_electron_2015,james_model_2026}. This region of converging field lines is often called the magnetosphere-ionosphere (MI) gap, is where intense particle energization happens that eventually forms the aurora \citep{tian_evidence_2026,dorelli_science_2026}. In global models, the $j_\parallel$ at the boundary drives a portion of the particle precipitation model \citep{zhang_electron_2015,mukhopadhyay_global_2022} which is how the observed magnetospheric energy input to the ionosphere is ultimately calculated \citep{bagheri_comparison_2023}. The standard quasi-stationary approach ignores physics of the auroral region, including Alfvenic wave dynamics, development of turbulence, and cross-scale coupling. Hence, the solutions to current global magnetospheric models have a resolution dependence because of the development of mesoscale structures \citep{gunzkofer_impact_2026}. We expect the development of turbulence in this region is connected to this.

The RMHD model derived in this study will be useful in modeling this region because it retains nonlinear mixing while simplifying other aspects of the model to make it computationally simpler. First, the fast and slow modes are ordered out of the dynamics, so the numerical stiffness of the problem is relaxed. Second, the equations are simplified, instead of 3D compressible MHD with 8 independent variables (for isotropic MHD) one solves 2.5D RMHD equations with 2 independent variables (field-aligned current, vorticity). Therefore, it becomes analytically tractable to perform a well posed mapping of field aligned current in the MI gap region and correctly include the effect of flux tube area scaling, changes in the Alfv\'en speed, wave reflection, and wave-wave interactions in a unified framework. Third, and most important, the RMHD equations retain finite vorticity which can drive turbulence unlike the simple current mapping approach.

Many of the aforementioned MI-coupling models (e.g. \citet{lysak_generalized_1993, spencer_analysis_2007}) use a circuit or linear wave model to connect energy flux from the magnetosphere to the ionosphere. RMHD can be seen a minimal nonlinear generalization of these models which maintains their core features (second order behavior, variable flux tube size, inhomogeneous background conditions, wave reflection) while including the dominant convective nonlinearity from the MHD equations. Now capable of modeling wave-wave interactions linking nearby flux tubes, the RMHD model of MI coupling predicts two phenomena that have basis in observations but until now were not present in models. First, there is an amplitude dependent behavior to RMHD. At a critical amplitude of wave energy, quantified in our study as a saturation current density, strong vorticity will turbulently cascade energy to smaller scales and break up a coherent field aligned current. We calculate this critical amplitude and find that it agrees with the maximum observed field align current observed by ground magnetometers. Second, this turbulence provides a natural way to form small scale waves which dissipate particles. Gradients in the background Alfv\'en speed can localize energy dissipation. We find that our estimate of this reflection driven turbulent dissipation is near the auroral acceleration region, and observed location of strong wave-particle interactions in the Earth's magnetosphere. These new features of MI coupling are important because they provide a conceptual bridge between two observational facts: Alfv\'en waves appear to power the aurora \citep{keiling_global_2003} and auroral acceleration often happens at kinetic perpendicular scales \citep{seyler_mathematical_1990}. Many studies link the two with the existing of a turbulent cascade \citep{chaston_turbulent_2008}, but do not detail how the cascade is formed. We estimate that a significant portion of incident Alfv\'enic Poynting flux in the Earth's magnetosphere would be transferred into such a cascade where a portion of it would contribute to the auroral particle flux. 

This article proceeds in two main tasks. First, following the methodology of  \citet{zank_equations_1992} (\textit{ZM92}) to systematically derive reduced MHD in dipolar coordinates with an inhomogeneous background for a low $\beta_p$ plasma, representative of the MI gap region. Although related questions have been derived (i.e. \cite{hunana_inhomogeneous_2010}), this article is the first to derive the locally incompressible system in the limit of an inhomogeneous background in dipole coordinates. Importantly, this result derives a set of necessary equilibrium conditions the system must obey in order for the equations to be valid instead of a priori assuming them. This includes constraints on components curvature that can be evaluated in the real magnetosphere. Secondly, these equations are analyzed to quantify the additional effects present in them which are not in circuit or linear wave models of MI-coupling. We identify two main effects. The first effect, saturation current density, is compared to a statistical distribution of field aligned current densities observed by the Active Magnetosphere and Planetary Electrodynamics Response Experiment (AMPERE) at an altitude of $\sim 800$ km and found to predict saturation current density values near the maximum observed ($\sim 5 \mu \text{A}/\text{m}^2$ predicted vs $4.6 \mu \text{A}/\text{m}^2$ observed). The second effect, reflection driven turbulent wave dissipation, predicts lossy energy transfer from the magnetosphere to the ionosphere near Alfv\'en speed gradients of at least $\sim 24\%$ but potentially up to $70\%$ of incident Poynting flux. This compares favorably to observations of auroral electron energy fluxes that were $46 \%$ of incident magnetospheric Alfv\'en Poynting flux \citep{tian_evidence_2026}. Our goal is to suggest that these phenomena are natural outcomes of maintaining turbulence generation mechanisms in models of the MI-gap, not necessarily to show perfect agreement with observations. Finally, we present a semi-empirical adjustment to the inner boundary conditions used in global magnetosphere MHD models which could be used to take these new effects into account.

We also use a novel methodology for the theoretical derivation task. Some of the derivation performed with a computer algebra program. There is an increase in the mathematical complexity of vector calculus operators when going from Cartesian coordinates to dipolar ones. We show that it is possible to perform some of the multiple scale/time analysis deterministically in a computer algebra program, much like the mathematics community has adopted computer verifiable proofs. This technique can trivially handle complex geometries.

Before proceeding to the derivation, we provide a brief outline of the CAS methodology used in the study.

\section{Computer Algebra System}
In order to treat the equations carefully, we have utilized a computer algebra software (CAS), \texttt{SymPy}, to perform part of the derivation, specifically to aid in collecting terms of common powers of $\varepsilon$. The CAS starts with a symbolic representation of the ideal MHD equations where $u$, $p$, $B$, and $\rho$ are symbolic functions of the variables $\alpha, \beta, \chi, t$ and the functions $A(\varepsilon \alpha), X(\varepsilon \chi)$. In CAS terminology, symbols are independent variables with no underlying representation as something else. Functions are composed of functions and symbols. The capital coordinate functions are defined as $A(\alpha) = \varepsilon \alpha$, where $\varepsilon$ is a symbol that represents a small quantity. This setup is necessary so that the chain rules in \texttt{SymPy} will automatically expand functions as follows $$\frac{\partial}{\partial \alpha} f(\alpha, A(\alpha)) = \frac{\partial f}{\partial \alpha} + \varepsilon \frac{\partial f}{\partial (\varepsilon \alpha)}.$$

Since $A$ is a function and not a symbol, \texttt{SymPy} will try to represent the argument of the partial derivative as the definition of $A$ instead of a symbol for the function $A$. After expanding out the equations and finding each order of $\varepsilon$ in each equation (which is trivial using the \texttt{collect} function), the resultant equations still contain many terms like $ {\partial f}/{\partial (\varepsilon \alpha)}$. While mathematically valid, these equations are incomprehensible. Our program implements a method that traverses the symbolic tree representation of each equation and replaces instances of $\varepsilon \alpha \to A$, $\varepsilon \beta \to B$, and $\varepsilon \chi \to X$, even as the argument of a partial derivative. At first $A$ and $X$ are treated as functions. Then they are replaced with symbols that make them independent of $\alpha$ and $\chi$ when rendered. This cleaning of the equations makes the symbolic representation comprehensible to humans.

The output is written to a LaTeX file that is presented in Appendixes \ref{appendix:cas} and \ref{appendix:slab_cas}. In all, there are 26 equations representing the ideal MHD equations at each order of $\varepsilon$. 13 of these have time derivatives involving the slow time scale, although many of these time derivatives drop out when density perturbations are set to zero. When an equation loses time dependence, then it is considered a ``constraint.'' From here, the equations and constraints are simplified by hand in a manner similar to \citet{zank_equations_1992}, hereafter \textit{ZM92}. The initial implementation of the CAS was generated with a large language model, although it went through significant revision before the final version. The main text makes reference to these results which are in Appendix \ref{appendix:cas}. Of course, one can also directly substitute the expansions in the MHD equations and collect terms of common power of $\varepsilon$ to get the same result as the supplemental material. Appendix \ref{appendix:slab_cas} also includes the same equations with the Lam\'e coefficients set to 1, so one can clearly see the simplification to Cartesian RMHD equation by equation. These expressions are what is used in \textit{ZM92}, so it is also clear where our result generalizes theirs.

\section{Derivation}
\noindent We begin with the equations of ideal MHD.
\begin{gather}
  \frac{\partial \rho}{\partial t} + \nabla \cdot \left(\rho \mathbf{u} \right) = 0 \\
  \rho \frac{\partial \mathbf{u}}{\partial t} + \rho (\mathbf{u} \cdot \nabla) \mathbf{u} = -\nabla \left(p + \frac{B^2}{2\mu_0}\right) + \frac{1}{\mu_0} \mathbf{B} \cdot \nabla \mathbf{B} \label{eqn:momentum eqn}\\
  \frac{\partial \mathbf{B}}{\partial t} = \nabla \times (\mathbf{v} \times \mathbf{B}) \\
  %\frac{\partial p}{\partial t} + (\mathbf{u} \cdot \nabla) p + \gamma p (\nabla \cdot \mathbf{u})= 0 \label{eqn:pressure eqn} \\
  \nabla \cdot \mathbf{B} = 0
\end{gather}
An equation of state relating $\rho$ and $p$ has been left unspecified. Already assumed in this formulation is an ideal Ohm's law due to the form the $\mathbf{J} \times \mathbf{B}$ force takes. Following \textit{ZM92}, this is first put in a form where a small parameter ($\varepsilon$) can be introduced in order to perform a multi-scale analysis on the MHD equations. Suppose scalars $p^*$, $\rho^*$, $B^*$, $\Phi^*$, and $u^*$ can be used to normalize $p$, $\rho$, $\mathbf{B}$, $\mathbf{u}$. Space and time are also nondimensionalized in such a way that $x^*/t^* = u^*$. Using the transformation $f \to f^* f$ for each variable such that the new unstarred version is nondimensional, (\ref{eqn:momentum eqn}) transforms to
\begin{equation}
  \rho \frac{\partial \mathbf{u}}{\partial t} + \rho (\mathbf{u} \cdot \nabla) \mathbf{u} = -\frac{\beta_p}{M_A^2} \nabla p +  \frac{1}{M_A^2} \left(-\nabla \frac{B^2}{2} + \mathbf{B} \cdot \nabla \mathbf{B}\right), \label{eqn:momentum eqn nondim}
\end{equation}where $\beta_p$ is the ratio of the sound to the Alfv\'en speed squared and $M_A = u^* / (B^* / \sqrt{\mu_0 \rho^*})$ is the Alfv\'enic Mach number. We shall assume that $M_A \sim \beta_p \sim \varepsilon$. This corresponds to a subsonic plasma in a strong magnetic field. Under these conditions, the momentum equation becomes (equivalent to \textit{ZM92} equation 2.4) 
\begin{equation}
\rho \frac{\partial \mathbf{u}}{\partial t} + \rho (\mathbf{u} \cdot \nabla) \mathbf{u} = -\frac{1}{\varepsilon} \nabla p + \frac{1}{\varepsilon^2} \left(-\nabla \frac{B^2}{2} + \mathbf{B} \cdot \nabla \mathbf{B}\right), \label{eqn:low beta momentum eqn nondim}
\end{equation}
A multiscale perturbation analysis of the ideal MHD equations is now performed following the methodology in \textit{ZM92}. The Kreiss theorem is invoked to determine the expansions for each variable directly determined from this form of the momentum equation via the power of $\varepsilon$ on each term. First the operators in space and time are expanded as follows
\begin{align}
  \frac{\partial}{\partial t} &\to \frac{\partial}{\partial t} + \frac{1}{\varepsilon} \frac{\partial}{\partial \tau} \\
  \nabla &\to \nabla_{\mathbf{x}} + \varepsilon \nabla_{\mathbf{X}} 
\end{align}
where $\mathbf{x} = \mathbf{x}(\alpha,\beta,\chi)$ represents a small wavelength perturbation, $\mathbf{X}=\mathbf{X}(A,X)$ represents a long wavelength perturbation, $t$ represents a slow time perturbation, and $\tau$ represents a fast time perturbation. $\alpha, \beta$, and, $\chi$ are the dipolar coordinates defined in the next section. On short scales they take the lowercase Greek form. On long scales they take the uppercase Latin form.

This key argument of the derivation is to assume that solutions only vary on the slow time. Dynamics that happen on the fast time can be ignored by assuming they are transient and finding what their equilibrium conditions are. For this problem, the slow time is taken as the convective timescale while fast time represents the acoustic and Alfv\'enic timescales. The low Mach number assumption means the convective timescale is the slowest. As long as the slow solutions obey the equilibrium conditions of the fast dynamics, which are found by setting all $\partial/\partial \tau$ terms to zero, then the equations of motion can be written for only the slow time. In this case, the slow time is taken on the advective time scale while the time scales associated with the fast and slow compressional modes are both considered fast.

Mathematically, in order for the total time derivative to be bounded as $\varepsilon \to 0$, the ordering of the field variables must follow from the powers of $\varepsilon$ in the momentum equation. This is the Kreiss theorem \citep{kreiss_problems_1980}. For this reason, the various variables are expanded as
\begin{align}
  \rho &= \rho_0(A, X) + \varepsilon \rho_1(\alpha, \beta, \chi, X, t) \label{eqn:rho-expansion}\\
  p &= p_0(A, X) + \varepsilon p_1(\alpha, \beta, \chi, X, t)\\
  \mathbf{u} &= u_\chi(X) e_{\chi} + \mathbf{u}(\alpha, \beta, \chi, X, t) \\
  \mathbf{B} &= B_0(A,X) e_{\chi} + \varepsilon \mathbf{b}(\alpha, \beta, \chi, X, t) \label{eqn:B-expansion}
\end{align}
In a dipole coordinate system, the magnetic field has a dependence on two coordinates $A$ and $X$ (defined in the next section). We have assumed that all perturbed quantities have a dependence on the long wavelength coordinates only along the magnetic field. The background quantities $\rho_0$ and $p_0$ depend on the long wavelength coordinates $A$ and $X$ only. $u_\chi$ depends only on $X$, which is necessary to avoid solutions with a sheared parallel flow background (distinct from a sheared perpendicular flow, which is retained to first order via $\mathbf{u}_\perp$). None of these background quantities depend on the short or long $\beta$, which equivalent to assuming axisymmetry around the polar axis on the background. We have chosen to break $u_\chi$ into a time independent part and a time dependent part at the same order in order to allow for an applied background flow in the plasma.

In order to maintain consistency between the magnetic field coordinate dependence and the coordinate system, the Lam\'e coefficients of all derivatives ($h_\alpha, h_\beta, h_\chi$) only depend on the long coordinates $A$ and $X$. In order for this to be valid, we define the radii of curvature of the coordinate system as the boundary between long and short length scales. As long as the perpendicular wavelengths of the RMHD perturbations are shorter than the radii of curvature of the magnetic field, the dependencies in (\ref{eqn:rho-expansion}) - (\ref{eqn:B-expansion}) are justified.

%When all of these equations are substituted in, this will impose several equations, one at each order, per original ideal MHD equation. Some of these will have time derivatives, while others will simply be constraints. As we will show, these constraints provide additional information about the solution. This logic shows that from simply the low Mach number and low beta assumption, a great number of consequences follow when considering the advective (slow) time scale.

From here, these expansions are substituted into the normalized MHD equations. The fast time derivative ($\partial / \partial \tau$) terms are set to zero in order to solve for the fast time equilibrium conditions. The equations at each power of $\varepsilon$ are computed in the supplementary material that is generated using a computer algebra system. The remaining task is to interpret the equations and apply the constraints imposed by the fast time equilibrium conditions. This is narrated in the text of this study with heavy reference to the supplementary material. Finally, it is shown that the slow time dynamical equations are the reduced MHD equations. These are further simplified with the conventional streamfunction representation of $\mathbf{u}_\perp$ and $\mathbf{b}_\perp$ while retaining the dipolar operators. 

\subsection{Divergence Free Coordinate System}
\label{sec:divergence free coordinate system}
This derivation will be performed in traditional dipole coordinates with the following orthogonal set of coordinates
\begin{equation}
  \alpha = \frac{\sin^2 \theta}{r}, \hspace{1em} \beta = \nu, \hspace{1em}  \chi = -\frac{\cos \theta}{r^2},
\end{equation}
where $r$ is the radius from the origin (set at the center of the Earth), $\theta$ is the co-latitude from the dipole axis, and $\nu$ is the azimuth around the dipolar axis. $\chi$ is a field-aligned coordinate. One can define the background magnetic field as $B_0 = M \nabla \chi$, where $M$ is a scalar defining the dipole strength. For the rest of this study, we will set $M=1$.

\begin{figure}
  \centering
  \includegraphics[width=3.375in]{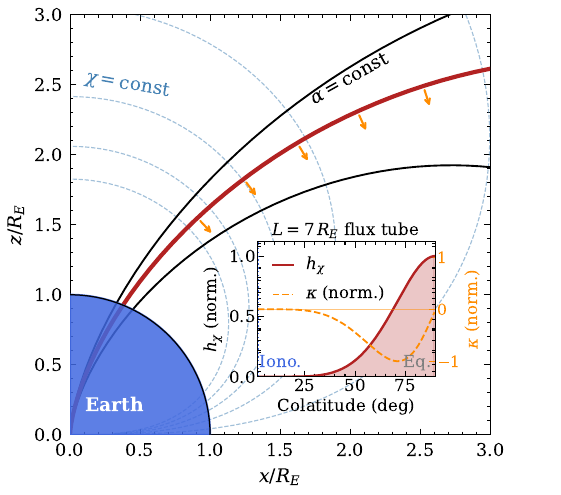}
  \caption{Diagram of dipole coordinate system. Lines of constant $\alpha$ and $\chi$ are shown. The magnetic field lines correspond to lines of constant $\alpha$, commonly known as L-shell. The inset plot shows the flux tube area ($h_\chi$) and curvature ($\kappa$) as a function of colatitude along the $L=7$ magnetic field line.}
  \label{fig:dipolediagram}
\end{figure}

The scale factors, also known as Lam\'e coefficients, are defined in dipole coordinates as \cite{kageyama_note_2006}. These can be interpreted as differential elements of flux tube area along each coordinate. Fig. \ref{fig:dipolediagram} shows the coordinate system and associated curvature ($\kappa$). One geometrical relation we wish to stress is that curvature of the field line and flux tube area are inherently related. The curvature is given by the gradients of the Lam\'e coefficients while the flux tube area is given by the Lam\'e coefficients themselves. Under a divergence free magnetic field, there is considerable structure that is imposed on the coordinate system.
\begin{align}
  h_\alpha &= {r^2}/{\left(\Theta \sin \theta\right)} \\
  h_\beta &= r \sin \theta \\
  h_\chi = h_X &= r^3/\Theta \\
  \Theta &= \sqrt{1+3\cos^2 \theta} \label{eqn:Theta def}
\end{align}

First, we will evaluate $\nabla \cdot \mathbf{B}_0$. The coordinate system has been defined such that the strength of the magnetic field is tied to the Lam\'e coefficients ($B_0 = 1/h_\chi$). This has the immediate consequence that
\begin{dmath}
  \frac{1}{B_0} \frac{\partial B_0}{\partial X} = -\frac{1}{h_\chi}\frac{\partial h_\chi}{\partial X} = -\left(\frac{1}{h_\alpha}\frac{\partial h_\alpha}{\partial X} + \frac{1}{h_\beta}\frac{\partial h_\beta}{\partial X}  \right). \label{eqn:divergence free coordinate system}
\end{dmath}
This is the same operation as (C 25) ($\nabla \cdot (\mathbf{B}_0+\mathbf{b})$) applied to just $\mathbf{B}_0$. Furthermore, we can recognize that that the $\chi$ Lam\'e coefficient is the flux tube area ($\mathcal{A}$), which makes (\ref{eqn:divergence free coordinate system}) a statement of flux tube area conservation. Solutions to it obey $\frac{\partial}{\partial X} \log \left(B_0 h_\chi \right)=\frac{\partial}{\partial X} \log \left(B_0 \mathcal{A} \right) = 0$. It is also a statement that the change in area of the perpendicular area of the coordinate system is balanced by the change in flux tube area.

(\ref{eqn:divergence free coordinate system}) has another consequence. \textit{The coordinate system itself is incompressible.} $h_\alpha h_\beta / h_\chi$ and $B_0 h_\alpha h_\beta$ are both constant. This is trivially verifiable with the dipole coordinate system. We wish to emphasize that this statement is true for \textit{any} magnetic field aligned coordinate system with flux tube area conservation simply because $\nabla \cdot \mathbf{B}_0 = 0$.
\subsection{Trivial Constraints}
Even under nontrivial geometry, the following are trivially true from the $\mathcal{O}(1)$ induction equations (C~5, 8, and 11) because the Lam\'e coefficients only depend on the long wavelength coordinates
\begin{align}
  \frac{\partial u_\alpha}{\partial \chi} &= 0 \\
  \frac{\partial u_\beta}{\partial \chi} &= 0 \\
  \frac{1}{h_\alpha}\frac{\partial u_\alpha}{\partial \alpha}+\frac{1}{h_\beta}\frac{\partial u_\beta}{\partial \beta} &= \nabla_\perp \cdot \mathbf{u}_\perp = 0. \label{eqn:O(1) induction chi}
\end{align}
This is by construction of our separation of long and short scales since it is assumed that the perpendicular fluid quantity perturbations are on scales smaller than the radii of curvature of the magnetic field and associated Lam\'e coefficients defined by the field geometry. Thus even in an arbitrary coordinate system under this expansion, the perpendicular velocity is locally incompressible in the perpendicular plane. (\ref{eqn:O(1) induction chi}) comes from an induction equation, not the continuity equation. The $\mathcal{O}(1)$ continuity equation (C 1) is
\begin{equation}
\nabla_\perp \cdot \mathbf{u}_\perp + \frac{1}{h_\chi}\frac{\partial u_\chi}{\partial \chi} = 0. \label{eqn:O(1) continuity eq}
\end{equation}
(\ref{eqn:O(1) continuity eq}) and (\ref{eqn:O(1) induction chi}) must both be satisfied by solutions that remain on the slowest time scale. Therefore, our subset of ideal MHD solutions must satisfy $\partial u_\chi / \partial \chi = 0$.

The $\mathcal{O}(\varepsilon^2)$ momentum equations (C~17, 21, and 24) can all be satisfied by either $\rho_1=0$ or a more complicated relation between the long scale derivatives and the components of $\mathbf{u}$. Solutions that satisfy any of these three constraints will be valid long time solutions. The $\rho_1 = 0$ case leads to reduced MHD and satisfies these equations, so we shall take $\rho_1 = 0$ as an additional constraint.

Using the constancy of $B_0$ geometrically corrected along a flux tube ($\partial(h_\chi B) /\partial A= 0$), the $\mathcal{O}(\varepsilon^{-1})$ momentum equations (C~14,18) yield the following constraints:
\begin{align}
  \frac{1}{h_\chi} \frac{\partial b_\alpha}{\partial \chi} &= \frac{1}{h_\alpha} \frac{\partial b_\chi}{\partial \alpha} \label{eqn:O(1/eps) mom alpha}\\
  \frac{1}{h_\chi} \frac{\partial b_\beta}{\partial \chi} &= \frac{1}{h_\beta}\frac{\partial b_\chi}{\partial \beta} \label{eqn:O(1/eps) mom beta}
\end{align}
These are statements that the perpendicular perturbed current is zero in both orthogonal directions. Even though there are derivatives of Lam\'e coefficients present at this order, they cancel with the variation of the background magnetic field. We will leave these relations as they are. They are used in deriving the pressure balance.

\subsection{Divergence Free Magnetic Field}
The derivation now proceeds to a second divergence free condition, one on the total magnetic field. The divergence-free condition on the total magnetic field yields two constraints, one from each order of $\varepsilon$ in $\nabla \cdot \mathbf{B} = 0$. The first happens at $\mathcal{O}(\varepsilon)$ (C25). It states that
\begin{dmath}
\frac{1}{B_0} \left(\frac{1}{h_\alpha}\frac{\partial b_\alpha}{\partial \alpha}+\frac{1}{h_\beta}\frac{\partial b_\beta}{\partial \beta}+\frac{1}{h_\chi}\frac{\partial b_\chi}{\partial \chi} \right) = -\frac{1}{h_\chi} \left(\frac{1}{B_0} \frac{\partial B_0}{\partial X} +\frac{1}{h_\alpha}\frac{\partial h_\alpha}{\partial X} + \frac{1}{h_\beta}\frac{\partial h_\beta}{\partial X} \right)=0. \label{eqn:O(eps) div B}
\end{dmath}
In the absence of a curved field line ($\partial h/\partial X = 0$), a homogeneous background magnetic field ($\partial B_0 / \partial X=0$), and no short wavelength perturbations of $b_\parallel = b_\chi$ along the field line ($\partial b_\chi/\partial \chi =0$), (\ref{eqn:O(eps) div B}) would reduce to $\nabla_\perp \cdot \mathbf{b} = 0$, which would allow writing $\mathbf{b}$ as the gradient of a scalar streamfunction. However, neither condition is trivially true for a curved magnetic field that has a changing value along the field line. What we find instead is that the right hand side of (\ref{eqn:O(eps) div B}) is zero by (\ref{eqn:divergence free coordinate system}). Therefore, the result of this is that on the short scales $\nabla \cdot \mathbf{b}=0$. For a generic discussion, see Appendix \ref{appendix:curvilinear constraint}. Under these circumstances, this can be written in short as
\begin{dmath}
\frac{\partial b_\chi}{\partial \chi} = - \nabla_\perp \cdot \mathbf{b} \label{eqn:O(eps) div B 2)}
\end{dmath}

The second constraint from $\nabla \cdot \mathbf{B} = 0$ is from the $\mathcal{O}(\varepsilon^2)$ expression (C26). Rewriting some of the curvature terms using (\ref{eqn:divergence free coordinate system}) and the incompressibility of the coordinate system, it states
\begin{dmath}
  \frac{\partial b_\chi}{\partial X} + \left(b_\alpha \frac{1}{h_\alpha}\frac{\partial}{\partial A} \log h_\chi h_\beta + b_\chi \frac{1}{B_0}\frac{\partial B_0}{\partial X} \right) = 0.
\end{dmath}
The long wavelength variations in $b_\chi$ are modified by the change in background magnetic field along and against the magnetic field lines. We shall leave these constraints here for now. We will eventually show that $b_\chi \neq b_\chi(\chi)$, which also shows that $\nabla_\perp \cdot \mathbf{b} = 0$, even under nonzero background field curvature.

\subsection{Continuity Equation}
Since $\rho_1$ is now zero, the $\mathcal{O}(\varepsilon)$ continuity equation (C2) is
\begin{dmath}
  \rho_0 \frac{1}{h_\chi}\frac{\partial u_\chi}{\partial X} + \frac{1}{h_\chi} u_\chi \left(\frac{\partial \rho_0}{\partial X} +\rho_0 \frac{1}{h_\alpha} \frac{\partial h_\alpha}{\partial X}+\rho_0 \frac{1}{h_\beta} \frac{\partial h_\beta}{\partial X} \right)+ \frac{1}{h_\alpha} u_\alpha \left(\frac{\partial \rho_0}{\partial A} + \rho_0 \frac{1}{h_\chi}\frac{\partial h_\chi}{\partial A}+\rho_0 \frac{1}{h_\beta}\frac{\partial h_\beta}{\partial A} \right )= 0. \label{eqn:O(eps) continuity}
\end{dmath}
In the Cartesian limit with a homogeneous $\rho_0$, one could conclude that $u_\chi$ must be a constant along $X$ as is done in \textit{ZM92}. In the curvilinear generalization, $\rho_0$ cannot be arbitrary and in fact must be inhomogeneous.

% Using (\ref{eqn:divergence free coordinate system}) from Sec. \ref{sec:divergence free coordinate system}, the second term of (\ref{eqn:O(eps) continuity}) can be interpreted physically. If the first and third terms are zero, then one possible solution is
% \begin{equation}
% \frac{1}{\rho_0} \frac{\partial \rho_0}{\partial X} - \frac{1}{B_0} \frac{\partial B_0}{\partial X} = \frac{\partial}{\partial X} \log \left(\rho_0/B_0\right) =0 \label{eqn:static parallel continuity}
% \end{equation}

% This corresponds to the well known property that plasma density scaled by flux tube area is conserved along a magnetic field line (\citet{gombosi_transport_1991} Eqn. 27b). As in \citet{khazanov_kinetic_2011}, one could imagine refining this estimate with corotation, ambipolar diffusion, and non-MHD effects such as high frequency waves, Coulomb collisions, and so forth. Essentially, $\rho_0$ must have long wavelength perturbations that are constant when multiplied by geometric factors along the flux tube.

The various terms in (\ref{eqn:O(eps) continuity}) can be independently zero, and hence the method of separation of variables can be used to solve the equation, if the Lam\'e coefficients within each partial derivative do not depend on the other variable. However, in this case they do have this dependence. Instead, we shall suppose a weaker condition that some of the Lam\'e derivatives are small. Under the following conditions, one can use separation of variables to solve (\ref{eqn:O(eps) continuity}).
\begin{equation}
  \label{eqn:lame_approx}
  \frac{1}{h_\alpha} \frac{\partial h_\chi}{\partial A} \ll 1 \hspace{2em}\text{and}\hspace{2em} \frac{1}{h_\chi h_\beta} \frac{\partial h_\beta}{\partial X} \ll 1.
\end{equation}

\begin{figure}
  \centering
  \includegraphics[width=5in]{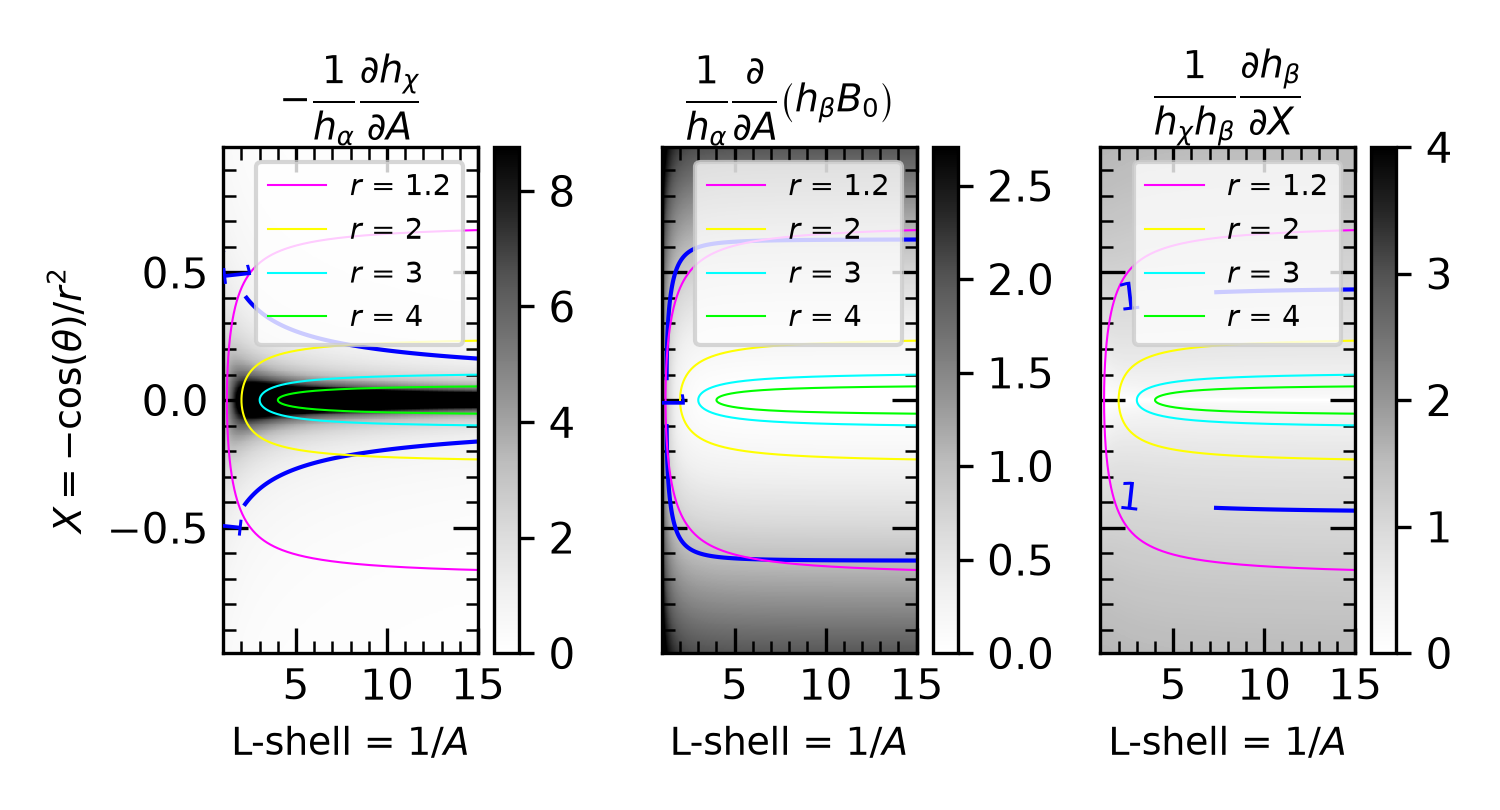}
  \caption{Curvature terms that are assumed to be small in this derivation. The grey shading indicates the magnitude of the term on the $A/X$ plane. The blue contour indicates where the quantity is unity. The colored contours indicate locations of constant spherical radius, $r$, in the $A/X$ plane. $X = 0$ is the equator. $X = \pm 1$ is the planet center. The curve $r=1.2$ approximates the surface.}
  \label{fig:neglect_curvature}
\end{figure}

To evaluate these assumptions, we shall directly compute these two quantities and plot them in the $A/X$ plane, as in Fig. \ref{fig:neglect_curvature}. The third quantity in the figure will be used later. In the region where $1.2<r<2$ and $L>2$, the first quantity is less than one. $\frac{1}{h_\chi h_\beta} \frac{\partial h_\beta}{\partial X}\ll 1$ for $r>1.5$. Therefore, (\ref{eqn:lame_approx}) is a reasonable assumption in the MI gap region. 

Using these assumptions, we will solve (\ref{eqn:O(eps) continuity}) by assuming the $\partial/\partial X$ and $\partial/\partial A$ terms can be satisfied independently. To begin with, the third term is zero if
\begin{equation}
\frac{\partial}{\partial A} \log \left(\rho_0 h_\chi h_\beta\right) =0. \label{eqn:cross field continuity}
\end{equation}
$h_\chi h_\beta$ is the scale factor of the differential area element cutting across field lines in the $\alpha$ direction (i.e. increasing L-shell in some parlance). In dipole coordinates this area is $h_\beta h_\chi = r^4 \sin \theta /\sqrt{1+3\cos^2 \theta}$. If $\rho^*$ is the value of the plasma density at the equator at a radius of $1 R_E$, then we find an expression for the density
\begin{equation}
  \frac{\rho_0}{\rho^*} = \frac{\sqrt{1+3\cos^2 \theta}}{r^4 \sin\theta}.
  \label{eqn:cross field result}
\end{equation}

Now, we solve for the condition that makes the first two terms of (\ref{eqn:O(eps) continuity}) zero. In this case, (\ref{eqn:O(eps) continuity}) becomes the following equilibrium condition
\begin{equation}
\frac{\partial}{\partial X} \log \left(\rho_0 u_\chi h_\chi \right) =0, \label{eqn:alt field aligned continuity X}
\end{equation}
assuming that $\partial u_\chi/\partial X \neq 0$. If $\partial u_\chi/\partial X = 0$, then the solution is
\begin{equation}
\frac{\partial}{\partial X} \log \left(\rho_0 h_\chi \right) =0. \label{eqn:static field aligned continuity X}
\end{equation}
These are statements that the parallel momentum flux along field a line is conserved. Finally, we combine these two conditions to find that the density is found by
\begin{equation}
  \label{eqn:density}
  \rho u_\chi h_\chi h_\beta \sim \text{const}.
\end{equation}

At this point, one may see the analogy between our RMHD derivation and fluid behavior of cold ionospheric plasma on closed magnetic flux tubes. The Earth’s plasmasphere is an example of a cold plasma on dipolar field lines \citep{lemaire_earths_1998}. In the completely refilled plasmasphere (with ionospheric source) under quiescent geomagnetic conditions this is approximately the density profile that has been measured experimentally. \citet{sheeley_empirical_2001} has empirical models developed from spacecraft observations that scale $\rho\sim r^{-3.6}, r^{-4}$, and $r^{-4.8}$ across field lines (the equatorial density profile) and are modulated by local time, geomagnetic activity, and other effects. From our model (\ref{eqn:cross field result}), $\rho(A) \sim (h_\beta h_\chi)^{-1} \sim r^{-4}$. Similar power law distributions of equatorial plasma density have been obtained with fluid-type specialized plasmaspheric model SAMI3 \citep{huba_modeling_2013}. While our goal is not to generate a plasmaspheric density model, we see that there is a correspondence between the constraints derived in this approach and steady-state behaviors of the background density observed in nature on closed dipolar flux tubes.

%In seeking a long time incompressible solutions to the ideal MHD equations, we find that in a dipole coordinate system aligned with the background magnetic field, \textit{there is only one permissible background density that prevents an equilibrium parallel flow}. That is, the validity of $u_\chi = 0$ and the choice of background density are tantamount linked to each other. In a straight magnetic field configuration, this stable configure is a homogeneous background, while in a dipole configuration, this background is more complex.

% Now show
Now, we shall take the time ($\partial / \partial t$) derivative of the the (\ref{eqn:O(eps) continuity}) under this solution.
\begin{dmath}
  \rho_0 \frac{1}{h_\chi}\frac{\partial ^2 u_\chi}{\partial X \partial t} + \frac{1}{h_\chi} \frac{\partial u_\chi}{\partial t} \left(\frac{\partial \rho_0}{\partial X} +\rho_0 \frac{1}{h_\chi} \frac{\partial h_\chi}{\partial X}\right) = 0.
\end{dmath}
This equation can be satisfied if $\partial u_\chi / \partial t=0$. This suggests that the parallel velocity must be static on the convective timescale due to setting $\rho_1 = 0$. In principle, if the equations were expanded under a fractional power of $\varepsilon$, this condition would be relaxed and the acoustic dynamics along the field line would be present on the convective timescale ($\partial u_\chi / \partial t \neq 0$, $\partial \rho_1 / \partial t \neq 0$) \citep{hunana_inhomogeneous_2010}. However, our motivation is to resolve the Alfv\'enic dynamics in the MI gap, so that particular refinement of this work is not necessary.   

\subsection{Pressure balance}
We now seek to show that these background equilibrium conditions imply a pressure balance within a flux tube. Under the assumptions derived thus far (including using (\ref{eqn:O(1/eps) mom alpha}) and (\ref{eqn:O(1/eps) mom beta}), $\partial u_\chi/\partial \chi=0$, $\partial u_\chi/\partial t=0$, and that the partial of $B_0$ balances the partial of $h_\chi$), the $\mathcal{O}(1)$ $\chi$ momentum equation (C22) can be written as
\begin{dmath}
\rho_0 (u_\perp \cdot \nabla_\perp) u_\chi   = -\frac{1}{h_\chi} \left(\frac{\partial p_1}{\partial \chi} + \frac{\partial p_0}{\partial X} \right). \label{eqn:O(1) chi mom}
\end{dmath}
Due to (\ref{eqn:O(1/eps) mom alpha}) and (\ref{eqn:O(1/eps) mom beta}), the $\hat{b}\cdot \nabla B$ term that is typically on the right hand side of the parallel momentum equation is zero. The ordering of $\mathbf{b}$ relative to $\mathbf{u}$ due to the Kreiss theorem causes the slow mode propagation coupling $u_\parallel$ to $b_\parallel$ to appear as a constraint at the next order. On slow time scales, $u_\chi$ is governed only by the pressure gradients.

We have already shown that $\partial u_\chi /\partial \chi = 0$. Therefore, (\ref{eqn:O(1) chi mom}) has the form of an equation where the left hand side does not depend on a coordinate ($\chi$), where the right hand side does. In order for this to be consistent, then the right hand side must not depend on the coordinate. This leads to the conclusion that either $\partial p_1 / \partial \chi =$ a constant or $\partial p_1 / \partial \chi =0$. We will take the solution that  $\partial p_1 / \partial \chi =0$.

Using these assumptions, we can now use (\ref{eqn:alt field aligned continuity X}) to provide some of the properties of $u_\chi$. We find that
\begin{dmath}
\rho_0 (u_\perp \cdot \nabla_\perp)  u_\chi  = -\frac{1}{h_\chi}\frac{\partial }{\partial X}\left(p_0\right). \label{eqn:O(1) chi mom 2}
\end{dmath}

Suppose now that $u_\chi$ has been chosen to match the density in the continuity equation. This is the same choice that gave (\ref{eqn:alt field aligned continuity X}). The $\partial/\partial X$ derivative can be taken for both sides yielding
\begin{equation}
\frac{\partial}{\partial X} \left( \frac{1}{h_\chi} \frac{\partial p_0}{\partial X} \right) = 0 \label{eqn:pressure potential constraint}
\end{equation}
This constraint is a byproduct of the low $\beta_p$ assumption. In order for the perturbation to not have any short wavelength field aligned perturbation, the background pressure must be negligible and geometrically adjusted constant. We have shown that $u_\chi$ has lost dependence on variables $u_\chi(\alpha,\beta,\chi,X,t) \to u_\chi(\alpha,\beta)$. Therefore, we can assume that $p_0 \to 0$ and $u_\chi \to 0$ because both variables are dynamically irrelevant. 

To proceed with the derivation, the $\mathcal{O}(\varepsilon)$ $\chi$ momentum equation (C23) is invoked and the constraints derived thus far, including the restrictions on the background quantities, are applied to it. When performing this reduction, the choice of seeking incompressible solutions ($\rho_1=0$) removes the time dependence and we are left with the following constraint
\begin{dmath}
 \frac{\partial}{\partial X}\left(\frac{b_{\alpha}^{2}+b_{\beta}^{2}}{2\mu_0} + p_1 \right) + \frac{\rho_0 u_\chi}{h_\alpha} \frac{\partial (h_\alpha u_\chi)}{\partial X} =\mathcal{K}_1/\mu_0 +\mathcal{K}_2.\label{eqn:pressure balance}
\end{dmath}
\begin{dmath}
\mathcal{K}_1 = -\frac{1}{h_\alpha} b_\alpha^2 \frac{\partial h_\alpha}{\partial X}+\frac{1}{h_\alpha}b_\alpha b_\chi \frac{\partial h_\chi}{\partial A} - \frac{1}{h_\beta} b_\beta^2 \frac{\partial h_\beta}{\partial X} \label{eqn:magnetic centrifugal pressure}
\end{dmath}
\begin{dmath}
\mathcal{K}_2 = \frac{1}{h_\alpha} \rho_0 u_\alpha^2 \frac{\partial h_\alpha}{\partial X}-\frac{1}{h_\alpha}\rho_0 u_\alpha u_\chi \frac{\partial h_\chi}{\partial A} + \frac{1}{h_\beta}\rho_0 u_\beta^2 \frac{\partial h_\beta}{\partial X} \label{eqn:hydrodynamic centrifugal pressure}
\end{dmath}
$\mathcal{K}_1$ and $\mathcal{K}_2$ are the contributions to this pressure balance from field line curvature. When $\mathcal{K}_1=\mu_0 \mathcal{K}_2=0$, this is equivalent to the slab RMHD pressure balance. They describe an effective centrifugal pressure per distance along the field line. Just as a spinning centrifugal pump that is used for chemical separation will apply a pressure to the liquid contents of its vial, a plasma flowing across the field line will feel a centrifugal force based on the location radius of curvature of the field line (the Lam\'e coefficient scale lengths). In the case that $b_\alpha$ and $u_\alpha$ are in phase with each other, as they are for small amplitude shear Alfv\'en waves, the right hand side of (\ref{eqn:pressure balance}) is zero, indicating that along a field line the magnetic centrifugal pressure cancels the hydrodynamic one. In fact, this is generally true, even without the polar region assumption. Terms of the same Lam\'e coefficient for $\mathcal{K}_1$ and $\mathcal{K}_2$ have opposite signs. Therefore, any Els\"asser type solution to the derived RMHD equations will see this type of cancellation.

\subsection{Parallel Induction Equation}
We now consider the dynamics of $b_\chi$. In the previous section, we showed that the dynamics of $u_\chi$ simplify substantially. Now we will apply the results of this derivation to this point to the equations of motion for $b_\chi$. First, the $\mathcal{O}(\varepsilon)$ $\chi$ induction equation with the curvature terms combined and canceled with the partials of $B_0$ is (C12)
\begin{dmath}
\left[\frac{\partial }{\partial t} +  (u_\perp \cdot \nabla_\perp)  \right] b_\chi  + \frac{1}{h_\chi} u_\chi \frac{\partial b_\chi}{\partial \chi} = (b_\perp \cdot \nabla_\perp) u_\chi - \underbrace{u_\alpha \frac{1}{h_\alpha} \frac{\partial}{\partial A} \left( h_\beta B_0 \right)}_{\nabla \cdot u}. \label{eqn:O(eps) chi induction}
\end{dmath}
There is a coupling between the parallel slow mode and the shear Alfv\'enic dynamics when the field line is curved due to the curvilinear terms in $\nabla \cdot \mathbf{u}$ even when $\rho_1=0$ \citep{dungey_hydromagnetic_1968,southwood_curvature_1985}. The second term on the right hand side is precisely this term. When the field line is straight, this term is zero. Since \ref{eqn:O(1) chi mom 2} has no dependence on $b_\chi$, the right hand side of \ref{eqn:O(eps) chi induction} are purely source terms that induce $b_\chi$.

More information is needed to make sense of (\ref{eqn:O(eps) chi induction}). At higher order and using the identity that $\partial(\log h_A + \log h_B)/\partial X = \partial/\partial X \log h_\chi$ (c.f. Sec. \ref{sec:divergence free coordinate system}) the $\mathcal{O}(\varepsilon^2)$ $\chi$ induction equation (C13) is 
\begin{dmath}
{b_\chi u_\chi} \frac{\partial}{\partial X} \log \left(B_0 b_\chi \right) + b_\alpha u_\chi \frac{1}{h_\alpha} \frac{\partial h_\chi}{\partial A} + b_\chi u_\alpha \frac{h_\chi}{h_\beta}\frac{\partial h_\beta}{\partial A} =0. \label{eqn:O(eps2) chi induction}
\end{dmath}
As written, this cannot be valid for an arbitrary coordinate system due to the aforementioned coupling between the slow and shear Alfv\'en mode when the field line is curved. At this point, we shall make an additional assumption that our equations are valid in a region of the dipole where $\frac{1}{h_\alpha} \frac{\partial h_\chi}{\partial A} \ll 1$ and $\frac{1}{h_\alpha} \frac{\partial}{\partial A} \left( h_\beta B_0 \right) \ll 1$. The first assumption was already used with the continuity equation. From Fig. \ref{fig:neglect_curvature} one can see that in almost all locations above the planetary surface, $\frac{1}{h_\alpha} \frac{\partial}{\partial A} \left( h_\beta B_0 \right) \ll 1$. $\frac{1}{h_\alpha} \frac{\partial h_\chi}{\partial A} \ll 1$ above the planetary surface, but becomes nonzero at the equator where curvature couples slow and Alfv\'en modes. We define our region as ionosphere-ward of this. \citet{david_generalized_2025} calls such a coordinate an ``expanded box'' in the context of solar wind turbulence. There can still be changes in flux tube area, but the flux tube does not curve faster than it expands.

Now suppose that $\frac{1}{h_\alpha} \frac{\partial h_\chi}{\partial A}=0$, which is substantiated outside of the equator in Fig. \ref{fig:neglect_curvature}. In this case, (\ref{eqn:O(eps2) chi induction}) is
\begin{dmath}
{b_\chi} u_\chi \frac{1}{B_0}\frac{\partial B_0}{\partial X}+u_\chi B_0 \frac{\partial b_\chi}{\partial X} + b_\chi u_\alpha \frac{h_\chi}{h_\beta}\frac{\partial h_\beta}{\partial A}=0.
\end{dmath}
$u_\chi$ and $u_\alpha$ are nonzero variables, and the terms in the parentheses do not naively cancel each other out. One solution to this constraint is that $b_\chi$ is zero. Following from this, (\ref{eqn:O(eps) chi induction}) is identically zero. We shall take this solution as the one for our reduced set of equations. \citet{davies_alfvenfast_2024} and \citet{davies_alfven_2025} performed studies of linear $b_\chi$ coupling to shear Alfv\'en modes including in a dipole geometry and found that the coupling is indeed small even under dipole field line geometry and $v_A$ gradients. These studies validate choosing the $b_\chi=0$ solution in the dipole region outside of the equator.    

\subsection{Reduced MHD Fluid Equations}
Consider the state of affairs of this point. All velocities have lost dependence on $\chi$. $u_\chi$ can be nonzero in a way that allows for an arbitrary $\rho_0$, but $u_\chi=0$. The pressure balance resembles an extension to the traditional RMHD pressure balance. The magnetic field perturbations have simplified to $b_\chi = 0$ and $\nabla_\perp \cdot \mathbf{b}_\perp=0$. If one examines the higher order induction and momentum equations, one will find them satisfied under the neglection of the curvature term. We can now derive the dynamical equations of motion.

Moving on to the momentum equations, under the previous assumptions, include $b_\chi=u_\chi=p_0=0$, the $\mathcal{O}(1)$ momentum equations in the $\alpha$ and $\beta$ directions (C~15 and 19) are
\begin{dmath}
 \rho_{0} \frac{\partial u_{\alpha}}{\partial t} + \underbrace{\frac{\rho_0}{h_{\alpha}} u_{\alpha} \frac{\partial u_{\alpha}}{\partial \alpha}+\frac{\rho_0}{h_{\beta}} u_{\beta} \frac{\partial u_{\alpha}}{\partial \beta}}_{\text{Advection}}  - \underbrace{\frac{1}{\mu_0}\left(B_{0}^2 \frac{\partial b_{\alpha}}{\partial X} +\frac{1}{h_{\alpha}} b_{\beta} \frac{\partial b_{\beta}}{\partial \alpha}- \frac{1}{h_{\beta}} b_{\beta} \frac{\partial b_{\alpha}}{\partial \beta} \right)}_{\text{Magnetic curvature + tension}}- \underbrace{\frac{1}{\mu_0 h_{\alpha}} B_{0}^2 b_{\alpha} \frac{\partial h_{\alpha}}{\partial X}}_{\text{Centrifugal}} + \underbrace{\frac{1}{h_{\alpha}} \frac{\partial p_{1}}{\partial \alpha}}_{\text{Pressure}} = 0 \label{eqn:momentum alpha}
\end{dmath}
\begin{dmath}
\rho_{0} \frac{\partial u_{\beta}}{\partial t}+\underbrace{\frac{1}{h_{\beta}} \rho_{0} u_{\beta} \frac{\partial u_{\beta}}{\partial \beta} + \frac{1}{h_{\alpha}} \rho_{0} u_{\alpha} \frac{\partial u_{\beta}}{\partial \alpha}}_{\text{Advection}} - \underbrace{\frac{1}{\mu_0}\left(B_{0}^2 \frac{\partial b_{\beta}}{\partial X} + \frac{1}{h_{\beta}} b_{\alpha} \frac{\partial b_{\alpha}}{\partial \beta} - \frac{1}{h_{\alpha}} b_{\alpha} \frac{\partial b_{\beta}}{\partial \alpha}\right)}_{\text{Magnetic curvature + tension}}- \underbrace{\frac{1}{\mu_0 h_{\beta}} B_{0}^2 b_{\beta} \frac{\partial h_{\beta}}{\partial X}}_{\text{Centrifugal}} + \underbrace{\frac{1}{h_{\beta}} \frac{\partial p_{1}}{\partial \beta}}_{\text{Pressure}} = 0 \label{eqn:momentum beta}
\end{dmath}

A similar process can be computed for the induction equation. The $\mathcal{O}(\varepsilon)$ $\alpha$ and $\beta$ induction equation components are (C~6 and 9)
\begin{equation}
\frac{\partial b_{\alpha}}{\partial t} + \underbrace{\frac{1}{h_{\beta}} u_{\beta} \frac{\partial b_{\alpha}}{\partial \beta} + \frac{1}{h_{\alpha}} u_{\alpha} \frac{\partial b_{\alpha}}{\partial \alpha}}_{\text{Advection}} - \underbrace{B_{0}^2 \frac{\partial u_{\alpha}}{\partial X} - \frac{1}{h_{\alpha}} b_{\alpha} \frac{\partial u_{\alpha}}{\partial \alpha} - \frac{1}{h_{\beta}} b_{\beta} \frac{\partial u_{\alpha}}{\partial \beta}}_{u \times B} + \underbrace{\frac{1}{h_{\alpha}} B_{0}^2 u_{\alpha} \frac{\partial h_{\alpha}}{\partial X}}_{\text{Centrifugal}} = 0 \label{eqn:O(1) induction alpha}
\end{equation}
\begin{equation}
  \frac{\partial b_{\beta}}{\partial t}+ \underbrace{\frac{1}{h_{\beta}} u_{\beta} \frac{\partial b_{\beta}}{\partial \beta} + \frac{1}{h_{\alpha}} u_{\alpha} \frac{\partial b_{\beta}}{\partial \alpha}}_{\text{Advection}} - \underbrace{B_{0}^2 \frac{\partial u_{\beta}}{\partial X}  - \frac{1}{h_{\alpha}} b_{\alpha} \frac{\partial u_{\beta}}{\partial \alpha} - \frac{1}{h_{\beta}} b_{\beta} \frac{\partial u_{\beta}}{\partial \beta}}_{u \times B}  + \underbrace{\frac{1}{h_{\beta}} B_{0}^2 u_{\beta} \frac{\partial h_{\beta}}{\partial X}}_{\text{Centrifugal}} = 0 \label{eqn:O(1) induction beta}
\end{equation}

We state the pressure balance from before, now with the curvature terms ($\mathcal{K}_1$ and $\mathcal{K}_2$) zero because of the relation $u_i = \pm b_i / \sqrt{\mu_0 \rho_0}$ , and neglect the centrifugal term from $u_\chi$ which is consistent with our assumption that $\frac{1}{h_\alpha} \frac{\partial h_\alpha}{\partial A} \ll 1$ in the area of interest. One can see that if $u_\chi=0$, this reverts back to the ``traditional'' RMHD pressure balance. 
\begin{dmath}
 \frac{\partial}{\partial X}\left(\frac{b_{\alpha}^{2}+b_{\beta}^{2}}{2\mu_0 } + p_1  + \frac{\rho_0}{2} u_\chi^2 \right) = \frac{\rho_0}{2} u_\chi^2 \frac{1}{B_0} \frac{\partial B_0}{\partial X}
\end{dmath}

These five equations subject to the following constraints are the equations of reduced MHD.
\begin{equation}
b_\chi = p_0 = \rho_1 =\frac{\partial}{\partial t} u_\chi=\frac{\partial}{\partial t} \rho_0=\frac{\partial}{\partial A} u_\chi = 0
\end{equation}
\begin{equation}
\frac{1}{h_\alpha} \frac{\partial h_\chi}{\partial A} = \frac{1}{h_\alpha} \frac{\partial}{\partial A} \left(h_\beta B_0\right) = \frac{1}{h_\beta h_\chi} \frac{\partial h_\beta}{\partial X} =0
\end{equation}
\begin{equation}
\mathbf{u}_\perp = \pm \mathbf{b}_\perp / \sqrt{\mu_0 \rho_0}
\end{equation}
\begin{equation}
\nabla_\perp \cdot \mathbf{b}_\perp = \nabla_\perp \cdot \mathbf{u}_\perp = 0
\end{equation}
Under the circumstance that $\rho_0 \to 1$, $u_\chi \to 0$, and $h_i \to 1$, these equations are identical to those in \textit{ZM92}. A later work by the same group, \citet{hunana_inhomogeneous_2010} generalized \textit{ZM92} to include inhomogeneities and these equations match their $\beta_p \ll 1$ case. However, their result neglected the pressure balance, which is indeed modified, nor solved for the conditions on the inhomogeneities that give rise to this particular ``locally incompressible'' system.

Aside from the standard RMHD terms, (\ref{eqn:momentum alpha}) - (\ref{eqn:O(1) induction beta}) also contain a centrifugal term, where the momentum and induction equations have opposite signs. Each product of $B_0/h_i \partial h_i / \partial X$ has units of one over length and can be identified as a radius of curvature. These can be simply understood through the frozen in condition. $B_0/h_\alpha \partial h_\alpha / \partial X$ represents the differential change in flux tube area in the $\alpha$ direction, likewise for the $\beta$ term. For more details on the relation between curvature and flux tube area see Appendix \ref{appendix:curvilinear constraint}. A change in $b_\alpha$ (or $b_\beta$) creates a $u_\alpha$ (or $u_\beta$) in order to conserve magnetic flux along streamlines due to the frozen in condition. However, as the streamlines of $\mathbf{u}$ become misaligned with the background field $\mathbf{B}_0$, there is an additional change in magnetic flux (the centrifugal term in (\ref{eqn:O(1) induction alpha})). Therefore, the perturbation in $b_\alpha$ (or $b_\beta$) need to compensate for two sources of magnetic flux change. Likewise, the term in (\ref{eqn:momentum alpha}) represents the incremental change in $u_\alpha$ from a $b_\alpha$ accounting for the flux tube area change. The momentum and induction equations have opposite signed centrifugal terms. This is because if the flux tube expands, i.e. $B_0/h_\alpha \partial h_\alpha / \partial X > 0$, additional $b_\alpha$ adds with the curvature. Likewise, for a fixed $u_\alpha$, positive $B_0/h_\alpha \partial h_\alpha / \partial X > 0$ requires less $b_\alpha$ to match $u_\alpha$.

A crucial aspect in (\ref{eqn:momentum alpha}) - (\ref{eqn:O(1) induction beta}) is the lack of centrifugal terms in the perpendicular operators. Full MHD in dipole coordinates maintains these. If these terms were maintained, $\mathbf{u}_\perp$ and $\mathbf{b}_\perp$ might become compressible or at least couple to the parallel flow dynamics in the perpendicular plane. Certainly the two potential reduction, addressed in the next section, would become modified or impossible. It is in the limit of short perpendicular wavelengths, shorter than the radii of curvature of the field line, that these gradients can be neglected. Along the field-aligned coordinate, the parallel wavelengths can be of the order of field tube expansion. Therefore, those centrifugal terms must be included in that case. 

\subsection{Two-potential reduction}
Although these equations are substantially simplified from MHD, there is a further simplification that can be performed. As this point, it is possible to derive a two-potential system of equations directly from (\ref{eqn:momentum alpha}), (\ref{eqn:momentum beta}), (\ref{eqn:O(1) induction alpha}), and (\ref{eqn:O(1) induction beta}) by defining a suitable set of perpendicular operators, Poisson brackets, and parallel operators.

We will start with a choice of potential streamfunctions that are defined as
\begin{align}
  \frac{1}{h_\beta} \frac{\partial \phi}{\partial \beta} = u_\alpha &,\hspace{2em}  -\frac{1}{h_\alpha} \frac{\partial \phi}{\partial \alpha} = u_\beta \\
  \frac{1}{h_\beta} \frac{\partial \psi}{\partial \beta} = b_\alpha / \mu_0 &,\hspace{2em}  -\frac{1}{h_\alpha} \frac{\partial \psi}{\partial \alpha} =b_\beta / \mu_0
\end{align}

From here, the procedure is to the form the equations of motion for the vorticity and parallel current. The $\chi$ component of $\partial \omega_\chi/\partial t$, where $\omega$ is the vorticity defined by $\omega = \nabla \times u$ is
\begin{dmath}
\frac{\partial \omega_\chi}{\partial t} = \frac{1}{h_\alpha h_\beta} \left[\frac{\partial (h_\alpha \partial u_\alpha/\partial t)}{\partial \beta} - \frac{\partial (h_\beta \partial u_\beta/\partial t)}{\partial \alpha} \right]
\end{dmath}

With vector fields of this kind, there are some general rules that will help reducing the four equations to two. Recall that for little $\alpha$ and $\beta$, the Lam\'e coefficients do not depend on them. Therefore, when dealing with these operations, a simpler approach can be taken. For any vector field where $w = \nabla \times (f \hat{e}_\chi)$ and $v = \nabla \times (g \hat{e}_\chi)$, where $f$ and $g$ are scalar streamfunctions, $\partial_i= \frac{1}{h_i} \frac{\partial}{\partial i}$, and subscripts indicate $f_\alpha = \partial_\alpha f$
\begin{dmath}
 \hat{e}_\chi \cdot \nabla \times (v \cdot \nabla) w = f_{\alpha \beta} (g_{\alpha \alpha} - g_{\beta \beta}) - g_{\alpha \beta} (f_{\alpha \alpha} - f_{\beta \beta}) - g_\beta \partial_\alpha \left(f_{\alpha \alpha} + f_{\beta \beta} \right) +  g_\alpha \partial_\beta \left(f_{\alpha \alpha} + f_{\beta \beta} \right) 
\end{dmath}
\begin{dmath}
 \hat{e}_\chi \cdot \nabla \times (v \cdot \nabla) v =\{g, \nabla_\perp^2 g \}
\end{dmath}
\begin{dmath}
 \hat{e}_\chi \cdot \nabla \times \left[(v \cdot \nabla) w -(w \cdot \nabla) v \right] =\nabla_\perp^2 \{f,g \} 
\end{dmath}
\begin{dmath}
\hat{e}_\chi \cdot \left(\nabla \times v \right)= - \nabla_\perp^2 g,
\end{dmath}
where
\begin{dmath}
 \{f, g\} = \frac{1}{h_\alpha h_\beta} \left(\frac{\partial f}{\partial \alpha} \frac{\partial g}{\partial \beta} - \frac{\partial f}{\partial \beta} \frac{\partial g}{\partial \alpha} \right) 
\end{dmath}
and
\begin{equation}
 \nabla_\perp^2 f = \frac{1}{h_\beta^2} \frac{\partial^2 f}{\partial \beta^2} + \frac{1}{h_\alpha^2} \frac{\partial^2 f}{\partial \alpha^2}= \partial_\alpha^2 f + \partial_\beta^2 f.
\end{equation}
At this point in the derivation, the Lam\'e coefficients and the perpendicular partials in the equation commute, so much of the algebra is synonymous with Cartesian geometry. The notable exception is that terms with parallel derivatives cannot be treated as simply. Those must be treated more carefully. A step by step simplification of the simplification of the parallel and centrifugal terms is in Appendix \ref{sec:two pot parallel terms}. When these substitutions are made, the equations of motion become
% \begin{widetext}
\begin{dmath}
  \rho_0\left(\frac{\partial \omega}{\partial t} + \{\phi,\omega\}\right) =  \mu_0 \{\psi,j_\parallel \} + {B_0^3} \frac{\partial}{\partial X} (B_0^{-1} j_\parallel)
   \label{eqn:vorticity final}
 \end{dmath}

\begin{equation}
  \frac{\partial j_\parallel}{\partial t} + \nabla_\perp^2 \{\phi,\psi \} = \frac{B_0}{\mu_0} \nabla_\perp^2 \frac{\partial \left(B_0 \phi \right)}{\partial X}
    \label{eqn:jpar final}
\end{equation}
  
\begin{equation}
\omega = \nabla_\perp^2 \phi, \hspace{1em} j_\parallel = \nabla_\perp^2 \psi
\end{equation}

(\ref{eqn:jpar final}) and (\ref{eqn:vorticity final}) are equivalent representations of (\ref{eqn:momentum alpha}) - (\ref{eqn:O(1) induction beta}) using the potentials $\phi$ and $\psi$. The dependence of $h_\alpha$, $h_\beta$, and $h_\chi$ on $X$ prevents the parallel propagation term from having a simple form. The field aligned terms have a factor of $B_0^{-1}$ under the $\partial/\partial X$ to show that they scale with flux tube area. The differing powers of $B_0$ inside the $\partial/\partial X$ terms come from the sign difference in the centrifugal terms with the parallel derivative in the induction equation.

The typical derivation of RMHD performs an additional step on the induction equation. (\ref{eqn:jpar final}) can be written with the Laplacian outside of some of the operators
\begin{dmath}
\nabla_\perp^2 \left(\frac{\partial \psi}{\partial t} + \{\phi,\psi \} - \frac{B_0}{\mu_0}\frac{\partial \left(B_0 \phi\right)}{\partial X} \right)=0. \label{eqn:jpar final redux}
\end{dmath}
The term inside the parentheses is typically taken as zero and presented as the second equation in RMHD.

\subsection{Els\"asser reduction}
In general, the momentum and induction equations can be combined into a set of forward and backward propagating Els\"asser variables that make clear the nature of turbulence generation. Using the identity

\begin{equation}
  \label{eqn:lap_pois_identity}
  \nabla_\perp^2 \{f,g\} = \{\nabla_\perp^2 f, g\} + \{f,\nabla_\perp^2 g\} + 2 \left( \{f_\alpha,g_\alpha\}+\{f_\beta,g_\beta\} \right),
\end{equation}
one can rewrite (\ref{eqn:vorticity final}) and (\ref{eqn:jpar final}) as
\begin{dmath}
  \label{eqn:elsasser_potential}
  \frac{\partial}{\partial t} \Omega^{\pm} + \{\zeta^{\mp},\Omega^{\pm}\} + \sum_{j=\alpha,\beta}\{\zeta^{\mp}_j,\zeta^{\pm}_j\} =  {v_A}\left[v_A B_0 \frac{\partial}{\partial X} \left(v_A^{-1} \frac{\Omega^+ - \Omega^-}{2}\right) \pm \nabla_\perp^2 \frac{\partial}{\partial X} \left(B_0 \frac{\zeta^+ + \zeta^-}{2}\right) \right]
\end{dmath}
where $\zeta^{\pm} = \phi \pm \sqrt{\frac{\mu_0}{\rho_0}} \psi$ and $\Omega^{\pm} = \nabla_\perp^2 \zeta^{\pm} = \omega \pm \sqrt{\frac{\mu_0}{\rho_0}} j_\parallel$. The typical Els\"asser variables are defined as $\mathbf{z}_\perp^{\pm} = \nabla \times (\zeta^{\pm} \hat{e}_\chi)$. (\ref{eqn:elsasser_potential}) fully describes the dynamics of the system including reflections and source terms due to inhomogeneous Alfv\'en speeds and flux tube areas. This inhomogeneous nonlinear extension retains the essence of MHD turbulence, that the nonlinear terms only depend on the interaction between counterpropagating wavepackets. However, it also has the addition of source terms from the inhomogeneous background.

\section{Numerical Implementation}
\label{sec:numerical_impl}
Now we summarize the numerical implementation of the RMHD equations. To validate the analytical solutions, we solve the axisymmetric two dimensional RMHD equations in the $(\alpha,X)$ plane:
\begin{equation}
    \begin{aligned}
        \partial_t\omega
        &=
        \frac{v_A^2}{h_\chi}\,\partial_\chi\mathcal{J}+S_\omega,\\
        \partial_t\mathcal{J}
        &=
        \frac{1}{h_\chi}\,\partial_\chi\omega
        -
        \frac{1}{h_\chi}
        \partial_\chi\ln\left(\frac{h_\beta}{h_\alpha}\right)\omega,\\
        \frac{1}{h_\alpha^2}\,\partial_\alpha^2\phi
        &=\omega,\\
        \frac{1}{h_\alpha^2}\,\partial_\alpha^2\psi
        &=\mathcal{J},
    \end{aligned}
\end{equation}
where we have taken the axisymmetric perpendicular Laplacian as $(1/h_\alpha^2)\partial_\alpha^2$ (with $\partial_\beta = 0$). $\omega$, $\phi$, and $\psi$ are the vorticity, velocity stream function, and magnetic flux function, respectively, but we take $\mathcal{J} = j_\parallel/B_0$ as the parallel current density weighted by the background field. 
The background Alfvén speed is defined as $v_A^2 = B_0^2/\rho_0$ (in code units; defined below), and $S_\omega$ is the driving, time-dependent source term for the vorticity.
Since the axisymmetric system has no $\beta$ dependence, the nonlinear Poisson brackets vanish and the resulting system is linear.

We simulate the system using a solver that calls the Dedalus spectral code \citep{burns_dedalus_2020}, 
building the governing equations as a Dedalus initial-value problem in $(\alpha,X)$. 
We supply all metric coefficients as fixed fields, so that Dedalus constructs the spectral operators, assembles the linear systems, and advances the solution in time. 
A custom Python layer generates the grid, pre-computes the scale factors, background parameters ($B_0$, $\rho_0$) and their derivatives in closed form, supplies them as fixed coefficient fields, and constructs the time-dependent source.
This is done so that no scale factor or metric derivative is ever spectrally differentiated: each coefficient enters the discretization in its exact closed form, truncated only to the grid's modes.
In the Dedalus problem setup, both the metric term proportional to $\partial_\chi\ln(h_\beta / h_\alpha)$ and the source $S_\omega$ are treated explicitly while all other terms are treated implicitly.

For the grid spacing, we use Chebyshev spectral bases for the non-periodic $\alpha$ and $X$ coordinates.
Since each dipole field line has a different arc length, we normalize $X$ such that $-1$ is the north ionosphere footpoint and 0 is the equatorial boundary ($+1$ would be the southern footpoint) in order to maintain a rectangular domain.
Dedalus's third-order, four-stage implicit-explicit
Runge-Kutta scheme (RK443) is used for time stepping.

Since each implicit stage requires solving a large linear system, we accelerate the solves using the CuPy Python library \citep{okuta_cupy_2017}. We implement a class fitting the Dedalus solver interface and replace the scipy LU factorization and solve calls with their CuPy equivalents (which call NVIDIA cuSOLVER matrix functions).
The factorization is computed once and reused at every step since the system matrix does not change (as long as the timestep is constant).

The benchmark we employ imitates \citet{davies_alfven_2025}, which launched Alfvén waves within a dipole.
First, we follow \citet{kageyama_note_2006} and \citet{woodroffe_ultra-low_2012,sakaki_convective_2024}

\begin{equation}
    \alpha=\frac{\sin^2\theta}{r}
\end{equation}
and the stretched field-aligned coordinate
\begin{equation}
    \chi=-\frac{\operatorname{arcsinh}(a u)}
             {\operatorname{arcsinh}(a)},
    \qquad
    u=\frac{\cos\theta/r^2}{\sqrt{1-\alpha}},
\end{equation}
covering 64 points over $\alpha\in[1/8,1/4]$ (corresponding to field lines $L=\alpha^{-1}=4$ to $8$), and 128 points over $\chi\in[-1,0]$.
The normalization by $\sqrt{1-\alpha}$ makes the coordinates non-orthogonal.
Following \citet{sakaki_convective_2024}, we neglect the resulting cross terms in $\nabla_\perp^2$, which involve $\partial_\chi$ and are small compared with the $\partial_\alpha$ terms under the flute ordering $k_\parallel\ll k_\perp$.

A perfectly conducting ionospheric boundary is imposed through $\omega=0$ at $\chi=-1$, and equatorial reflection is imposed through $\mathcal{J}=0$ at $\chi=0$. 
In $\alpha$, the potentials ($\phi, ~\psi$) are imposed to be zero at $L=4$ and $L=8$.

Although \citet{sakaki_convective_2024} used $a = 1000$ for their $L\approx 5.8$--$16.4$ domain (their Fig. 1), we choose $a = 300$ such that $h_\chi$ is more uniform along the field lines in $L=4$--$8$.
The Chebyshev nodes cluster at both ends of each field line, giving a physical spacing between adjacent nodes of order $10^{-3}\,R_E$ near the line ends and $0.1$--$0.2\,R_E$ near mid-line.
Relative to arc length, the grid has a mild footpoint concentration along the longer field lines ($L\gtrsim6$), where $h_\chi$ is smallest at the ionosphere.

All quantities are expressed in code units: lengths in $R_E$, the magnetic field in units of $B_{\mathrm{ref}}$, density in units of $\rho_{\mathrm{ref}}$, speeds in $v_{\mathrm{ref}}=B_{\mathrm{ref}}/\sqrt{\mu_0\rho_{\mathrm{ref}}}$, time in $t_0=R_E/v_{\mathrm{ref}}$, and vorticity in $1/t_0$.
The background field is the unit dipole,
\begin{equation}
    B_0=\frac{\sqrt{1+3\cos^2\theta}}{r^3},
\end{equation}
so that $B_0=1/216$ at the $L=6$ equator.
Two density profiles are considered: the constant-Alfv\'en-speed case, $\rho_0=B_0^2$, which gives $v_A=1$ throughout the domain, and the radially varying case, $\rho_0=C/r^2$. 
We choose $C=1/1296$ such that $v_A=1$ at the $L=6$ magnetic equator; this results in $v_A \approx 67$ at the footpoint.

For the Alfvén wave driving, although \citet{davies_alfven_2025} used an equatorial boundary condition, we here define the driving source term $S_\omega$:
\begin{equation}
    S_\omega(\alpha,\chi,t)
    =
    \frac{W(\alpha)}{h_\chi(\alpha,\chi)}
    g\!\left(\chi;\sigma_\chi(\alpha)\right)\tau(t),
\end{equation}
where
\begin{equation}
    g\!\left(\chi;\sigma_\chi\right)
    =
    \frac{1}{\sigma_\chi\sqrt{2\pi}}
    \exp\left(-\frac{\chi^2}{2\sigma_\chi^2}\right)
\end{equation}
is a unit-area Gaussian centered at the magnetic equator. 
Its width is defined by
\begin{equation}
    \sigma_\chi(\alpha)
    =
    \frac{\sigma_s}{h_{\chi,\mathrm{eq}}(\alpha)},
    \qquad
    \sigma_s\simeq6.3\times10^{-3}\,R_E,
\end{equation}
so that the source has a fixed physical thickness across field lines. 
$\sigma_s$ is chosen such that the sheet is marginally resolved by the simulation grid, with approximately five nodes within $\pm2\sigma_\chi$ near the ring center and at least four nodes on every field line within the ring.

The spatial strength $W(\alpha)$ is constructed from the ring vortex of \citet{davies_alfven_2025}.
Their ring potential is a $C^2$ step in the equatorial distance $R$ from the ring center,
\begin{equation}
    \psi_D(R)=
    \begin{cases}
        R_a, & 0\leq R<R_a,\\[4pt]
        R-\dfrac{R_b-R_a}{2\pi}\sin\left(2\pi\dfrac{R-R_a}{R_b-R_a}\right), & R_a\leq R\leq R_b,\\[8pt]
        R_b, & R>R_b,
    \end{cases}
\end{equation}
with inner edge $R_a=0.2$ and ring radius $R_b=1$, centered at $r_g=6$ in the equatorial plane.
Since this ring is localized in $\beta$, we retain only its axisymmetric component and define the stream function
\begin{equation}
    \phi_{m0}(\alpha)
    =
    \left\langle
    \left(\frac{L}{r_g}\right)^3
    \left[\psi_D(R)-R_b\right]
    \right\rangle_\beta,
    \qquad
    R^2=L^2+r_g^2-2Lr_g\cos\beta,
\end{equation}
where $\langle\cdot\rangle_\beta$ denotes the average over $\beta$.
The factor $(L/r_g)^3=B_0(r_g)/B_0(L)$, exact at the equator where $B_0=L^{-3}$, compensates for the $1/B_0$ in the ring's $\mathbf{E}\times\mathbf{B}$ drift, $|\mathbf{u}|=|\nabla_\perp\psi_D|/B_0$ (\citet{davies_alfven_2025}, Eqs.~6, 15--17), so that our stream function, with $\mathbf{u}_\perp=\hat{e}_\chi\times\nabla_\perp\phi$, reproduces the ring's equatorial flow to within the term dropped by the flute ordering, $\left[\psi_D-R_b\right]\nabla_\perp(L/r_g)^3$, which amounts to $\simeq20\%$ of the peak at this ring scale.
Subtracting $R_b$ ensures that $\phi_{m0}$ vanishes outside the ring, so that the source is confined to $L\in[5,7]$ and is never felt by the $\alpha$ boundaries.
The source strength is then
\begin{equation}
    W(\alpha)
    =
    2\,v_{A,\mathrm{eq}}(\alpha)\,\phi_0\,\omega_{m0}(\alpha),
    \qquad
    \omega_{m0}
    =
    \left.\frac{1}{h_\alpha^2}\,\partial_\alpha^2\phi_{m0}\right|_{\chi=0},
\end{equation}
where $v_{A,\mathrm{eq}}$ is the Alfv\'en speed at the equator of each field line.
Since a sheet source of strength $W$ launches pulses of amplitude $W/(2v_A)$ in each direction along the field line, the pulse launched toward the ionosphere leaves the source sheet carrying vorticity $\phi_0\,\omega_{m0}\,\tau(t)$.
The amplitude $\phi_0$ is chosen such that the peak equatorial flow speed is $|\mathbf{u}|_{\mathrm{peak}}=10^{-3}\,v_{A,\mathrm{eq}}(L_c)$, where $L_c\simeq5.96$ is the grid line nearest the ring center.
In the radially varying case, $v_A$ at the flow peak ($L\simeq6.56$) is $0.84$, so the local ratio there is $|\mathbf{u}|/v_A\simeq1.2\times10^{-3}$.

Finally, the temporal envelope $\tau$ is a single raised-cosine pulse,
\begin{equation}
    \tau(t)=
    \begin{cases}
        \dfrac{1}{2}
        \left[
        1-\cos\left(\dfrac{2\pi t}{t_p}\right)
        \right],
        & 0\leq t\leq t_p,\\[6pt]
        0, & \text{otherwise},
    \end{cases}
\end{equation}
with $t_p$ the pulse length. 
For the simulations presented here $t_p = 1$ and the step size $\Delta t = 0.01$. Runs are simulated for approximately two bounces of the Alfv\'en wavepacket, which differs for different Alfv\'en speed profiles.

\section{Analysis and Discussion}
The current approach to the MI coupling problem which motivated this derivation can be seen as a limit of the RMHD equations when $\partial \omega / \partial t = \{\phi,\omega\} = \{\psi,j_\parallel\}=0$. In this section, we shall establish this limit and then discuss what the neglected physics are and the ramification they have on MI coupling. The two leading coupled global models of the Earth's magnetosphere have an inner boundary at 2-3 $R_E$ radius where $j_\parallel$ is mapped to the ionosphere using this conservation property \citep{yu_including_2010,gasparini_new_2025,james_model_2026}. Under the conditions above $\partial/\partial X \left(j_\parallel / B_0 \right)=0$. Using this condition, the pattern of static $j_\parallel$ is rectified with the ionospheric conductivity and potential and forms a boundary condition for the MHD solver. (\ref{eqn:vorticity final}) shows that this solution neglects the mixing and time dependent vorticity associated with the field aligned current. During geomagnetic storms, the traditional approach captures the large scale dynamics but neglects small scale $j_\parallel$ features that appear in the MI gap region (e.g. see Fig. 7 in \citet{yu_modeling_2015}). Field aligned current from such global models most likely forms the outer scale of the turbulence that is happening locally, which explains this incomplete agreement between models and observations.

From the Els\"asser reduction, we can point out that there is a close connection between this boundary condition and energy transfer from the magnetosphere to the ionosphere. The parallel Poynting flux is given by
\begin{equation}
  \label{eqn:poyning_reduction}
  S_\parallel = \frac{E_\perp B_\perp}{\mu_0} = B_0 \nabla \phi \cdot \nabla \psi= \frac{B_0 \sqrt{\rho_0}}{2} \left( |\nabla \zeta^+|^2 - |\nabla \zeta^-|^2 \right)   
\end{equation}
If the conductivity at the ionosphere is used as the only source of wave reflection, this is equivalent to assume $|\zeta^+| = \mathcal{R} |\zeta^-|$, where $\mathcal{R}$ is a reflection coefficient between $0-1$ set \textit{only} by the conductivity of the ionosphere. This sets $S_\parallel\sim(1-\mathcal{R}^2)|\zeta^-|^2 $. Therefore, we can conclude that that wave reflectivity very directly controls energy input from the magnetosphere to the ionosphere. An imbalance in the incoming and outgoing Poynting flux is a core feature of geomagnetic storm and substorm behavior \citep{keiling_assessing_2019}. Therefore, understanding the wave reflection properties in the inhomogeneous context has a strong ramification on MI-coupling. One feature of the RMHD equations we will show is that wave reflection can come from inhomogeneities of the Alfv\'en speed, not just the boundary condition.

The energy in each Els\"asser variables is \citep{perez_direct_2013}
\begin{equation}
  E^\pm = \int \frac{1}{4} \rho_0 |\nabla_\perp \zeta^\pm|^2 h_\chi h_\alpha h_\beta d \chi d\alpha d\beta.
  \label{eqn:elsasser_energy}
\end{equation}
The total energy of the system is $E = E^+ + E^-$. In our analysis, it is useful to decompose the energy of the system in this way since it captures the Poynting flux into ($E^-$) and out of ($E^+$) the ionosphere. This can be directly compared to observations, such as \citet{keiling_assessing_2019, chaston_energy_2024}. The energy density, meaning the integrand of (\ref{eqn:elsasser_energy}), is proportional to the density adjusted Els\"asser potential amplitude.

\begin{figure}
  \centering
  \includegraphics[width=4.3in]{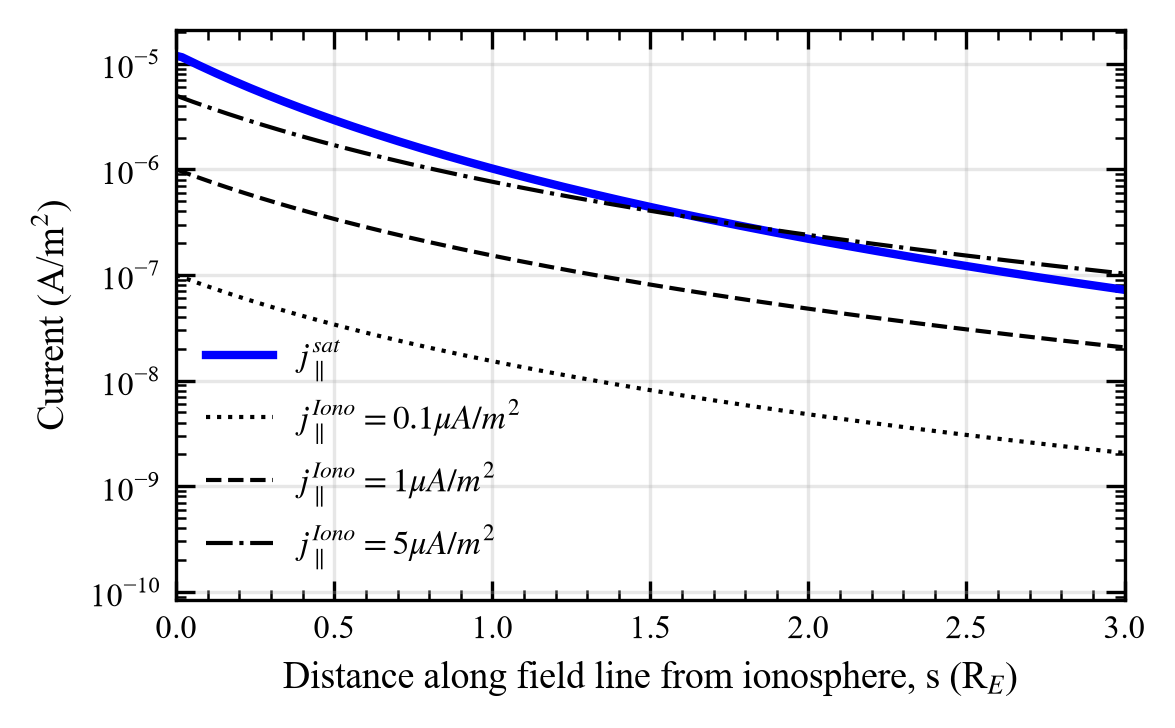}
  \caption{Saturation current for nonlinear wave-wave mixing to become important in RMHD equations. This value quantifies what is meant by a small amplitude solution. In this plot, the ionosphere starts at $s=0, r=1 R_E + 1000 \text{km}$. }
  \label{fig:saturation_current}
\end{figure}

The RMHD equations have very different behavior for large and small amplitude solutions. One method to quantify the threshold for large and small is to consider for what magnitudes of $\psi$, $\phi$, $\omega$, and $j_\parallel$ is the Poisson bracket term small relative to the parallel term. This condition happens when the following terms balance
\begin{equation}
  \{\psi, j_\parallel\} \sim \frac{B_0^3}{\mu_0} \frac{\partial}{\partial X} \left( \frac{j_\parallel}{B_0} \right).
  \label{eqn:saturation-balance}
\end{equation}
Assuming that $\{\psi, j_\parallel\}\sim k_\perp^2 \psi j_\parallel$ and $\frac{\partial}{\partial X} \left( \frac{j_\parallel}{B_0} \right) \sim \frac{j_\parallel}{B_0^2} \frac{\partial B_0}{\partial X}$ so the right hand size of (\ref{eqn:saturation-balance}) is $\frac{1}{\mu_0} j_\parallel B_0 \frac{\partial B_0}{\partial X}$, the current where the turbulence effects become important is approximately
\begin{equation}
  \label{eqn:saturation_current}
  j_\parallel^{sat} = \frac{1}{\mu_0} \frac{\partial B_0}{\partial s},
\end{equation}
where the independent variable has been changed to $s$. We will denote this quantity the ``saturation current density,'' because it describes an order of magnitude where linear wave behavior breaks down. For an L-shell of 9, this quantity is plotted for the dipole field in Fig. \ref{fig:saturation_current}. L-shells greater than 3 have nearly identical results. To compare with that, $j_\parallel\sim B_0$ profiles are plotted in dashed lines with values at the ionosphere of $0.1\mu \text{A}/\text{m}^2$, $1 \mu \text{A}/\text{m}^2$, and $5 \mu \text{A} / \text{m}^2$. These values are representative of FAC amplitudes during quiet times and different magnitude substorms respectively \cite{fu_distributions_2025}. The threshold $j_\parallel^{sat}$ will be lowest at the equator, however due to flux tube area conservation actual $j_\parallel$ profiles will be highest in amplitude at the ionosphere. Therefore, the point where the threshold $j_\parallel^{sat}$ is crossed will happen between the two. 

The 5 $\mu \text{A} / \text{m}^2$ FAC case crosses over the saturation current threshold predicted by (\ref{eqn:saturation_current}). The nonlinear term in the RMHD equations would then become important. These amplitudes are consistent with strongly driven substorms that result in local modifications to the convection pattern \cite{wang_substorm_2026}. 5 $\mu \text{A} / \text{m}^2$ is also consistent with the maximum observed FAC (4.6 $\mu \text{A} / \text{m}^2$) in the AMPERE data set \cite{carter_mean_2023}. The aforementioned SWMF--AMPERE comparison had a maximum FAC of 2.5 $\mu \text{A} / \text{m}^2$ \cite{yu_modeling_2015}. All of this suggests that turbulent energy dissipation could be a meaningful limitation on FAC amplitude during geomagnetic storms. The field line mapping approach used in global MHD models is substantiated during quiet times (low FAC), but becomes meaningfully modified during storm times. In further analysis, we shall use $j_\parallel^{sat}$ to delineate between large and small amplitude solutions.

\subsection{Single Wavepacket Propagation}
\label{sec:single wavepacket}
First, consider a single inward propagating wavepacket ($\zeta^+ = 0$). This is the most basic solution to the RMHD equations. Neglecting the nonlinear terms due to assuming a small amplitude ($j_\parallel < j_\parallel^{sat}$) (\ref{eqn:elsasser_potential}) reduces to

\begin{equation}
  \label{eqn:elsasser_potential_unidirectional}
  \frac{\partial \Omega^-}{\partial t} = {v_A}\left[v_A B_0 \frac{\partial}{\partial X} \left(v_A^{-1} {\Omega^-}\right) + \nabla_\perp^2 \frac{\partial}{\partial X} \left(B_0 {\zeta^-}\right)\right] , \hspace{1em} \nabla_\perp^2 \zeta^- = \Omega^-.
\end{equation}
The solution to this equation is
\begin{equation}
  \label{eqn:downward-uniform-solution}
  \zeta^- = \frac{C(X-B_0 v_A t)}{k_\perp(X)} \sqrt{\frac{v_A(X)}{B_0(X)}} e^{ i(k_\alpha \alpha + k_\beta \beta)},
\end{equation}
where $k_\perp = \sqrt{(k_\alpha/h_\alpha)^2 + (k_\beta/h_\beta)^2}$ and $C(X-B_0 v_A t)$ is some envelope function. This solution is the WKB solution for Alfv\'en waves. It depends on $\rho_0^{-1/4}(X)$, in agreement with \citep{barnes_theory_1992}. (\ref{eqn:poyning_reduction}) shows that $S_\parallel/B_0$ is constant for this solution, which corresponds to the flux tube area adjusted Poynting flux being constant. (\ref{eqn:elsasser_energy}) can be used to estimate the total energy of this system as $E^-(t) \propto \int C^2(u) du$, where $u = X-B_0 v_A t$. As expected, this is a constant. 

\subsection{Reflection Driven Turbulence}
\label{sec:rdt}
\begin{figure}
  \centering
  \includegraphics[width=4in]{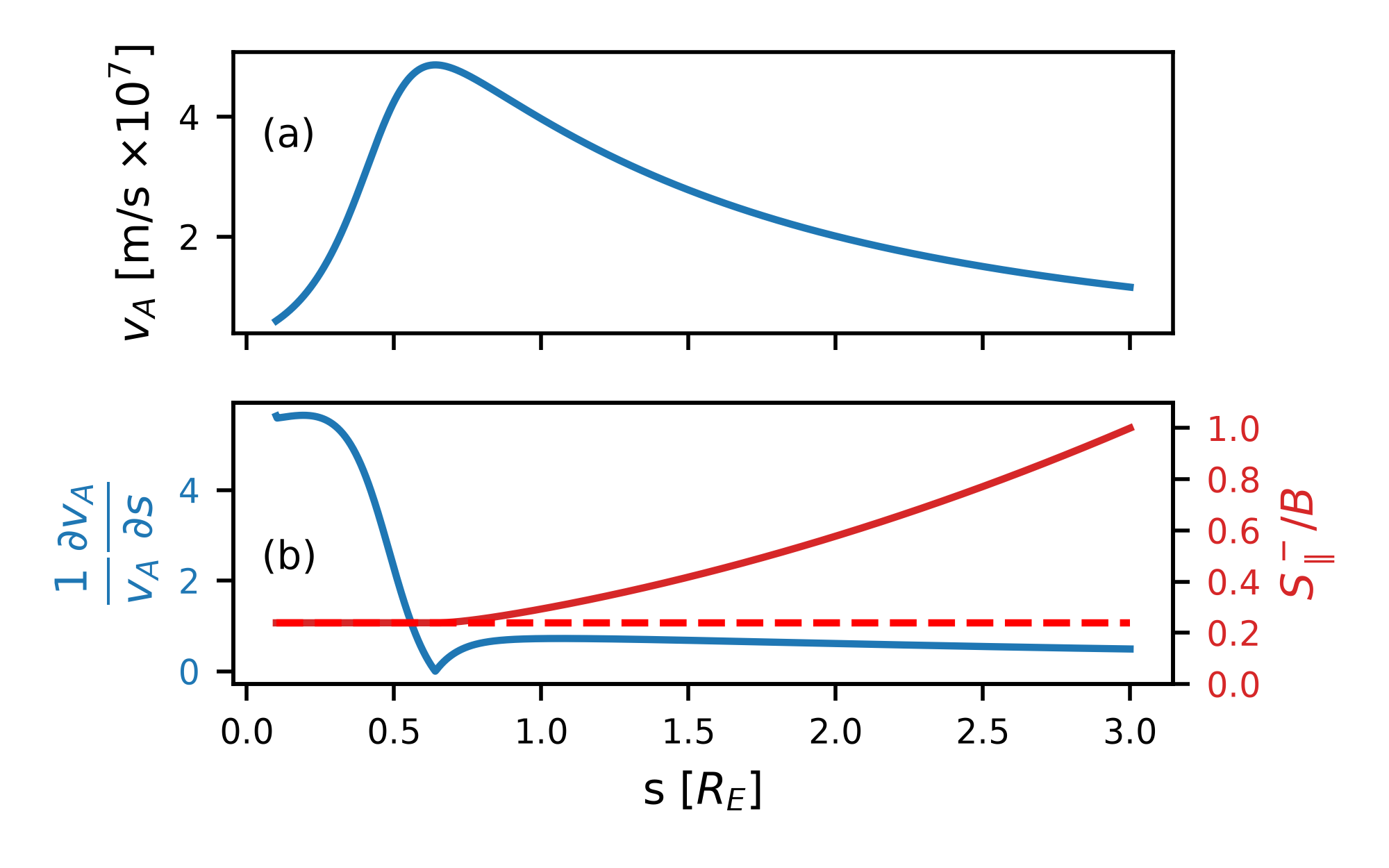}
  \caption{Wave attenuation of ionosphere propagating wave power via reflection driven turbulence near auroral field lines. $R$ indicates distance along the field line in Earth radii above the ionosphere ($s=0$). Wave power is normalized to its value at $s=3$. The dashed line is shows $S_\parallel^-/B$  at the Alfv\'en speed maxima. }
  \label{fig:rdt}
\end{figure}

Let us generalize the approach in the previous section to approximate the nonlinear terms in steady state. The effect of neglecting the nonlinear terms is to ignore wave-wave interactions that cascade energy to smaller perpendicular scales. One way to treat this simply is to assume that this cascade serves as a large-scale energy dissipation mechanism. Anywhere there is a large $\zeta^\pm \cdot \zeta^{\mp}$ amplitude, there is an effective dissipation of energy from that scale to smaller scales.
As discussed in \citep{heinemann_nonwkb_1980, squire_transport_2026}, the $\partial_X v_A^{-1} \zeta^- $ term in (\ref{eqn:elsasser_potential}) acts as a source of $\zeta^+$ which can interact with $\zeta^-$ to dissipate energy to smaller scales via turbulence even if there is no $\zeta^+$ originally present. Under these circumstances, we seek an approximate solution to (\ref{eqn:elsasser_potential}) under the circumstances that $\zeta^+ \ll \zeta^-$, $\zeta^+$ is generated by the inhomogeneities in the Alfv\'en speed, and any self-generated $\zeta^+$ is immediately cascaded to smaller scales. The amplitude of the inhomogeneity generated $\zeta^+$ is set by the balance $\frac{1}{2 \mu_0} v_A^2  B_0 \partial / \partial X \left(1/v_A\right) \Omega^- \sim  \{\zeta^+,\Omega^-\}$ in the $\zeta^+$ equation. For a single perpendicular wavelength driven $\Omega^-$, this gives an amplitude of $\zeta^+ \sim (v_A/k_\perp)^2 B_0 \partial/\partial X\left(1/v_A \right) / 2 /\mu_0$. In the $\zeta^-$ equation, the $\{\zeta^-, \Omega^+\}$ term becomes nonnegligible. However, we have written $\zeta^+$ as a function of $v_A$, so this can be substituted in. Therefore the $\Omega^+$ contribution to the $\Omega^-$ nonlinear term is

\begin{equation}
  \{\zeta^-, \Omega^+\} \sim - \frac{1}{2 \mu_0} k_\perp^2 \zeta^- B_0 \frac{\partial v_A}{\partial X},
  \label{eqn:rdt-assumption}
\end{equation}
where $k_\perp$ defines the outer scale of the turbulent cascade. Now including this term in (\ref{eqn:elsasser_potential_unidirectional}) revises the unidirectional propagation of a wave to satisfy the following equation with a third term that is the approximate turbulent dissipation of the wave due to Alfv\'en wave inhomogeneities. 
\begin{dmath}
  \label{eqn:elsasser_potential_unidirectional_rdt}
  v_A B_0 \frac{\partial}{\partial X} \left(v_A^{-1} {\Omega^-}\right) + \nabla_\perp^2 \frac{\partial}{\partial X} \left(B_0 {\zeta^-}\right) -k_\perp^2 \zeta^- B_0 \frac{1}{v_A}\frac{\partial v_A}{\partial X}=0.
\end{dmath}
This can be solved in a similar manner as above. One finds that the $v_A$ dependence of the solution has dropped out and the solution is
\begin{equation}
  \zeta^- = \frac{C(X)}{k_\perp(X)} B^{-1/2}_0(X)  e^{ i(k_\alpha \alpha + k_\beta \beta)}.
  \label{eqn:rdt_solution}
\end{equation}
This solution does not conserve Poynting flux in the same way as the WKB solution. In this case, $S_\parallel / B_0 \sim 1/v_A$. This indicates that as the Alfv\'en speed increases along the field line from the magnetosphere to the ionosphere more and more of the initial Alfv\'enic energy is turbulently dissipated. This quantified in Fig. \ref{fig:rdt}. The density model used in \citep{kletzing_electron_1994} and a dipole magnetic field strength is used to calculate the Alfv\'en speed. The red line in subplot b shows the fraction of wave energy that remains normalized to its energy at $s=3$ calculated using (\ref{eqn:poyning_reduction}) and (\ref{eqn:rdt_solution}). Since $v_A$ is not monotonic in this profile, below the Alfv\'en speed peak, the energy is assumed to be constant. Despite conservative modeling the region below the Alfv\'en maximum, by the time the wave is incident on the ionosphere its energy has been reduced to $1/5$ of its original value via turbulent dissipation.

Let us estimate the proportion of power in the flux tube that has been turbulently dissipated by this mechanism. The wave energy per unit length in a flux tube is
\begin{equation}
  \label{eqn:energy-per-unit-length}
  \frac{\partial E}{\partial s} = \frac{S_\parallel}{B_0 v_A}
\end{equation}
Thus, the proportion of total energy dissipated in a flux tube can be estimated to be
\begin{equation}
  \label{eqn:energy-tube}
  E/E_0 = \left(\int \frac{S^{\text{gradient}}_\parallel}{B_0 v_A} ds\right) {\Large{/}} \left(\int \frac{S^{\text{uniform}}_\parallel}{B_0 v_A} ds\right)
\end{equation}
Using (\ref{eqn:rdt_solution}) and (\ref{eqn:downward-uniform-solution}) to estimate the Poynting fluxes, assuming the constants are the same in both equation, and assuming that $v_A(s) = v_0 (s/s_0)^-p$ for some positive $p$ and initial speed $v_0$
\begin{equation}
  \label{eqn:energy-tube2}
  E/E_0 = \frac{\int_{s_0}^{s_{max}} v_A^{-2} ds}{\int_{s_0}^{s_{max}} v_A^{-1} ds} = \frac{p+1}{2p+1} \frac{\left(v_A(s_{max})/v_0\right)^{1+2p}-1}{\left(v_A(s_{max})/v_0\right)^{1+2p}+1}.
\end{equation}
For the profile in Fig. \ref{fig:rdt}, $\rho \sim s^{-1.55}$ and $B_0 \sim s^{-3}$, $v_A \sim s^{-2.225}$, and $v_A^{max}/v_0 \sim 4$, and therefore $E/E_0 \sim 0.59$. Thus, in steady state, we expect approximately $40\%$ of energy to be dissipated in that particular flux tube. In the limit of $p\to \infty$, the fraction of dissipated power asymptotes to $1/2$. 

In order to validate this analytical estimate, we conduct an idealized numerical experiment using the numerical implementation of (\ref{eqn:vorticity final}) and (\ref{eqn:jpar final}) discussed in Sec. \ref{sec:numerical_impl}. We simulate a single downward directed Alfv\'en wavepacket initialized at the equator under two circumstances: a uniform background $v_A$ and a $v_A$ which falls off with field line distance. In both cases, a reflecting boundary is used at the ionosphere and the equator. This is representative of a magnetosphere with a uniformly high conductivity ionosphere that is symmetrically mirrored in the northern and southern hemisphere. The simulation is only performed in two dimensions, it is assumed that $\partial/\partial \beta$ of each quantity is zero. Thus this experiment ignores any effects from the Poisson brackets, but includes the effect of Alfv\'en speed gradients, which is the specific effect we are testing.

\begin{figure}
  \centering
  \includegraphics[width=5.2in]{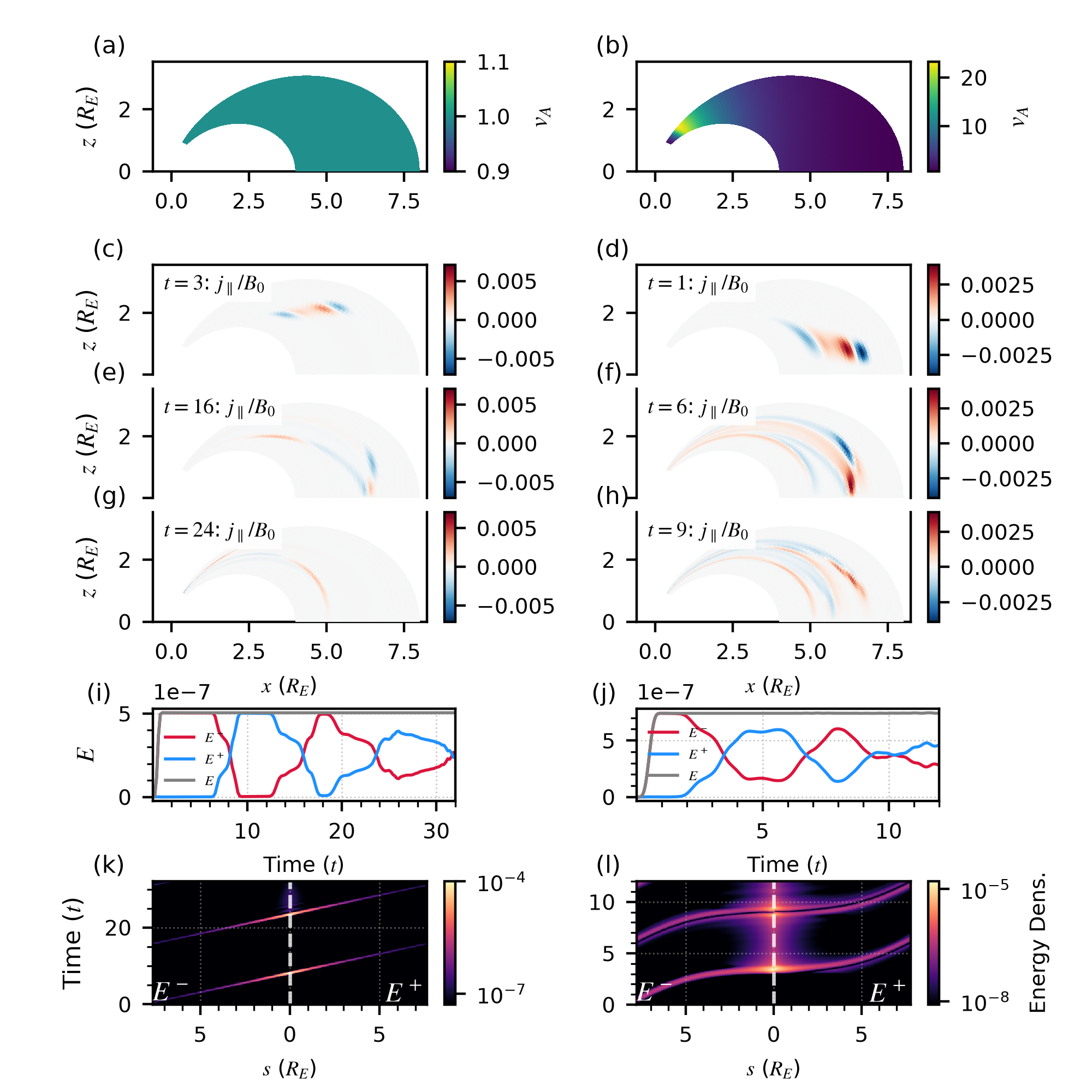}
  \caption{Numerical experiment demonstrating the effect of Alfv\'en speed gradients on the propagation of Alfv\'en waves in a dipole using the RMHD equations. The left columns has a uniform $v_A$ while the right column has a nonuniform $v_A$. The top row shows the normalized Alfv\'en speed. The second, third, and fourth rows show $j_\parallel/B_0$ at different time steps in the simulation. The second last row shows the energy in each direction of Els\"asser variable. The bottom column shows the energy density in each Els\"asser variable along the field line with maximum amplitude (fixed $\alpha$).}
  \label{fig:experiment_vagrad}
\end{figure}

The nonuniform Alfv\'en speed profile uses a density representative of auroral field lines. Specifically, we use a normalized version of the model in \citet{kletzing_electron_1994}. The model is
\begin{equation}
  \label{eqn:kletzing_density_model}
  n_K(s) = 6 \times 10^4 e^{-(s-318)/383} + 1.34 \times 10^7 s^{-1.55},
\end{equation}
where $s$ has units of kilometers, and $n_k$ has units of per cubic centimeter. Our normalized model used in Fig. \ref{fig:rdt} and the right column of Fig. \ref{fig:experiment_vagrad} is
\begin{equation}
  \label{eqn:normalized_kletzing_density}
  \rho_0(s) = n_K(s) / n_K(s= 1 R_E).
\end{equation}

The results of the simulation can be seen in Fig. \ref{fig:experiment_vagrad}. Fig. \ref{fig:experiment_vagrad}a and \ref{fig:experiment_vagrad}b show the Alfv\'en speed normalized to the Alfv\'en speed at 1 $R_E$. The left column has a uniform speed, while the right column has a nonuniform speed with the profile in Fig. \ref{fig:rdt}. These are idealized profiles for demonstrating the principle of gradient induced reflection. Figs.  \ref{fig:experiment_vagrad}c-h show snapshots of $j_\parallel/B_0$ at the beginning of the simulation, near the first bounce time, and near the end from top to bottom respectively. Each simulation has been run for approximately two bounce times (how long it takes a wave packet to go from equator, reflect off the ionosphere, and return to the equator). Due to the different Alfv\'en speed profiles, these bounce times are different between the two cases so the simulations have different end times. 

The numerical results are cast in Els\"asser variables $\Omega^+$ and $\Omega^-$. From each of these, the energy in each Els\"asser mode are estimated as \cite{perez_direct_2013}
\begin{equation}
  E^\pm = \int \frac{1}{4} \rho_0 h_\alpha^2 |\Omega^\pm|^2 h_\chi h_\alpha h_\beta d \chi d\alpha d\beta.
  \label{eqn:elsasser_energy_numerical}
\end{equation}
This expression for the energy is equivalent to the volume integral of (\ref{eqn:poyning_reduction}) if $k_\perp \sim h_\alpha$, hence $h_\alpha |\Omega^\pm| \sim |\nabla \zeta^\pm|$. For these two dimensional where the component of $k_\perp$ along $\beta$ has been ignored and $k_\alpha$ is constant, this is an exact comparison as shown in Sec. \ref{sec:single wavepacket}. The energy in each of these variables is shown in Figs. \ref{fig:experiment_vagrad}i and \ref{fig:experiment_vagrad}j. The sum of $E^+$ and $E^-$, shown as the grey line, is conserved. However, the proportion of energy in each mode is very different between the two cases. In the uniform $v_A$ profile case, the energy profile resembles square waves. The slope of the transition region comes from the time-of-flight dispersion from each field line having a different length, a feature which has been observed before \cite{davies_alfven_2025}. The nonuniform case generates reflections in the domain, not just at the boundary. The downward energy loses energy much more rapidly before it can reach the boundary.

The difference between Figs. \ref{fig:experiment_vagrad}i and \ref{fig:experiment_vagrad}j allows us to estimate the energy loss to Alfv\'en speed gradients. After the first bounce, all energy goes into $E^+$ in the uniform case. However, in the nonuniform case after the first bounce $E^+=6\times 10^{-7}$, $E^-=1.9 \times 10^{-7}$, and $E=7.9 \times 10^{-7}$. The leftover $E^-$ is not the original flux, but rather the result of many gradient reflections which would be dissipated if this simulation was 3d. From this, we can estimate that $\sim 24\%$ of the downward wave energy would be lost to gradient driven reflections in one transit of this field line. The value for $a$ in Fig. \ref{fig:rdt} is 0.24, which matches our numerical simulation.

The dynamics of this process can be seen in further detail in Figs. \ref{fig:experiment_vagrad}k and \ref{fig:experiment_vagrad}l. The energy density (integrand of (\ref{eqn:elsasser_energy_numerical})) along the maximum amplitude field line is shown in time and field line coordinate. The upward energy plot is flipped about $X$ and shown next to the downward energy one. Because the reflecting boundary condition exists on both sides, one can clearly see that $E^-$ energy becomes $E^+$ energy when it encounters a boundary. The main wavepacket is visible in both uniform and nonuniform cases. It has an amplitude that scales with flux tube area. In the nonuniform case, there are shaded regions between the main wavepacket. These are gradient induced reflections. One can see that they maximize in the region near the ionosphere, where the gradient is strongest. There is also a small first reflection that propagates alongside the main wavepacket near the equator. One can see that it was created at $\sim t=3$ when the initial wavepacket first entered the Alfv\'en speed gradient.

\begin{figure}
  \centering
  \includegraphics[width=4in]{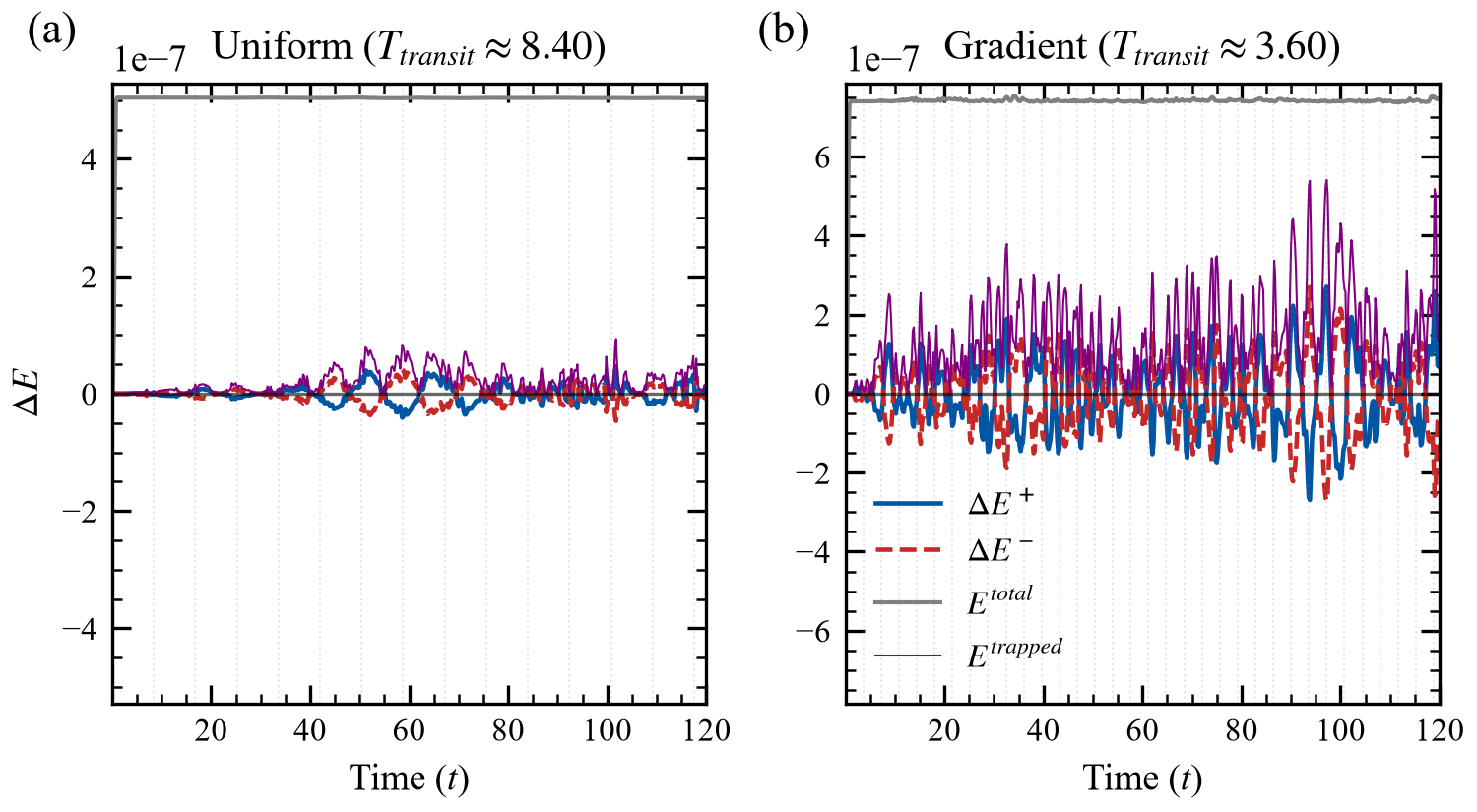}
  \caption{Deviations in Els\"asser energies from those predicted by WKB theory, i.e. (\ref{eqn:wkb-energy-down}) - (\ref{eqn:transit-time}). Simulations are run for many bounce times to observe the long term energy transfer. Vertical dashed lines indicate intervals of average bounce time.}
  \label{fig:iar}
\end{figure}

In Fig. \ref{fig:experiment_vagrad}l, the energy density has a tendency to linger near the ionosphere due to the contribution of Alfv\'en speed gradients. Previous studies have proposed the existence of an ionospheric Alfv\'enic resonator, wherein the localized region of high Alfv\'en velocity forms a resonant cavity \cite{lysak_generalized_1993, sydorenko_nonlinear_2008, woodroffe_ultra-low_2012, potapov_morphology_2021,lysak_kinetic_2023}. Alfv\'en speed gradients are also considered to be the source of the resonator, although they are not typically modeled in an Els\"asser variable formulation. To examine our results for evidence of such a phenomena, consider Fig. \ref{fig:iar}. It shows the Els\"asser energy in the simulation with the energy predicted by WKB theory present. If Alfv\'en wavepackets were modeled by WKB theory with no gradient driven reflection the energy density integrated along each field line is
\begin{equation}
  \label{eqn:wkb-energy-down}
  \mathcal{P}^-(\alpha) =\int U^- h_\chi dX= \sum_{n=0}^{\infty} (-1)^{n} \mathcal{P}_0(t - nT) \, \mathcal{H}(t - nT),
\end{equation}
\begin{equation}
  \label{eqn:wkb-energy-up}
  \mathcal{P}^+(\alpha) = \int U^+ h_\chi dX = \sum_{n=0}^{\infty} (-1)^{n-1} \mathcal{P}_0(t - nT) \, \mathcal{H}(t - nT),
\end{equation}
where
\begin{equation}
  \label{eqn:transit-time}
  T(\alpha) = \int \frac{1}{v_A} ds,
\end{equation}
is the transit time calculated per field line, $\mathcal{P}_0$ is the total energy per field line at time $t$, and $\mathcal{H}$ is the Heaviside step function. Integrating over $\alpha$, the energy is $E^\pm_{WKB} = \int \mathcal{P}^\pm h_\alpha d\alpha$. This approximates the alternating step functions which have a field line distance dispersion in Fig. \ref{fig:experiment_vagrad}i. In Fig. \ref{fig:iar}a, $\Delta E$ is an order of magnitude less than than the simulation, so this is a good approximation in the uniform $v_A$ case. In the nonuniform case, there is a considerable deviation from this simple bouncing picture. Roughly every bounce time, $\Delta E$ jumps up and then slowly decays. This coincides with when the main wavepacket interacts with the Alfv\'en speed gradient. 

Fig. \ref{fig:iar} demonstrates how an ionospheric Alfv\'enic resonator could work in RMHD. Wavepackets lose a proportion of their energy in regions of $v_A$ gradients to reflections. These reflections rapidly generate other reflections, so the lost energy remains localized near the gradient. Every time the main wavepacket transits the resonator, another proportion of its energy is lost to energizing the resonator. The proportion of energy lost to the resonator depends on the wavelength of the wavepacket relative to the length scale of the Alfv\'en speed gradient, so the proportion of energy lost in any given transit is highly variable. As seen from Fig. \ref{fig:iar}b, the loss efficiency varies from 10-70\% of the initial energy.

Our initial estimate from steady state reflection driven turbulence phenomenology using (\ref{eqn:energy-tube2}) was a 40\% energy loss. The simulations demonstrate that the time dependent analog of this analysis which includes the entire reflection coefficient, but neglects the nonlinear cascade shows a highly variable reflective efficiency. When averaged over time, the mean trapped energy in Fig. \ref{fig:iar}b is 18\% of the total. While neither method fully captures the precise magnitude of the energy interplay between the initial wavepacket and the proportion which becomes reflected and dissipated by the speed gradients, they agree that reflection driven turbulence can absorb sizable fractions of incident wave power.

\subsection{Revised Boundary Condition}
Given this analysis, it becomes clear that the assumption that $j_\parallel \sim B$ ignores turbulent energy transfer. A steady $j_\parallel$ is composed of $\Omega^+$ and $\Omega^-$, both of which could lose energy to turbulence. It should be noted that all models use $j_\parallel$ to both calculate Joule heating and particle precipitation inputs to the ionosphere without accounting for the energy lost to either mechanism \citep{zhang_electron_2015,mukhopadhyay_global_2022}. This is a related problem where the steady state $j_\parallel$ is used to drive other phenomena without removing the energy from $j_\parallel$.

There is also a known difference in ionospheric energy input and $j_\parallel$ structuring that depends on high vs coarse spatial resolution runs of the same code \citep{gunzkofer_impact_2026}. The higher resolution runs more finely resolve mesoscale structures which reduces Joule heating by localizing strong $j_\parallel$. We interpret this result as partially resolving the effect which we are describing.

Although a full time dependent simulation is necessary to validate these, we propose a minimal modification to the inner boundary of global MHD models under quasi-steady state conditions that accounts for this energy transfer from the large scale to small scale $j_\parallel$. Note, the ionospheric physics that constitute the lower boundary are assumed to be in steady state. At high enough time resolutions, this assumption too will break down \citep{laundal_global_2025}. Suppose that after many Alfv\'en transit times, the downward flux is reduced a factor $a$ and upward flux by $b$ such that

\begin{equation}
  \frac{\Omega^-_I}{B_I} = a\frac{\Omega^-_M}{B_M}, \hspace{0.5em} \text{  and  } \hspace{0.5em} \frac{\Omega^+_M}{B_M} = b \frac{\Omega^+_I}{B_I},
\end{equation}
where $a = \sqrt{v_A^I / v_A^*}$ and $b = \sqrt{v_A^M / v_A^*}$, $v_A^*$ is the maximum Alfv\'en speed along a field line, $v_A^I$ is the Alfv\'en speed at the ionosphere and $v_A^M$ is the Alfv\'en speed at the magnetosphere. The scaling of these ratios as the square root of the difference in Alfv\'en speed comes from (\ref{eqn:rdt_solution}), which is itself a steady-state solution to the RMHD equations under a reflection driven turbulence assumption (\ref{eqn:rdt-assumption}). The values of $a$ and $b$ vary from field line to field line depending on the magnetic field and density profile. For the profile in Fig. \ref{fig:rdt}, $a\sim 0.24$ and $b \sim 0.12$. Adding and subtracting $\Omega^+$ and $\Omega^-$ yields

\begin{equation}
  j_\parallel^I = \frac{(\frac{1}{b} + a)}{2} \frac{B_I}{B_M} \left[j_\parallel^M + \sqrt{\frac{\rho_I}{\mu_0}} \frac{(\frac{1}{b} - a)}{(\frac{1}{b} + a)} \omega^M \right].
  \label{eqn:fac_transfer}
\end{equation}
when $a=b=1$, indicating that there is no turbulent dissipation, this returns to $j_\parallel \sim B$. this $j_\parallel$ is input to some steady-state ionospheric potential solver (i.e. Eqn. 1 of \citep{zhang_electron_2015}) to get $\phi^I$. This is then projected back up to the magnetosphere via
\begin{equation}
    \phi^M = \frac{(b+ \frac{1}{a})}{2} \frac{B_M}{B_I} \left[\phi^I + \frac{(b-\frac{1}{a} )}{(b+\frac{1}{a})} \sqrt{\frac{\mu_0}{\rho^I}} \psi^I \right].
 \label{eqn:phi_I2M}
\end{equation}
Taken together, (\ref{eqn:fac_transfer}) and (\ref{eqn:phi_I2M}) are sufficient to substitute into traditional global magnetosphere models and an electrostatic ionospheric potential solver.  It is now required that $\omega^M$ and $\psi^I$ be taken into account when doing this mapping. $a$ and $b$ must be generated from empirical density magnetic field models of the MI gap. The density will depend strongly on the outflow characteristics of the system, photoionization, recombination, and other cold plasma sinks and sources. Despite these shortcomings, this approach is an improvement on the current state of affairs.

\subsection{Future Work and Comparison to other MI Coupling Models}
Future developments can be done in a few different directions. First, ionospheric conductivity is not resolved in the model so Ohmic dissipation of power not included. Secondly, we note that current implementation of RMHD is based on isotropic ideal single fluid MHD equations and therefore lacks mechanisms for parallel electric field formation, a crucial piece of physics in this region. The development of turbulence is probably connected to the formation of parallel electric fields since many particle precipitation events are associated with kinetic scale Alfv\'en waves \citep{stasiewicz_small_2000,stasiewicz_identification_2000,chaston_how_2007,chaston_auroral_2021}. It is important to contextualize where RMHD is valid. At Earth, the short scales in our model ($\alpha, \beta$) are smaller than the radii of curvature of the field line (potentially several $R_E$) but larger than kinetic scales (sub km). Thus there is approximately three decades of validity from $\sim 1-1000$ km. Turbulence formation at field line scales may be the first step for energy in order to form kinetic scale perpendicular scale lengths to ultimately dissipate energy into the particles. One can envision an extension of this work by developing multi-fluid and/or anisotropic RMHD formulation to describe inertial Alfv\'en waves and mirror force with Knight-type quasistationary potential drop \citep{kataoka_small-scale_2021}. This would allow to introduce different electron acceleration mechanisms that produce both Alfv\'enic and quasistationary auroral forms, and, importantly, interactions between those. Examples of such approach that involve extended RMHD in the slab geometry could be found in \citet{seyler_mathematical_1990} and \citet{chaston_smallscale_2010}.  Calculation of electron precipitation will allow to include the ionospheric feedback on RMHD turbulence development. Finally, coupling of RMHD with global codes will allow to provide the solution in the MI gap region, and understand turbulence effects for the magnetosphere-ionosphere coupling.

There a rich literature investigating the feedback instability as a turbulence generator via a secondary Kelvin-Helmholtz instability. This work could enhance or modulate these results. \citet{watanabe_feedback_2010} derives a similar set of RMHD equations with an equivalent normalization that can be seen by substituting $\phi \to B_0 \phi$ and $\psi\to  B_0 \psi$ into (\ref{eqn:vorticity final}) and (\ref{eqn:jpar final}). \citet{sakaki_convective_2024} states the equations from \citet{watanabe_feedback_2010} in a dipole coordinate system, although then simulates their instability with the linearized versions which neglects any turbulence generation by Alfv\'en speed inhomogeneities. In linear models, the feedback instability is enhanced by speed gradient effects \citep{lysak_feedback_1991}, so one would expect the turbulence generation to also be modified. Most recently, \citet{sakaki_nonlinear_2026} demonstrated the growth of vortex structures consistent with auroral curls in simulations. Observations of auroral turbulence from incoherent scatter radars have noted that turbulence appears to be generated in the magnetosphere \citep{ivarsen_turbulence_2024}. This suggests that the turbulence generation occurs in the magnetosphere, and not uniformly along the field line as suggested by these studies. 

In a situation where all nonlinear terms in (\ref{eqn:vorticity final}) and (\ref{eqn:jpar final}) are zero, which is equivalent  neglecting the Poisson brackets because the magnitude of $\phi$ and $\psi$ are small, the equations of motion reduce to
\begin{equation}
\frac{\partial \omega}{\partial t} = \mu_0 v_A^2 B_0 \frac{\partial }{\partial X} \left(\frac{j_\parallel}{B_0} \right), \label{eqn:1d vort}
\end{equation}
and
\begin{equation}
\frac{\partial j_\parallel}{\partial t} = \frac{B_0}{\mu_0} \frac{\partial}{\partial X} \left(B_0 \omega \right),\label{eqn:1d curr}
\end{equation}
where $v_A$ is the Alfv\'en speed ($v_A^2 = B_0^2 / \mu_0 \rho_0$). These equations describes a one-dimensional wave propagation of varying speed and flux tube area. There are many MI coupling studies which adopt variations of this model, sometimes with kinetic extensions and ionospheric coupling or perhaps changing the dependent variables from $\omega, j_\parallel$ to $\phi, \psi$ or $E_\perp, B_\perp$, or neglecting the factors of $B_0$ inside the $d/dX$ term, but critically neglect the nonlinear term \citep{knudsen_alfven_1992,lysak_generalized_1993,lysak_magnetosphere-ionosphere_2004,miles_alfvenic_2018,woodroffe_ultra-low_2012}. Our work has shown that these models must be refined at large amplitude. Specifically, we estimate the saturation current density to be somewhere between 3-5 $\mu \text{A}/\text{m}^2$ along auroral dipole field lines at Earth. At that value, the nonlinear terms become nonnegligible and strong turbulence begins to develop. Future work can refine this value. 

\section{Conclusions}
In this study, the equations of Reduced MHD were derived and examined in the context of a dipole coordinate system. Our derivation began by assuming a low $M_A$ low $\beta$ plasma with perturbations that are stretched along magnetic field lines ($k_\perp \gg k_\parallel$). Under those assumptions we derived that the plasma must satisfies the following conditions on convective time scales to be described by RMHD:
\begin{itemize}
  \item There are no time dependent parallel magnetic field nor parallel velocity perturbations. 
  \item The velocity perturbations are coupled to the magnetic perturbation such that $u_\perp = \pm b_\perp / \sqrt{\mu_0 \rho_0}$.
  \item The Lam\'e coefficients have no dependence on the short coordinates (i.e. the smallest radius of curvature is larger than the largest short scale).
  \item The flux tube expands more than it curves, justifying neglecting the coupling of the slow and shear Alfv\'en modes. This is quantified by terms in Fig. \ref{fig:neglect_curvature}.
  \item In the plane perpendicular to the magnetic field $\mathbf{b}_\perp$ and $\mathbf{u}_\perp$ are divergence free. 
  \end{itemize}
  
The meaning of incompressibility takes on a nuanced meaning which can only be understood by deriving it as a constraint. In the strictest sense of the word ($\nabla \cdot \mathbf{u}=0$), an incompressible plasma cannot have an inhomogeneous Alfv\'en speed because the density cannot vary. However, the magnetospheric Alfv\'en speed changes along a field line, so incompressibility must be relaxed to accurately describe this region. Our analysis shows that a weaker sense of incompressibility ($\nabla_\perp \cdot \mathbf{u}_\perp = 0$, $\rho_1=0$, and $\partial u_\chi / \partial X \neq 0$) is possible if the density profile ($\rho_0(A,X)$) is supported by a nonuniform static background parallel flow ($u_\chi(X)$). The presence of an inhomogeneous Alfv\'en speed, a core feature of the magnetospheric application in mind, implies this parallel flow. By its nature, this background flow is compressible and must come from some non-Alfv\'enic physics of the system that acts on time scales slower than the shear Alfv\'enic dynamics. To logically use our equations, one must assume that process does not depend on the Alfv\'enic dynamics and can simply be imposed on our system by setting the density profile. 

Under those circumstances, which are substantiated, our final equations, (\ref{eqn:jpar final}) and (\ref{eqn:vorticity final}) or equivalently (\ref{eqn:elsasser_potential}), describe RMHD with geometric factors accounting for a varying Alfv\'en speed and magnetic field flux tube area. This allows for realistic studies of turbulence with inhomogeneous backgrounds and highly varying flux tube areas such as the polar regions of the Earth's magnetosphere. From the RMHD equations, we predict two behaviors that ought to occur in the MI coupling problems that are extensions of this RMHD equations from their homogeneous isotropic counterpart. First, the gradients in the Alfv\'en speed ought to partially reflected downward going Poynting flux. This effect is strongest in the region 1-2 $R_E$. Secondly, at large amplitudes, turbulence is generated from wave-wave mixing. This was quantified to be approximately when FACs have a magnitude near 5 $\mu \text{A}/\text{m}^2$ at the ionosphere. These values are coincident with FACs during geomagnetic storms, indicating that our new effect becomes important then. When combined, these two effects will turbulently dissipate energy from global FAC patterns into small scale structures preferentially during strong driving. Using the reflection driven turbulence phenomenology, we have provided an estimate of turbulent dissipation in the MI coupling and updated the standard boundary condition used in global models with a simplified boundary condition which can be used in future work.

\citet{tian_evidence_2026} suggested a wave to particle energy conversion rate of at least 46\% in the auroral acceleration region is possible. This figure is the proportion of downward electron energy flux below the region as a fraction of incident Poynting flux. \citet{keiling_global_2003} demonstrates that the return Poynting flux is substantially smaller than the incident under similar circumstances, corroborating the existence of this energy conversion process. These observations raise the question, how can incident Alfv\'enic Poynting flux dissipate in the magnetosphere? Our analysis predicts that a reflection driven turbulence dissipation mechanism could convert a substantial fraction of the large scale wave energy into small scale waves in a similar region without any ionospheric conductivity gradients, although those will also contribute. The steady state analytical prediction estimates $40\%$ in steady state. Our time dependent solution shows conversion rates of $24\%$ after only two Alfv\'en transit times with much higher percentages at later times. From there, we hypothesis the waves become kinetic scale and a portion of this turbulently dissipated energy can energize particles. Whatever this global energy conversion rate is, it ought to be considered in global MHD models as a dissipation mechanism between the lower boundary condition and the ionosphere. We provide a minimal modification to the existing one which emulates this process.

Finally, we have performed this derivation with a novel computer algebra system (CAS) assisted method. This allows the trivial computation of perturbation theory under arbitrarily complex geometries. This method allowed us to be explicit in our operator definitions while reducing our risk of human error. The full expansions at each order of $\varepsilon$ before simplification with the constraints has been provided in Appendix \ref{appendix:cas} so the reader can follow our steps.  

\section{Acknowledgments}
I.M.D., C.B., N.Y.B., and J.C.D. were supported by the National Aeronautics and Space Administration under grant NNH21ZDA001N-LWSSC, Living With a Star Strategic Capabilities. N.Y.B. was partially funded by the NASA Internal Scientist Funding Model. I.M.D. thanks George Khazanov and Austin Brenner for discussing this work. A large language model was used to generate the initial code that was used to generate the figures and CAS. The source code for the CAS can be found with the LaTeX source. Its output is in the appendix. 

\appendix
\section{Differential Geometry of Flux Tube Area}
\label{appendix:curvilinear constraint}
Suppose a quantity ($q$) obeys the following constraint where $\mathcal{K}$ is some curvature term that involves partials of the Lam\'e coefficients such as in (\ref{eqn:O(eps) div B}) or (\ref{eqn:O(eps) continuity})
\begin{equation}
\frac{\partial}{\partial X}\left(q \right) = \mathcal{K} q.
\end{equation}
This can be solved through traditional methods to find the constraint that
\begin{equation}
\log q = \int \mathcal{K} dX. \label{eqn:flux tube conservation}
\end{equation}
This type of constraint would appear to have a very close connection to flux tube conservation laws. If $\mathcal{K}$ defines how the field line curves in space, then it must be related to the area change of a flux tube. This would give (\ref{eqn:flux tube conservation}) the physical interpretation of a $q_2 / q_1 = \exp\left(\int_{X_1}^{X_2} \mathcal{K} dX \right ) = \mathcal{A}_1/\mathcal{A}_2$  along a field line.
\begin{figure}
  \centering
  \includegraphics[width=5in]{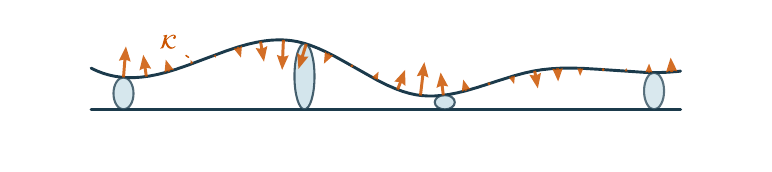}
  \caption{The relation $\exp\left(\int_{X_1}^{X_2} \mathcal{K} dX \right ) = \mathcal{A}_1/\mathcal{A}_2$ visualized along a flux tube. If one integrates the curvature (dark orange arrows) along the top curved magnetic field line, the flux tube area, visualized by the ellipses, changes in response.}
  \label{fig:flux_tube_curv_area_relation}
\end{figure}

To the author's knowledge, a specific relation between the integral of curvature along a flux tube to the change in flux tube area is new in plasma physics. Consider first a situation with two parallel field lines that do not curve. The curvature along the field line is zero, so there is no change in flux tube area. Now consider an instance where a straight field line and an arbitrarily curved field lines are roughly next to each other. Such a situation is shown in figure \ref{fig:flux_tube_curv_area_relation}. The integral of signed curvature determines the relative closeness of the two magnetic field lines. If these two lines define the boundary of a flux tube, then it is obvious that the flux tube will shrink or grow with the curvature of the magnetic field line. The consequence of this relation is a prescriptive way for calculating flux tube area changes. Given the Lam\'e coefficients and their gradients, it is possible to directly compute the changes in flux tube area along a magnetic field line.

If $\mathcal{K} = \frac{1}{h} \frac{\partial h}{\partial X}$ for some Lam\'e coefficient $h$, as is the case in this work, the solution to (\ref{eqn:flux tube conservation}) becomes even simpler. It is just
\begin{equation}
\log\left(q / h \right) = 0.
\end{equation}
The prescription of a field aligned coordinate system, and associated field aligned Lam\'e coefficients, lead to a simple situation relating curvature to flux tube area. A very general conclusion that can be drawn is \textit{when magnetic fields are allowed to curve, the background quantities must depend on the coordinate system in a way that conserves flux tube area}. The converse, a curved magnetic field geometry with constant background conditions, would lead to an nonphysical situation.

This relation is purely based on the geometry of the coordinate system. There is very little physics specific to magnetic fields entailed in this relation. For this reason, we can see close parallels in other fields. In structural dynamics, there are two results known as the First and Second Mohr or Moment-Area Theorems. The first states that the change in slope of a beam is given by the integral of the applied moment along the beam length. The second states that the vertical deflection is given by the higher order moment of the first integral. For a given beam cross section and support locations, the levelness of the beam at either end can be used to solve for the deflection for a given applied load. We can identify the second theorem with the integral relation above.

%It is not rigorously proved here, but there is probably a close connection to the first and second Mohr theorems and the Jacobi/geodesic deviation equation.
%See \url{https://arxiv.org/pdf/physics/0606044} for more on dipole coordinates

\section{Two-potential reduction of parallel terms}
\label{sec:two pot parallel terms}
In order to go from the component form expression of the perpendicular momentum and induction equation to the two-potential version, it is necessary to substitute in the potentials and then take the parallel component of the curl. Due to the dependence of the Lam\'e coefficients on the coordinates, this is an onerous process and is summarized here. First, with the momentum equation
\begin{dmath}
  \hat{e}_\chi \cdot \nabla \times \left[B_0^2 \left(\partial_X b_\alpha + b_\alpha \partial_X \log h_\alpha  \right)\hat{e}_\alpha  + B_0^2 \left(\partial_X b_\beta + b_\beta \partial_X \log h_\beta \right) \hat{e}_\beta \right] =
\end{dmath}
\begin{dmath}
B_0^2 \left[\frac{1}{h_\beta}\frac{\partial}{\partial X} \left(\frac{1}{h_\beta}\frac{\partial^2 \psi}{\partial \beta^2} \right) + \frac{1}{h_\alpha h_\beta^2}\frac{\partial^2 \psi}{\partial \beta^2}\frac{\partial h_\alpha}{\partial X} + \frac{1}{h_\alpha}\frac{\partial}{\partial X} \left(\frac{1}{h_\alpha}\frac{\partial^2 \psi}{\partial \alpha^2} \right)+ \frac{1}{h_\alpha^2 h_\beta}\frac{\partial^2 \psi}{\partial \alpha^2}\frac{\partial h_\beta}{\partial X} \right] =
\end{dmath}
\begin{dmath}
B_0^2 \left[\frac{1}{h_\alpha h_\beta}\frac{\partial}{\partial X} \left(\frac{h_\alpha}{h_\beta}\frac{\partial^2 \psi}{\partial \beta^2} \right)  + \frac{1}{h_\alpha h_\beta}\frac{\partial}{\partial X} \left(\frac{h_\beta}{h_\alpha}\frac{\partial^2 \psi}{\partial \alpha^2} \right) \right] =
\end{dmath}
These two terms can be combined. The identity $h_\alpha h_\beta = h_\chi = B_0^{-1}$ can then be used to simplify these expressions to
\begin{dmath}
\frac{B_0^2}{h_\chi}\frac{\partial}{\partial X} \left(h_\chi \nabla_\perp^2 \psi \right) = {B_0^3}\frac{\partial}{\partial X} \left(B_0^{-1} \nabla_\perp^2 \psi \right)
\end{dmath}

Similarly, with the induction equation
\begin{dmath}
  \hat{e}_\chi \cdot \nabla \times \left[B_0^2 \left(-\partial_X u_\alpha + u_\alpha \partial_X \log h_\alpha  \right)\hat{e}_\alpha  + B_0^2 \left(-\partial_X u_\beta + u_\beta \partial_X \log h_\beta \right) \hat{e}_\beta \right] =
\end{dmath}
\begin{dmath}
B_0^2 \left[\frac{1}{h_\beta}\frac{\partial}{\partial X} \left(\frac{1}{h_\beta}\frac{\partial^2 \phi}{\partial \beta^2} \right) - \frac{1}{h_\alpha h_\beta^2}\frac{\partial^2 \phi}{\partial \beta^2}\frac{\partial h_\alpha}{\partial X} + \frac{1}{h_\alpha}\frac{\partial}{\partial X} \left(\frac{1}{h_\alpha}\frac{\partial^2 \phi}{\partial \alpha^2} \right)- \frac{1}{h_\alpha^2 h_\beta}\frac{\partial^2 \phi}{\partial \alpha^2}\frac{\partial h_\beta}{\partial X} \right] =
\end{dmath}
and then using integration by parts
\begin{dmath}
B_0^2 \left[\frac{\partial}{\partial X} \left(\nabla_\perp^2 \phi \right) + \frac{1}{h_\beta^3}\frac{\partial^2 \phi}{\partial \beta^2} \frac{\partial h_\beta}{\partial X} - \frac{1}{h_\alpha h_\beta^2}\frac{\partial^2 \phi}{\partial \beta^2}\frac{\partial h_\alpha}{\partial X} + \frac{1}{h_\alpha^3}\frac{\partial^2 \phi}{\partial \alpha^2} \frac{\partial h_\alpha}{\partial X}- \frac{1}{h_\alpha^2 h_\beta}\frac{\partial^2 \phi}{\partial \alpha^2}\frac{\partial h_\beta}{\partial X} \right] =
\end{dmath}
and combining curvature terms
\begin{dmath}
B_0^2 \left[\frac{\partial}{\partial X} \left(\nabla_\perp^2 \phi \right) + \left(\frac{1}{h_\alpha}\frac{\partial h_\alpha}{\partial X} - \frac{1}{h_\beta}\frac{\partial h_\beta}{\partial X}\right) \left(\frac{1}{h_\alpha^2}\frac{\partial^2 \phi}{\partial \alpha^2} - \frac{1}{h_\beta^2}\frac{\partial^2 \phi}{\partial \beta^2} \right) \right].
\end{dmath}
Expanding the first term back out and combining, one finds
\begin{dmath}
B_0^2 \left[\nabla_\perp^2  \frac{\partial \phi}{\partial X} - \left(\frac{1}{h_\alpha}\frac{\partial h_\alpha}{\partial X} + \frac{1}{h_\beta}\frac{\partial h_\beta}{\partial X}\right) \left(\frac{1}{h_\alpha^2}\frac{\partial^2 \phi}{\partial \alpha^2} + \frac{1}{h_\beta^2}\frac{\partial^2 \phi}{\partial \beta^2} \right) \right].
\end{dmath}
Since $\nabla^2_\perp B_0 = B_0 \nabla_\perp^2$, this simplifies to
\begin{dmath}
B_0^2 \left[\nabla_\perp^2 \phi + \nabla_\perp^2 B_0^{-1} \frac{\partial B_0}{\partial X} \phi \right]= \nabla_\perp^2 B_0  \frac{\partial \left(B_0 \phi \right)}{\partial X}= 0.
\end{dmath}

\section{Computer Algebra System Output}
\label{appendix:cas}
\subsection*{Continuity O($1$)}
\begin{dmath}
\label{cas:Continuity O(1)}
\frac{1}{h_{\chi}} \rho_{0} \frac{\partial u_{\chi}}{\partial \chi} + \frac{1}{h_{\beta}} \rho_{0} \frac{\partial u_{\beta}}{\partial \beta} + \frac{1}{h_{\alpha}} \rho_{0} \frac{\partial u_{\alpha}}{\partial \alpha} = 0
\end{dmath}
\subsection*{Continuity O($\varepsilon$)}
\begin{dmath}
\label{cas:Continuity O(2)}
\frac{\partial \rho_{1}}{\partial t} + \frac{1}{h_{\chi}} \rho_{0} \frac{\partial u_{\chi}}{\partial X} + \frac{1}{h_{\chi}} \rho_{1} \frac{\partial u_{\chi}}{\partial \chi} + \frac{1}{h_{\chi}} u_{\chi} \frac{\partial \rho_{0}}{\partial X} + \frac{1}{h_{\chi}} u_{\chi} \frac{\partial \rho_{1}}{\partial \chi} + \frac{1}{h_{\beta}} \rho_{1} \frac{\partial u_{\beta}}{\partial \beta} + \frac{1}{h_{\beta}} u_{\beta} \frac{\partial \rho_{1}}{\partial \beta} + \frac{1}{h_{\beta} h_{\chi}} \rho_{0} u_{\chi} \frac{\partial h_{\beta}}{\partial X} + \frac{1}{h_{\alpha}} \rho_{1} \frac{\partial u_{\alpha}}{\partial \alpha} + \frac{1}{h_{\alpha}} u_{\alpha} \frac{\partial \rho_{0}}{\partial A} + \frac{1}{h_{\alpha}} u_{\alpha} \frac{\partial \rho_{1}}{\partial \alpha} + \frac{1}{h_{\alpha} h_{\chi}} \rho_{0} u_{\alpha} \frac{\partial h_{\chi}}{\partial A} + \frac{1}{h_{\alpha} h_{\chi}} \rho_{0} u_{\chi} \frac{\partial h_{\alpha}}{\partial X} + \frac{1}{h_{\alpha} h_{\beta}} \rho_{0} u_{\alpha} \frac{\partial h_{\beta}}{\partial A} = 0
\end{dmath}
\subsection*{Div u($1$)}
\begin{dmath}
\label{cas:Div u(1)}
\frac{1}{h_{\chi}} \frac{\partial u_{\chi}}{\partial \chi} + \frac{1}{h_{\beta}} \frac{\partial u_{\beta}}{\partial \beta} + \frac{1}{h_{\alpha}} \frac{\partial u_{\alpha}}{\partial \alpha} = 0
\end{dmath}
\subsection*{Div u($\varepsilon$)}
\begin{dmath}
\label{cas:Div u(2)}
\frac{1}{h_{\chi}} \frac{\partial u_{\chi}}{\partial X} + \frac{1}{h_{\beta} h_{\chi}} u_{\chi} \frac{\partial h_{\beta}}{\partial X} + \frac{1}{h_{\alpha} h_{\chi}} u_{\alpha} \frac{\partial h_{\chi}}{\partial A} + \frac{1}{h_{\alpha} h_{\chi}} u_{\chi} \frac{\partial h_{\alpha}}{\partial X} + \frac{1}{h_{\alpha} h_{\beta}} u_{\alpha} \frac{\partial h_{\beta}}{\partial A} = 0
\end{dmath}
\subsection*{Induction Alpha O($1$)}
\begin{dmath}
\label{cas:Induction Alpha O(1)}
- \frac{1}{h_{\chi}} B_{0} \frac{\partial u_{\alpha}}{\partial \chi} = 0
\end{dmath}
\subsection*{Induction Alpha O($\varepsilon$)}
\begin{dmath}
\label{cas:Induction Alpha O(2)}
- \frac{1}{h_{\chi}} B_{0} \frac{\partial u_{\alpha}}{\partial X} + \frac{1}{h_{\alpha} h_{\chi}} B_{0} u_{\alpha} \frac{\partial h_{\alpha}}{\partial X} + \frac{1}{h_{\chi}} b_{\alpha} \frac{\partial u_{\chi}}{\partial \chi} + \frac{1}{h_{\beta}} b_{\alpha} \frac{\partial u_{\beta}}{\partial \beta} - \frac{1}{h_{\beta}} b_{\beta} \frac{\partial u_{\alpha}}{\partial \beta} - \frac{1}{h_{\chi}} b_{\chi} \frac{\partial u_{\alpha}}{\partial \chi} + \frac{\partial b_{\alpha}}{\partial t} + \frac{1}{h_{\chi}} u_{\chi} \frac{\partial b_{\alpha}}{\partial \chi} + \frac{1}{h_{\beta}} u_{\beta} \frac{\partial b_{\alpha}}{\partial \beta} + \frac{1}{h_{\alpha}} u_{\alpha} \frac{\partial b_{\alpha}}{\partial \alpha} = 0
\end{dmath}
\subsection*{Induction Alpha O($\varepsilon^2$)}
\begin{dmath}
\label{cas:Induction Alpha O(3)}
\frac{1}{h_{\chi}} b_{\alpha} \frac{\partial u_{\chi}}{\partial X} + \frac{1}{h_{\beta} h_{\chi}} b_{\alpha} u_{\chi} \frac{\partial h_{\beta}}{\partial X} + \frac{1}{h_{\alpha} h_{\chi}} b_{\alpha} u_{\alpha} \frac{\partial h_{\chi}}{\partial A} + \frac{1}{h_{\alpha} h_{\beta}} b_{\alpha} u_{\alpha} \frac{\partial h_{\beta}}{\partial A} - \frac{1}{h_{\chi}} b_{\chi} \frac{\partial u_{\alpha}}{\partial X} + \frac{1}{h_{\alpha} h_{\chi}} b_{\chi} u_{\alpha} \frac{\partial h_{\alpha}}{\partial X} + \frac{1}{h_{\chi}} u_{\chi} \frac{\partial b_{\alpha}}{\partial X} = 0
\end{dmath}
\subsection*{Induction Beta O($1$)}
\begin{dmath}
\label{cas:Induction Beta O(1)}
- \frac{1}{h_{\chi}} B_{0} \frac{\partial u_{\beta}}{\partial \chi} = 0
\end{dmath}
\subsection*{Induction Beta O($\varepsilon$)}
\begin{dmath}
\label{cas:Induction Beta O(2)}
- \frac{1}{h_{\chi}} B_{0} \frac{\partial u_{\beta}}{\partial X} + \frac{1}{h_{\beta} h_{\chi}} B_{0} u_{\beta} \frac{\partial h_{\beta}}{\partial X} - \frac{1}{h_{\alpha}} b_{\alpha} \frac{\partial u_{\beta}}{\partial \alpha} + \frac{1}{h_{\chi}} b_{\beta} \frac{\partial u_{\chi}}{\partial \chi} + \frac{1}{h_{\alpha}} b_{\beta} \frac{\partial u_{\alpha}}{\partial \alpha} - \frac{1}{h_{\chi}} b_{\chi} \frac{\partial u_{\beta}}{\partial \chi} + \frac{\partial b_{\beta}}{\partial t} + \frac{1}{h_{\chi}} u_{\chi} \frac{\partial b_{\beta}}{\partial \chi} + \frac{1}{h_{\beta}} u_{\beta} \frac{\partial b_{\beta}}{\partial \beta} + \frac{1}{h_{\alpha}} u_{\alpha} \frac{\partial b_{\beta}}{\partial \alpha} = 0
\end{dmath}
\subsection*{Induction Beta O($\varepsilon^2$)}
\begin{dmath}
\label{cas:Induction Beta O(3)}
\frac{1}{h_{\alpha} h_{\beta}} b_{\alpha} u_{\beta} \frac{\partial h_{\beta}}{\partial A} + \frac{1}{h_{\chi}} b_{\beta} \frac{\partial u_{\chi}}{\partial X} + \frac{1}{h_{\alpha} h_{\chi}} b_{\beta} u_{\alpha} \frac{\partial h_{\chi}}{\partial A} + \frac{1}{h_{\alpha} h_{\chi}} b_{\beta} u_{\chi} \frac{\partial h_{\alpha}}{\partial X} - \frac{1}{h_{\chi}} b_{\chi} \frac{\partial u_{\beta}}{\partial X} + \frac{1}{h_{\beta} h_{\chi}} b_{\chi} u_{\beta} \frac{\partial h_{\beta}}{\partial X} + \frac{1}{h_{\chi}} u_{\chi} \frac{\partial b_{\beta}}{\partial X} = 0
\end{dmath}
\subsection*{Induction Chi O($1$)}
\begin{dmath}
\label{cas:Induction Chi O(1)}
\frac{1}{h_{\beta}} B_{0} \frac{\partial u_{\beta}}{\partial \beta} + \frac{1}{h_{\alpha}} B_{0} \frac{\partial u_{\alpha}}{\partial \alpha} = 0
\end{dmath}
\subsection*{Induction Chi O($\varepsilon$)}
\begin{dmath}
\label{cas:Induction Chi O(2)}
\frac{1}{h_{\beta} h_{\chi}} B_{0} u_{\chi} \frac{\partial h_{\beta}}{\partial X} + \frac{1}{h_{\alpha} h_{\chi}} B_{0} u_{\chi} \frac{\partial h_{\alpha}}{\partial X} + \frac{1}{h_{\alpha} h_{\beta}} B_{0} u_{\alpha} \frac{\partial h_{\beta}}{\partial A} - \frac{1}{h_{\alpha}} b_{\alpha} \frac{\partial u_{\chi}}{\partial \alpha} - \frac{1}{h_{\beta}} b_{\beta} \frac{\partial u_{\chi}}{\partial \beta} + \frac{1}{h_{\beta}} b_{\chi} \frac{\partial u_{\beta}}{\partial \beta} + \frac{1}{h_{\alpha}} b_{\chi} \frac{\partial u_{\alpha}}{\partial \alpha} + \frac{\partial b_{\chi}}{\partial t} + \frac{1}{h_{\chi}} u_{\chi} \frac{\partial B_{0}}{\partial X} + \frac{1}{h_{\chi}} u_{\chi} \frac{\partial b_{\chi}}{\partial \chi} + \frac{1}{h_{\beta}} u_{\beta} \frac{\partial b_{\chi}}{\partial \beta} + \frac{1}{h_{\alpha}} u_{\alpha} \frac{\partial B_{0}}{\partial A} + \frac{1}{h_{\alpha}} u_{\alpha} \frac{\partial b_{\chi}}{\partial \alpha} = 0
\end{dmath}
\subsection*{Induction Chi O($\varepsilon^2$)}
\begin{dmath}
\label{cas:Induction Chi O(3)}
\frac{1}{h_{\alpha} h_{\chi}} b_{\alpha} u_{\chi} \frac{\partial h_{\chi}}{\partial A} + \frac{1}{h_{\beta} h_{\chi}} b_{\chi} u_{\chi} \frac{\partial h_{\beta}}{\partial X} + \frac{1}{h_{\alpha} h_{\chi}} b_{\chi} u_{\chi} \frac{\partial h_{\alpha}}{\partial X} + \frac{1}{h_{\alpha} h_{\beta}} b_{\chi} u_{\alpha} \frac{\partial h_{\beta}}{\partial A} + \frac{1}{h_{\chi}} u_{\chi} \frac{\partial b_{\chi}}{\partial X} = 0
\end{dmath}
\subsection*{Momentum Alpha O($1/\varepsilon$)}
\begin{dmath}
\label{cas:Momentum Alpha O(0)}
\frac{1}{h_{\alpha} h_{\chi}} \frac{B_{0}^{\mathtt{\text{2}}} \frac{\partial h_{\chi}}{\partial A}}{\mu_{0}} - \frac{1}{h_{\chi}} \frac{B_{0} \frac{\partial b_{\alpha}}{\partial \chi}}{\mu_{0}} + \frac{1}{h_{\alpha}} \frac{B_{0} \frac{\partial B_{0}}{\partial A}}{\mu_{0}} + \frac{1}{h_{\alpha}} \frac{B_{0} \frac{\partial b_{\chi}}{\partial \alpha}}{\mu_{0}} = 0
\end{dmath}
\subsection*{Momentum Alpha O(1)}
\begin{dmath}
\label{cas:Momentum Alpha O(1)}
\rho_{0} \frac{\partial u_{\alpha}}{\partial t} + \frac{1}{h_{\chi}} \rho_{0} u_{\chi} \frac{\partial u_{\alpha}}{\partial \chi} + \frac{1}{h_{\beta}} \rho_{0} u_{\beta} \frac{\partial u_{\alpha}}{\partial \beta} + \frac{1}{h_{\alpha}} \rho_{0} u_{\alpha} \frac{\partial u_{\alpha}}{\partial \alpha} + \frac{1}{h_{\alpha}} \frac{\partial p_{1}}{\partial \alpha} - \frac{1}{h_{\alpha} h_{\chi}} \frac{B_{0} b_{\alpha} \frac{\partial h_{\alpha}}{\partial X}}{\mu_{0}} + \frac{1}{h_{\alpha} h_{\chi}} \frac{2 B_{0} b_{\chi} \frac{\partial h_{\chi}}{\partial A}}{\mu_{0}} - \frac{1}{h_{\chi}} \frac{B_{0} \frac{\partial b_{\alpha}}{\partial X}}{\mu_{0}} - \frac{1}{h_{\beta}} \frac{b_{\beta} \frac{\partial b_{\alpha}}{\partial \beta}}{\mu_{0}} + \frac{1}{h_{\alpha}} \frac{b_{\beta} \frac{\partial b_{\beta}}{\partial \alpha}}{\mu_{0}} - \frac{1}{h_{\chi}} \frac{b_{\chi} \frac{\partial b_{\alpha}}{\partial \chi}}{\mu_{0}} + \frac{1}{h_{\alpha}} \frac{b_{\chi} \frac{\partial B_{0}}{\partial A}}{\mu_{0}} + \frac{1}{h_{\alpha}} \frac{b_{\chi} \frac{\partial b_{\chi}}{\partial \alpha}}{\mu_{0}} = 0
\end{dmath}
\subsection*{Momentum Alpha O($\varepsilon$)}
\begin{dmath}
\label{cas:Momentum Alpha O(2)}
\rho_{1} \frac{\partial u_{\alpha}}{\partial t} + \frac{1}{h_{\chi}} \rho_{0} u_{\chi} \frac{\partial u_{\alpha}}{\partial X} + \frac{1}{h_{\chi}} \rho_{1} u_{\chi} \frac{\partial u_{\alpha}}{\partial \chi} + \frac{1}{h_{\beta}} \rho_{1} u_{\beta} \frac{\partial u_{\alpha}}{\partial \beta} + \frac{1}{h_{\alpha}} \rho_{1} u_{\alpha} \frac{\partial u_{\alpha}}{\partial \alpha} + \frac{1}{h_{\alpha} h_{\chi}} \rho_{0} u_{\alpha} u_{\chi} \frac{\partial h_{\alpha}}{\partial X} - \frac{1}{h_{\alpha} h_{\chi}} \rho_{0} u_{\chi}^{\mathtt{\text{2}}} \frac{\partial h_{\chi}}{\partial A} - \frac{1}{h_{\alpha} h_{\beta}} \rho_{0} u_{\beta}^{\mathtt{\text{2}}} \frac{\partial h_{\beta}}{\partial A} - \frac{1}{h_{\alpha} h_{\chi}} \frac{b_{\alpha} b_{\chi} \frac{\partial h_{\alpha}}{\partial X}}{\mu_{0}} + \frac{1}{h_{\alpha} h_{\beta}} \frac{b_{\beta}^{\mathtt{\text{2}}} \frac{\partial h_{\beta}}{\partial A}}{\mu_{0}} + \frac{1}{h_{\alpha} h_{\chi}} \frac{b_{\chi}^{\mathtt{\text{2}}} \frac{\partial h_{\chi}}{\partial A}}{\mu_{0}} - \frac{1}{h_{\chi}} \frac{b_{\chi} \frac{\partial b_{\alpha}}{\partial X}}{\mu_{0}} = 0
\end{dmath}
\subsection*{Momentum Alpha O($\varepsilon^2$)}
\begin{dmath}
\label{cas:Momentum Alpha O(3)}
\frac{1}{h_{\chi}} \rho_{1} u_{\chi} \frac{\partial u_{\alpha}}{\partial X} + \frac{1}{h_{\alpha} h_{\chi}} \rho_{1} u_{\alpha} u_{\chi} \frac{\partial h_{\alpha}}{\partial X} - \frac{1}{h_{\alpha} h_{\chi}} \rho_{1} u_{\chi}^{\mathtt{\text{2}}} \frac{\partial h_{\chi}}{\partial A} - \frac{1}{h_{\alpha} h_{\beta}} \rho_{1} u_{\beta}^{\mathtt{\text{2}}} \frac{\partial h_{\beta}}{\partial A} = 0
\end{dmath}
\subsection*{Momentum Beta O($1/\varepsilon$)}
\begin{dmath}
\label{cas:Momentum Beta O(0)}
- \frac{1}{h_{\chi}} \frac{B_{0} \frac{\partial b_{\beta}}{\partial \chi}}{\mu_{0}} + \frac{1}{h_{\beta}} \frac{B_{0} \frac{\partial b_{\chi}}{\partial \beta}}{\mu_{0}} = 0
\end{dmath}
\subsection*{Momentum Beta O(1)}
\begin{dmath}
\label{cas:Momentum Beta O(1)}
\rho_{0} \frac{\partial u_{\beta}}{\partial t} + \frac{1}{h_{\chi}} \rho_{0} u_{\chi} \frac{\partial u_{\beta}}{\partial \chi} + \frac{1}{h_{\beta}} \rho_{0} u_{\beta} \frac{\partial u_{\beta}}{\partial \beta} + \frac{1}{h_{\beta}} \frac{\partial p_{1}}{\partial \beta} + \frac{1}{h_{\alpha}} \rho_{0} u_{\alpha} \frac{\partial u_{\beta}}{\partial \alpha} - \frac{1}{h_{\beta} h_{\chi}} \frac{B_{0} b_{\beta} \frac{\partial h_{\beta}}{\partial X}}{\mu_{0}} - \frac{1}{h_{\chi}} \frac{B_{0} \frac{\partial b_{\beta}}{\partial X}}{\mu_{0}} + \frac{1}{h_{\beta}} \frac{b_{\alpha} \frac{\partial b_{\alpha}}{\partial \beta}}{\mu_{0}} - \frac{1}{h_{\alpha}} \frac{b_{\alpha} \frac{\partial b_{\beta}}{\partial \alpha}}{\mu_{0}} - \frac{1}{h_{\chi}} \frac{b_{\chi} \frac{\partial b_{\beta}}{\partial \chi}}{\mu_{0}} + \frac{1}{h_{\beta}} \frac{b_{\chi} \frac{\partial b_{\chi}}{\partial \beta}}{\mu_{0}} = 0
\end{dmath}
\subsection*{Momentum Beta O($\varepsilon$)}
\begin{dmath}
\label{cas:Momentum Beta O(2)}
\rho_{1} \frac{\partial u_{\beta}}{\partial t} + \frac{1}{h_{\chi}} \rho_{0} u_{\chi} \frac{\partial u_{\beta}}{\partial X} + \frac{1}{h_{\chi}} \rho_{1} u_{\chi} \frac{\partial u_{\beta}}{\partial \chi} + \frac{1}{h_{\beta}} \rho_{1} u_{\beta} \frac{\partial u_{\beta}}{\partial \beta} + \frac{1}{h_{\beta} h_{\chi}} \rho_{0} u_{\beta} u_{\chi} \frac{\partial h_{\beta}}{\partial X} + \frac{1}{h_{\alpha}} \rho_{1} u_{\alpha} \frac{\partial u_{\beta}}{\partial \alpha} + \frac{1}{h_{\alpha} h_{\beta}} \rho_{0} u_{\alpha} u_{\beta} \frac{\partial h_{\beta}}{\partial A} - \frac{1}{h_{\alpha} h_{\beta}} \frac{b_{\alpha} b_{\beta} \frac{\partial h_{\beta}}{\partial A}}{\mu_{0}} - \frac{1}{h_{\beta} h_{\chi}} \frac{b_{\beta} b_{\chi} \frac{\partial h_{\beta}}{\partial X}}{\mu_{0}} - \frac{1}{h_{\chi}} \frac{b_{\chi} \frac{\partial b_{\beta}}{\partial X}}{\mu_{0}} = 0
\end{dmath}
\subsection*{Momentum Beta O($\varepsilon^2$)}
\begin{dmath}
\label{cas:Momentum Beta O(3)}
\frac{1}{h_{\chi}} \rho_{1} u_{\chi} \frac{\partial u_{\beta}}{\partial X} + \frac{1}{h_{\beta} h_{\chi}} \rho_{1} u_{\beta} u_{\chi} \frac{\partial h_{\beta}}{\partial X} + \frac{1}{h_{\alpha} h_{\beta}} \rho_{1} u_{\alpha} u_{\beta} \frac{\partial h_{\beta}}{\partial A} = 0
\end{dmath}
\subsection*{Momentum Chi O(1)}
\begin{dmath}
\label{cas:Momentum Chi O(1)}
\rho_{0} \frac{\partial u_{\chi}}{\partial t} + \frac{1}{h_{\chi}} \rho_{0} u_{\chi} \frac{\partial u_{\chi}}{\partial \chi} + \frac{1}{h_{\chi}} \frac{\partial p_{0}}{\partial X} + \frac{1}{h_{\chi}} \frac{\partial p_{1}}{\partial \chi} + \frac{1}{h_{\beta}} \rho_{0} u_{\beta} \frac{\partial u_{\chi}}{\partial \beta} + \frac{1}{h_{\alpha}} \rho_{0} u_{\alpha} \frac{\partial u_{\chi}}{\partial \alpha} - \frac{1}{h_{\alpha} h_{\chi}} \frac{B_{0} b_{\alpha} \frac{\partial h_{\chi}}{\partial A}}{\mu_{0}} + \frac{1}{h_{\chi}} \frac{b_{\alpha} \frac{\partial b_{\alpha}}{\partial \chi}}{\mu_{0}} - \frac{1}{h_{\alpha}} \frac{b_{\alpha} \frac{\partial B_{0}}{\partial A}}{\mu_{0}} - \frac{1}{h_{\alpha}} \frac{b_{\alpha} \frac{\partial b_{\chi}}{\partial \alpha}}{\mu_{0}} + \frac{1}{h_{\chi}} \frac{b_{\beta} \frac{\partial b_{\beta}}{\partial \chi}}{\mu_{0}} - \frac{1}{h_{\beta}} \frac{b_{\beta} \frac{\partial b_{\chi}}{\partial \beta}}{\mu_{0}} = 0
\end{dmath}
\subsection*{Momentum Chi O($\varepsilon$)}
\begin{dmath}
\label{cas:Momentum Chi O(2)}
\rho_{1} \frac{\partial u_{\chi}}{\partial t} + \frac{1}{h_{\chi}} \rho_{0} u_{\chi} \frac{\partial u_{\chi}}{\partial X} + \frac{1}{h_{\chi}} \rho_{1} u_{\chi} \frac{\partial u_{\chi}}{\partial \chi} + \frac{1}{h_{\chi}} \frac{\partial p_{1}}{\partial X} + \frac{1}{h_{\beta}} \rho_{1} u_{\beta} \frac{\partial u_{\chi}}{\partial \beta} - \frac{1}{h_{\beta} h_{\chi}} \rho_{0} u_{\beta}^{\mathtt{\text{2}}} \frac{\partial h_{\beta}}{\partial X} + \frac{1}{h_{\alpha}} \rho_{1} u_{\alpha} \frac{\partial u_{\chi}}{\partial \alpha} - \frac{1}{h_{\alpha} h_{\chi}} \rho_{0} u_{\alpha}^{\mathtt{\text{2}}} \frac{\partial h_{\alpha}}{\partial X} + \frac{1}{h_{\alpha} h_{\chi}} \rho_{0} u_{\alpha} u_{\chi} \frac{\partial h_{\chi}}{\partial A} + \frac{1}{h_{\alpha} h_{\chi}} \frac{b_{\alpha}^{\mathtt{\text{2}}} \frac{\partial h_{\alpha}}{\partial X}}{\mu_{0}} - \frac{1}{h_{\alpha} h_{\chi}} \frac{b_{\alpha} b_{\chi} \frac{\partial h_{\chi}}{\partial A}}{\mu_{0}} + \frac{1}{h_{\chi}} \frac{b_{\alpha} \frac{\partial b_{\alpha}}{\partial X}}{\mu_{0}} + \frac{1}{h_{\beta} h_{\chi}} \frac{b_{\beta}^{\mathtt{\text{2}}} \frac{\partial h_{\beta}}{\partial X}}{\mu_{0}} + \frac{1}{h_{\chi}} \frac{b_{\beta} \frac{\partial b_{\beta}}{\partial X}}{\mu_{0}} = 0
\end{dmath}
\subsection*{Momentum Chi O($\varepsilon^2$)}
\begin{dmath}
\label{cas:Momentum Chi O(3)}
\frac{1}{h_{\chi}} \rho_{1} u_{\chi} \frac{\partial u_{\chi}}{\partial X} - \frac{1}{h_{\beta} h_{\chi}} \rho_{1} u_{\beta}^{\mathtt{\text{2}}} \frac{\partial h_{\beta}}{\partial X} - \frac{1}{h_{\alpha} h_{\chi}} \rho_{1} u_{\alpha}^{\mathtt{\text{2}}} \frac{\partial h_{\alpha}}{\partial X} + \frac{1}{h_{\alpha} h_{\chi}} \rho_{1} u_{\alpha} u_{\chi} \frac{\partial h_{\chi}}{\partial A} = 0
\end{dmath}
\subsection*{Div B O($\varepsilon$)}
\begin{dmath}
\label{cas:Div B O(2)}
\frac{1}{h_{\beta} h_{\chi}} B_{0} \frac{\partial h_{\beta}}{\partial X} + \frac{1}{h_{\alpha} h_{\chi}} B_{0} \frac{\partial h_{\alpha}}{\partial X} + \frac{1}{h_{\chi}} \frac{\partial B_{0}}{\partial X} + \frac{1}{h_{\chi}} \frac{\partial b_{\chi}}{\partial \chi} + \frac{1}{h_{\beta}} \frac{\partial b_{\beta}}{\partial \beta} + \frac{1}{h_{\alpha}} \frac{\partial b_{\alpha}}{\partial \alpha} = 0
\end{dmath}
\subsection*{Div B O($\varepsilon^2$)}
\begin{dmath}
\label{cas:Div B O(3)}
\frac{1}{h_{\alpha} h_{\chi}} b_{\alpha} \frac{\partial h_{\chi}}{\partial A} + \frac{1}{h_{\alpha} h_{\beta}} b_{\alpha} \frac{\partial h_{\beta}}{\partial A} + \frac{1}{h_{\beta} h_{\chi}} b_{\chi} \frac{\partial h_{\beta}}{\partial X} + \frac{1}{h_{\alpha} h_{\chi}} b_{\chi} \frac{\partial h_{\alpha}}{\partial X} + \frac{1}{h_{\chi}} \frac{\partial b_{\chi}}{\partial X} = 0
\end{dmath}

\newpage
\section{Computer Algebra System Output, $h \to 1$}
\label{appendix:slab_cas}
\subsection*{Continuity O($1$)}
\begin{dmath}
\rho_{0} \frac{\partial u_{\alpha}}{\partial \alpha} + \rho_{0} \frac{\partial u_{\beta}}{\partial \beta} + \rho_{0} \frac{\partial u_{\chi}}{\partial \chi} = 0
\end{dmath}
\subsection*{Continuity O($\varepsilon$)}
\begin{dmath}
\rho_{0} \frac{\partial u_{\chi}}{\partial X} + \rho_{1} \frac{\partial u_{\alpha}}{\partial \alpha} + \rho_{1} \frac{\partial u_{\beta}}{\partial \beta} + \rho_{1} \frac{\partial u_{\chi}}{\partial \chi} + u_{\alpha} \frac{\partial \rho_{0}}{\partial A} + u_{\alpha} \frac{\partial \rho_{1}}{\partial \alpha} + u_{\beta} \frac{\partial \rho_{1}}{\partial \beta} + u_{\chi} \frac{\partial \rho_{0}}{\partial X} + u_{\chi} \frac{\partial \rho_{1}}{\partial \chi} + \frac{\partial \rho_{1}}{\partial t} = 0
\end{dmath}
\subsection*{Div u($1$)}
\begin{dmath}
\frac{\partial u_{\alpha}}{\partial \alpha} + \frac{\partial u_{\beta}}{\partial \beta} + \frac{\partial u_{\chi}}{\partial \chi} = 0
\end{dmath}
\subsection*{Div u($\varepsilon$)}
\begin{dmath}
\frac{\partial u_{\chi}}{\partial X} = 0
\end{dmath}
\subsection*{Induction Alpha O($1$)}
\begin{dmath}
- B_{0} \frac{\partial u_{\alpha}}{\partial \chi} = 0
\end{dmath}
\subsection*{Induction Alpha O($\varepsilon$)}
\begin{dmath}
- B_{0} \frac{\partial u_{\alpha}}{\partial X} + b_{\alpha} \frac{\partial u_{\beta}}{\partial \beta} + b_{\alpha} \frac{\partial u_{\chi}}{\partial \chi} - b_{\beta} \frac{\partial u_{\alpha}}{\partial \beta} - b_{\chi} \frac{\partial u_{\alpha}}{\partial \chi} + u_{\alpha} \frac{\partial b_{\alpha}}{\partial \alpha} + u_{\beta} \frac{\partial b_{\alpha}}{\partial \beta} + u_{\chi} \frac{\partial b_{\alpha}}{\partial \chi} + \frac{\partial b_{\alpha}}{\partial t} = 0
\end{dmath}
\subsection*{Induction Alpha O($\varepsilon^2$)}
\begin{dmath}
b_{\alpha} \frac{\partial u_{\chi}}{\partial X} - b_{\chi} \frac{\partial u_{\alpha}}{\partial X} + u_{\chi} \frac{\partial b_{\alpha}}{\partial X} = 0
\end{dmath}
\subsection*{Induction Beta O($1$)}
\begin{dmath}
- B_{0} \frac{\partial u_{\beta}}{\partial \chi} = 0
\end{dmath}
\subsection*{Induction Beta O($\varepsilon$)}
\begin{dmath}
- B_{0} \frac{\partial u_{\beta}}{\partial X} - b_{\alpha} \frac{\partial u_{\beta}}{\partial \alpha} + b_{\beta} \frac{\partial u_{\alpha}}{\partial \alpha} + b_{\beta} \frac{\partial u_{\chi}}{\partial \chi} - b_{\chi} \frac{\partial u_{\beta}}{\partial \chi} + u_{\alpha} \frac{\partial b_{\beta}}{\partial \alpha} + u_{\beta} \frac{\partial b_{\beta}}{\partial \beta} + u_{\chi} \frac{\partial b_{\beta}}{\partial \chi} + \frac{\partial b_{\beta}}{\partial t} = 0
\end{dmath}
\subsection*{Induction Beta O($\varepsilon^2$)}
\begin{dmath}
b_{\beta} \frac{\partial u_{\chi}}{\partial X} - b_{\chi} \frac{\partial u_{\beta}}{\partial X} + u_{\chi} \frac{\partial b_{\beta}}{\partial X} = 0
\end{dmath}
\subsection*{Induction Chi O($1$)}
\begin{dmath}
B_{0} \frac{\partial u_{\alpha}}{\partial \alpha} + B_{0} \frac{\partial u_{\beta}}{\partial \beta} = 0
\end{dmath}
\subsection*{Induction Chi O($\varepsilon$)}
\begin{dmath}
- b_{\alpha} \frac{\partial u_{\chi}}{\partial \alpha} - b_{\beta} \frac{\partial u_{\chi}}{\partial \beta} + b_{\chi} \frac{\partial u_{\alpha}}{\partial \alpha} + b_{\chi} \frac{\partial u_{\beta}}{\partial \beta} + u_{\alpha} \frac{\partial B_{0}}{\partial A} + u_{\alpha} \frac{\partial b_{\chi}}{\partial \alpha} + u_{\beta} \frac{\partial b_{\chi}}{\partial \beta} + u_{\chi} \frac{\partial B_{0}}{\partial X} + u_{\chi} \frac{\partial b_{\chi}}{\partial \chi} + \frac{\partial b_{\chi}}{\partial t} = 0
\end{dmath}
\subsection*{Induction Chi O($\varepsilon^2$)}
\begin{dmath}
u_{\chi} \frac{\partial b_{\chi}}{\partial X} = 0
\end{dmath}
\subsection*{Momentum Alpha O($1/\varepsilon$)}
\begin{dmath}
\frac{B_{0} \frac{\partial B_{0}}{\partial A}}{\mu_{0}} - \frac{B_{0} \frac{\partial b_{\alpha}}{\partial \chi}}{\mu_{0}} + \frac{B_{0} \frac{\partial b_{\chi}}{\partial \alpha}}{\mu_{0}} = 0
\end{dmath}
\subsection*{Momentum Alpha O(1)}
\begin{dmath}
\rho_{0} u_{\alpha} \frac{\partial u_{\alpha}}{\partial \alpha} + \rho_{0} u_{\beta} \frac{\partial u_{\alpha}}{\partial \beta} + \rho_{0} u_{\chi} \frac{\partial u_{\alpha}}{\partial \chi} + \rho_{0} \frac{\partial u_{\alpha}}{\partial t} + \frac{\partial p_{1}}{\partial \alpha} - \frac{B_{0} \frac{\partial b_{\alpha}}{\partial X}}{\mu_{0}} - \frac{b_{\beta} \frac{\partial b_{\alpha}}{\partial \beta}}{\mu_{0}} + \frac{b_{\beta} \frac{\partial b_{\beta}}{\partial \alpha}}{\mu_{0}} + \frac{b_{\chi} \frac{\partial B_{0}}{\partial A}}{\mu_{0}} - \frac{b_{\chi} \frac{\partial b_{\alpha}}{\partial \chi}}{\mu_{0}} + \frac{b_{\chi} \frac{\partial b_{\chi}}{\partial \alpha}}{\mu_{0}} = 0
\end{dmath}
\subsection*{Momentum Alpha O($\varepsilon$)}
\begin{dmath}
\rho_{0} u_{\chi} \frac{\partial u_{\alpha}}{\partial X} + \rho_{1} u_{\alpha} \frac{\partial u_{\alpha}}{\partial \alpha} + \rho_{1} u_{\beta} \frac{\partial u_{\alpha}}{\partial \beta} + \rho_{1} u_{\chi} \frac{\partial u_{\alpha}}{\partial \chi} + \rho_{1} \frac{\partial u_{\alpha}}{\partial t} - \frac{b_{\chi} \frac{\partial b_{\alpha}}{\partial X}}{\mu_{0}} = 0
\end{dmath}
\subsection*{Momentum Alpha O($\varepsilon^2$)}
\begin{dmath}
\rho_{1} u_{\chi} \frac{\partial u_{\alpha}}{\partial X} = 0
\end{dmath}
\subsection*{Momentum Beta O($1/\varepsilon$)}
\begin{dmath}
- \frac{B_{0} \frac{\partial b_{\beta}}{\partial \chi}}{\mu_{0}} + \frac{B_{0} \frac{\partial b_{\chi}}{\partial \beta}}{\mu_{0}} = 0
\end{dmath}
\subsection*{Momentum Beta O(1)}
\begin{dmath}
\rho_{0} u_{\alpha} \frac{\partial u_{\beta}}{\partial \alpha} + \rho_{0} u_{\beta} \frac{\partial u_{\beta}}{\partial \beta} + \rho_{0} u_{\chi} \frac{\partial u_{\beta}}{\partial \chi} + \rho_{0} \frac{\partial u_{\beta}}{\partial t} + \frac{\partial p_{1}}{\partial \beta} - \frac{B_{0} \frac{\partial b_{\beta}}{\partial X}}{\mu_{0}} + \frac{b_{\alpha} \frac{\partial b_{\alpha}}{\partial \beta}}{\mu_{0}} - \frac{b_{\alpha} \frac{\partial b_{\beta}}{\partial \alpha}}{\mu_{0}} - \frac{b_{\chi} \frac{\partial b_{\beta}}{\partial \chi}}{\mu_{0}} + \frac{b_{\chi} \frac{\partial b_{\chi}}{\partial \beta}}{\mu_{0}} = 0
\end{dmath}
\subsection*{Momentum Beta O($\varepsilon$)}
\begin{dmath}
\rho_{0} u_{\chi} \frac{\partial u_{\beta}}{\partial X} + \rho_{1} u_{\alpha} \frac{\partial u_{\beta}}{\partial \alpha} + \rho_{1} u_{\beta} \frac{\partial u_{\beta}}{\partial \beta} + \rho_{1} u_{\chi} \frac{\partial u_{\beta}}{\partial \chi} + \rho_{1} \frac{\partial u_{\beta}}{\partial t} - \frac{b_{\chi} \frac{\partial b_{\beta}}{\partial X}}{\mu_{0}} = 0
\end{dmath}
\subsection*{Momentum Beta O($\varepsilon^2$)}
\begin{dmath}
\rho_{1} u_{\chi} \frac{\partial u_{\beta}}{\partial X} = 0
\end{dmath}
\subsection*{Momentum Chi O(1)}
\begin{dmath}
\rho_{0} u_{\alpha} \frac{\partial u_{\chi}}{\partial \alpha} + \rho_{0} u_{\beta} \frac{\partial u_{\chi}}{\partial \beta} + \rho_{0} u_{\chi} \frac{\partial u_{\chi}}{\partial \chi} + \rho_{0} \frac{\partial u_{\chi}}{\partial t} + \frac{\partial p_{0}}{\partial X} + \frac{\partial p_{1}}{\partial \chi} - \frac{b_{\alpha} \frac{\partial B_{0}}{\partial A}}{\mu_{0}} + \frac{b_{\alpha} \frac{\partial b_{\alpha}}{\partial \chi}}{\mu_{0}} - \frac{b_{\alpha} \frac{\partial b_{\chi}}{\partial \alpha}}{\mu_{0}} + \frac{b_{\beta} \frac{\partial b_{\beta}}{\partial \chi}}{\mu_{0}} - \frac{b_{\beta} \frac{\partial b_{\chi}}{\partial \beta}}{\mu_{0}} = 0
\end{dmath}
\subsection*{Momentum Chi O($\varepsilon$)}
\begin{dmath}
\rho_{0} u_{\chi} \frac{\partial u_{\chi}}{\partial X} + \rho_{1} u_{\alpha} \frac{\partial u_{\chi}}{\partial \alpha} + \rho_{1} u_{\beta} \frac{\partial u_{\chi}}{\partial \beta} + \rho_{1} u_{\chi} \frac{\partial u_{\chi}}{\partial \chi} + \rho_{1} \frac{\partial u_{\chi}}{\partial t} + \frac{\partial p_{1}}{\partial X} + \frac{b_{\alpha} \frac{\partial b_{\alpha}}{\partial X}}{\mu_{0}} + \frac{b_{\beta} \frac{\partial b_{\beta}}{\partial X}}{\mu_{0}} = 0
\end{dmath}
\subsection*{Momentum Chi O($\varepsilon^2$)}
\begin{dmath}
\rho_{1} u_{\chi} \frac{\partial u_{\chi}}{\partial X} = 0
\end{dmath}
\subsection*{Div B O($\varepsilon$)}
\begin{dmath}
\frac{\partial B_{0}}{\partial X} + \frac{\partial b_{\alpha}}{\partial \alpha} + \frac{\partial b_{\beta}}{\partial \beta} + \frac{\partial b_{\chi}}{\partial \chi} = 0
\end{dmath}
\subsection*{Div B O($\varepsilon^2$)}
\begin{dmath}
\frac{\partial b_{\chi}}{\partial X} = 0
\end{dmath}

\bibliographystyle{jpp}
\bibliography{refs}

@article{atkinson_auroral_1970,
  title = {Auroral Arcs: {{Result}} of the Interaction of a Dynamic Magnetosphere with the Ionosphere},
  shorttitle = {Auroral Arcs},
  author = {Atkinson, G.},
  year = 1970,
  journal = {Journal of Geophysical Research (1896-1977)},
  volume = {75},
  number = {25},
  pages = {4746--4755},
  issn = {2156-2202},
  doi = {10.1029/JA075i025p04746},
  urldate = {2026-10-16},
  copyright = {Copyright \copyright{} 1970 by the American Geophysical Union.},
  langid = {english}
}

@article{bagheri_comparison_2023,
  title = {Comparison of Empirical Models of Ionospheric Heating to Global Simulations},
  author = {Bagheri, Fatemeh and Lopez, Ramon E.},
  year = 2023,
  month = jun,
  journal = {Frontiers in Astronomy and Space Sciences},
  volume = {10},
  publisher = {Frontiers},
  issn = {2296-987X},
  doi = {10.3389/fspas.2023.1170390},
  urldate = {2026-10-27},
  langid = {english}
}

@article{barnes_theory_1992,
  title = {Theory of Magnetohydrodynamic Waves: {{The WKB Approximation}} Revisited},
  shorttitle = {Theory of Magnetohydrodynamic Waves},
  author = {Barnes, Aaron},
  year = 1992,
  month = aug,
  journal = {Journal of Geophysical Research},
  volume = {97},
  number = {A8},
  pages = {105--112},
  publisher = {American Geophysical Union (AGU)},
  doi = {10.1029/92JA00996},
  urldate = {2026-03-20},
  langid = {english}
}

@article{burns_dedalus_2020,
  title = {Dedalus: {{A}} Flexible Framework for Numerical Simulations with Spectral Methods},
  shorttitle = {Dedalus},
  author = {Burns, Keaton J. and Vasil, Geoffrey M. and Oishi, Jeffrey S. and Lecoanet, Daniel and Brown, Benjamin P.},
  year = 2020,
  month = apr,
  journal = {Physical Review Research},
  volume = {2},
  number = {2},
  pages = {023068},
  publisher = {American Physical Society},
  doi = {10.1103/PhysRevResearch.2.023068},
  urldate = {2026-10-19}
}

@article{carter_mean_2023,
  title = {Mean {{Energy Flux}}, {{Associated Derived Height-Integrated Conductances}}, and {{Field-Aligned Current Magnitudes Evolve Differently During}} a {{Substorm}}},
  author = {Carter, J. A. and Milan, S. E. and Forsyth, C. and Lester, M. E. and Walach, M.-T. and Gjerloev, J. and Paxton, L. J. and Anderson, B. J.},
  year = 2023,
  journal = {Journal of Geophysical Research: Space Physics},
  volume = {128},
  number = {2},
  pages = {e2022JA030942},
  issn = {2169-9402},
  doi = {10.1029/2022JA030942},
  urldate = {2026-10-13},
  langid = {english}
}

@article{chaston_auroral_2021,
  title = {An {{Auroral Alfv\'en Wave Cascade}}},
  author = {Chaston, C. C.},
  year = 2021,
  month = mar,
  journal = {Frontiers in Astronomy and Space Sciences},
  volume = {8},
  publisher = {Frontiers},
  issn = {2296-987X},
  doi = {10.3389/fspas.2021.618429},
  urldate = {2025-10-06},
  langid = {english}
}

@article{chaston_energy_2024,
  title = {Energy {{Transport}} and {{Conversion Above}} a {{Bright Discrete Auroral Arc}}},
  author = {Chaston, C. C.},
  year = 2024,
  journal = {Journal of Geophysical Research: Space Physics},
  volume = {129},
  number = {12},
  pages = {e2024JA032870},
  issn = {2169-9402},
  doi = {10.1029/2024JA032870},
  urldate = {2026-07-17},
  langid = {english}
}

@article{chaston_how_2007,
  title = {How Important Are Dispersive {{Alfv\'en}} Waves for Auroral Particle Acceleration?},
  author = {Chaston, C. C. and Carlson, C. W. and McFadden, J. P. and Ergun, R. E. and Strangeway, R. J.},
  year = 2007,
  journal = {Geophysical Research Letters},
  volume = {34},
  number = {7},
  issn = {1944-8007},
  doi = {10.1029/2006GL029144},
  urldate = {2025-03-19}
}

@article{chaston_smallscale_2010,
  title = {Small-Scale Auroral Current Sheet Structuring},
  author = {Chaston, C. C. and Seki, K.},
  year = 2010,
  journal = {Journal of Geophysical Research: Space Physics},
  volume = {115},
  number = {A11},
  issn = {2156-2202},
  doi = {10.1029/2010JA015536},
  urldate = {2026-10-28},
  copyright = {Copyright 2010 by the American Geophysical Union.},
  langid = {english}
}

@article{chaston_turbulent_2008,
  title = {The {{Turbulent Alfv\'enic Aurora}}},
  author = {Chaston, C. C. and Salem, C. and Bonnell, J. W. and Carlson, C. W. and Ergun, R. E. and Strangeway, R. J. and McFadden, J. P.},
  year = 2008,
  month = apr,
  journal = {Physical Review Letters},
  volume = {100},
  number = {17},
  pages = {175003},
  doi = {10.1103/PhysRevLett.100.175003},
  urldate = {2025-07-15}
}

@article{david_generalized_2025,
  title = {Generalized Expanding-Box Formulations of Reduced Magnetohydrodynamics in the Solar Wind},
  author = {David, V. and Chandran, B. D. G. and Meyrand, R. and Squire, J. and Yerger, E. L.},
  year = 2025,
  month = apr,
  journal = {Journal of Plasma Physics},
  volume = {91},
  number = {2},
  pages = {E68},
  issn = {0022-3778, 1469-7807},
  doi = {10.1017/S0022377825000145},
  urldate = {2026-07-06},
  langid = {english}
}

@article{davies_alfven_2025,
  title = {Alfv\'en Wave Propagation in a Dipole Magnetic Field},
  author = {Davies, R. and Wright, A. N. and Elsden, T.},
  year = 2025,
  month = dec,
  journal = {Physics of Plasmas},
  volume = {32},
  number = {12},
  pages = {122101},
  issn = {1070-664X},
  doi = {10.1063/5.0287720},
  urldate = {2026-09-23}
}

@article{davies_alfvenfast_2024,
  title = {Alfv\'en-Fast Wave Coupling in a {{3D}} Non-Uniform Medium},
  author = {Davies, R. and Wright, A. N.},
  year = 2024,
  month = jul,
  journal = {Physics of Plasmas},
  volume = {31},
  number = {7},
  pages = {072114},
  issn = {1070-664X},
  doi = {10.1063/5.0210894},
  urldate = {2026-10-13}
}

@article{desjardin_steepening_2026,
  title = {Steepening of Inertial {{Alfv\'en}} Waves},
  author = {DesJardin, Ian and Dorelli, John and Iii, Lynn Wilson and Khazanov, George and Shuster, Jason},
  year = 2026,
  month = apr,
  journal = {Journal of Plasma Physics},
  volume = {92},
  number = {2},
  pages = {E26},
  issn = {0022-3778, 1469-7807},
  doi = {10.1017/S0022377826101330},
  urldate = {2026-03-05},
  langid = {english}
}

@article{di_mare_new_2024,
  title = {New {{Regime}} of {{Inertial Alfv\'en Wave Turbulence}} in the {{Auroral Ionosphere}}},
  author = {Di Mare, Francesca and Howes, Gregory G.},
  year = 2024,
  month = jul,
  journal = {Physical Review Letters},
  volume = {133},
  number = {4},
  pages = {045201},
  doi = {10.1103/PhysRevLett.133.045201},
  urldate = {2025-08-27}
}

@article{dorelli_science_2026,
  title = {Science {{Opportunities}} in the {{TRACERS Back Orbit}}: {{Ionospheric Signatures}} of {{Magnetotail Reconnection}}},
  shorttitle = {Science {{Opportunities}} in the {{TRACERS Back Orbit}}},
  author = {Dorelli, J. C. and Friedel, R. and Chen, L.-J. and Connor, H. and Miles, D. and Fuselier, S. and Petrinec, S. and Kletzing, C. A. and Angelopoulos, V. and Bonnell, J. and Bounds, S. and Buzulukova, N. and Chaston, C. C. and Christopher, I. and Da Silva, D. and Gjerloev, J. and Halekas, J. and Hesse, M. and Hospodarsky, G. and Liu, Y.-H. and Lotko, W. and Oieroset, M. and Runov, A. and Sotirelis, T. and Strangeway, R.},
  year = 2026,
  month = mar,
  journal = {Space Science Reviews},
  volume = {222},
  number = {2},
  pages = {22},
  issn = {0038-6308, 1572-9672},
  doi = {10.1007/s11214-026-01270-1},
  urldate = {2026-08-05},
  langid = {english}
}

@incollection{dungey_hydromagnetic_1968,
  title = {Hydromagnetic {{Waves}}},
  booktitle = {Physics of {{Geomagnetic Phenomena}}},
  author = {Dungey, J. W.},
  year = 1968,
  pages = {913},
  publisher = {Academic Press},
  address = {New York, NY}
}

@article{fu_distributions_2025,
  title = {The {{Distributions}} and {{Temporal Variations}} of {{FACs During Substorms}} and {{Quiet Periods}}},
  author = {Fu, Haobo and Yue, Chao and Zong, Q.-G. and Zhou, Xuzhi and Wang, Shan and Liu, Jianjun and Hu, Zejun and Yang, Huigen and Wang, Yongfu and Yu, Xiangqian and Ye, Yuguang and Zou, Hong},
  year = 2025,
  month = aug,
  journal = {Journal of Geophysical Research: Space Physics},
  volume = {130},
  number = {8},
  pages = {e2025JA033976},
  publisher = {John Wiley \& Sons, Ltd},
  issn = {2169-9402},
  doi = {10.1029/2025JA033976},
  urldate = {2026-10-13},
  langid = {english}
}

@article{gasparini_new_2025,
  title = {A {{New Approach}} to {{Data-Model Comparisons}}: {{Using MAGE}} and {{Lompe}} to {{Unravel Ionosphere-Magnetosphere Electrodynamics}}},
  shorttitle = {A {{New Approach}} to {{Data-Model Comparisons}}},
  author = {Gasparini, S. and Kepko, L. and Sorathia, K. A. and Michael, A. and Merkin, V. G. and Laundal, K. M. and Norgren, C.},
  year = 2025,
  journal = {Journal of Geophysical Research: Space Physics},
  volume = {130},
  number = {10},
  pages = {e2025JA034033},
  issn = {2169-9402},
  doi = {10.1029/2025JA034033},
  urldate = {2026-07-28},
  langid = {english}
}

@article{gunzkofer_impact_2026,
  title = {Impact of {{Magnetosphere Forcing}} and {{Model Resolution}} on {{Joule Heating Estimates}} in {{WACCM-X During}} the 24 {{August}} 2005 {{Storm}}},
  author = {G{\"u}nzkofer, Florian and Liu, Hanli and Pham, Kevin and Liu, Huixin and Stober, Gunter and McInerney, Joseph M. and Themens, David R. and Borries, Claudia},
  year = 2026,
  journal = {Earth and Space Science},
  volume = {13},
  number = {7},
  pages = {e2025EA004941},
  issn = {2333-5084},
  doi = {10.1029/2025EA004941},
  urldate = {2026-10-27},
  langid = {english}
}

@article{hallinan_smallscale_1970,
  title = {Small-Scale Auroral Arc Distortions},
  author = {Hallinan, Thomas J. and Davis, T. Neil},
  year = 1970,
  month = dec,
  journal = {Planetary and Space Science},
  volume = {18},
  number = {12},
  pages = {1735--1744},
  issn = {0032-0633},
  doi = {10.1016/0032-0633(70)90007-3},
  urldate = {2026-10-16}
}

@article{heinemann_nonwkb_1980,
  title = {Non-{{WKB Alfv\'en}} Waves in the Solar Wind},
  author = {Heinemann, M. and Olbert, S.},
  year = 1980,
  month = mar,
  journal = {Journal of Geophysical Research},
  volume = {85},
  number = {A3},
  pages = {1311--1327},
  publisher = {American Geophysical Union (AGU)},
  doi = {10.1029/JA085iA03p01311},
  urldate = {2026-07-02},
  langid = {english}
}

@article{huba_modeling_2013,
  title = {Modeling the Plasmasphere with {{SAMI3}}},
  author = {Huba, J. and Krall, J.},
  year = 2013,
  journal = {Geophysical Research Letters},
  volume = {40},
  number = {1},
  pages = {6--10},
  issn = {1944-8007},
  doi = {10.1029/2012GL054300},
  urldate = {2026-07-30},
  copyright = {\copyright 2012. American Geophysical Union. All Rights Reserved.},
  langid = {english}
}

@article{hull_alfvenic_2016,
  title = {The ``{{Alfv\'enic}} Surge'' at Substorm Onset/Expansion and the Formation of ``{{Inverted Vs}}'': {{Cluster}} and {{IMAGE}} Observations},
  shorttitle = {The ``{{Alfv\'enic}} Surge'' at Substorm Onset/Expansion and the Formation of ``{{Inverted Vs}}''},
  author = {Hull, A. J. and Chaston, C. C. and Frey, H. U. and Fillingim, M. O. and Goldstein, M. L. and Bonnell, J. W. and Mozer, F. S.},
  year = 2016,
  journal = {Journal of Geophysical Research: Space Physics},
  volume = {121},
  number = {5},
  pages = {3978--4004},
  issn = {2169-9402},
  doi = {10.1002/2015JA022000},
  urldate = {2025-11-03},
  copyright = {\copyright 2016. American Geophysical Union. All Rights Reserved.}
}

@article{hunana_inhomogeneous_2010,
  title = {Inhomogeneous {{Nearly Incompressible Description}} of {{Magnetohydrodynamic Turbulence}}},
  author = {Hunana, P. and Zank, G. P.},
  year = 2010,
  month = jun,
  journal = {The Astrophysical Journal},
  volume = {718},
  number = {1},
  pages = {148},
  publisher = {The American Astronomical Society},
  issn = {0004-637X},
  doi = {10.1088/0004-637X/718/1/148},
  urldate = {2026-07-10},
  langid = {english}
}

@article{ivarsen_turbulence_2024,
  title = {Turbulence {{Around Auroral Arcs}}},
  author = {Ivarsen, Magnus F. and Huyghebaert, Devin R. and Gillies, Megan D. and {St-Maurice}, Jean-Pierre and Themens, David R. and Oppenheim, Meers and Gustavsson, Bj{\"o}rn J. and Billett, Daniel and Pitzel, Brian and Galeschuk, Draven and Donovan, Eric and Hussey, Glenn C.},
  year = 2024,
  journal = {Journal of Geophysical Research: Space Physics},
  volume = {129},
  number = {8},
  pages = {e2023JA032309},
  issn = {2169-9402},
  doi = {10.1029/2023JA032309},
  urldate = {2026-07-17},
  langid = {english}
}

@article{ivarsen_turbulence_2024-1,
  title = {Turbulence {{Embedded Into}} the {{Ionosphere}} by {{Electromagnetic Waves}}},
  author = {Ivarsen, Magnus F. and Gillies, Megan D. and Huyghebaert, Devin R. and {St-Maurice}, Jean-Pierre and Lozinsky, Adam and Galeschuk, Draven and Donovan, Eric and Hussey, Glenn C.},
  year = 2024,
  journal = {Journal of Geophysical Research: Space Physics},
  volume = {129},
  number = {8},
  pages = {e2023JA032310},
  issn = {2169-9402},
  doi = {10.1029/2023JA032310},
  urldate = {2026-07-30},
  copyright = {\copyright 2024. The Author(s).},
  langid = {english}
}

@article{james_model_2026,
  title = {Model {{Validation}} of {{M-I Coupling}} in {{SWMF}}},
  author = {James, Tre'Shunda and Glocer, Alex},
  year = 2026,
  journal = {Journal of Geophysical Research: Space Physics},
  volume = {131},
  number = {3},
  eprint = {https://agupubs.onlinelibrary.wiley.com/doi/pdf/10.1029/2025JA034284},
  pages = {e2025JA034284},
  doi = {10.1029/2025JA034284}
}

@article{kadomtsev_nonlinear_1973,
  title = {Nonlinear Helical Perturbations of a Plasma in the Tokamak},
  author = {Kadomtsev, B. B. and Pogutse, O. P.},
  year = 1973,
  month = aug,
  journal = {Zhurnal Eksperimentalnoi i Teoreticheskoi Fiziki},
  volume = {65},
  pages = {575--589},
  issn = {0044-4510},
  urldate = {2026-07-10}
}

@article{kageyama_note_2006,
  title = {A {{Note}} on the {{Dipole Coordinates}}},
  author = {Kageyama, Akira and Sugiyama, Tooru and Watanabe, Kunihiko and Sato, Tetsuya},
  year = 2006,
  month = mar,
  journal = {Computers \& Geosciences},
  volume = {32},
  number = {2},
  eprint = {physics/0408133},
  pages = {265--269},
  issn = {00983004},
  doi = {10.1016/j.cageo.2005.06.006},
  urldate = {2026-02-17},
  archiveprefix = {arXiv}
}

@article{kataoka_small-scale_2021,
  title = {Small-{{Scale Dynamic Aurora}}},
  author = {Kataoka, Ryuho and Chaston, Christopher C. and Knudsen, David and Lynch, Kristina A. and Lysak, Robert L. and Song, Yan and Rankin, Robert and Murase, Kiyoka and Sakanoi, Takeshi and Semeter, Joshua and Watanabe, Tomo-Hiko and Whiter, Daniel},
  year = 2021,
  journal = {Space Science Reviews},
  volume = {217},
  number = {1},
  pages = {17},
  issn = {0038-6308},
  doi = {10.1007/s11214-021-00796-w},
  urldate = {2026-07-28},
  pmcid = {PMC8550089},
  pmid = {34720215}
}

@article{kataoka_turbulent_2011,
  title = {Turbulent Microstructures and Formation of Folds in Auroral Breakup Arc},
  author = {Kataoka, Ryuho and Miyoshi, Yoshizumi and Sakanoi, Takeshi and Yaegashi, Ayumi and Shiokawa, Kazuo and Ebihara, Yusuke},
  year = 2011,
  journal = {Journal of Geophysical Research: Space Physics},
  volume = {116},
  number = {A1},
  issn = {2156-2202},
  doi = {10.1029/2010JA016334},
  urldate = {2026-01-08}
}

@article{keiling_assessing_2019,
  title = {Assessing the Global {{Alfv\'en}} Wave Power Flow into and out of the Auroral Acceleration Region during Geomagnetic Storms},
  author = {Keiling, Andreas and Thaller, Scott and Wygant, John and Dombeck, John},
  year = 2019,
  month = jun,
  journal = {Science Advances},
  volume = {5},
  number = {6},
  pages = {eaav8411},
  publisher = {American Association for the Advancement of Science},
  doi = {10.1126/sciadv.aav8411},
  urldate = {2026-07-13}
}

@article{keiling_global_2003,
  title = {The {{Global Morphology}} of {{Wave Poynting Flux}}: {{Powering}} the {{Aurora}}},
  shorttitle = {The {{Global Morphology}} of {{Wave Poynting Flux}}},
  author = {Keiling, A. and Wygant, J. R. and Cattell, C. A. and Mozer, F. S. and Russell, C. T.},
  year = 2003,
  month = jan,
  journal = {Science},
  volume = {299},
  number = {5605},
  pages = {383--386},
  publisher = {American Association for the Advancement of Science},
  doi = {10.1126/science.1080073},
  urldate = {2026-07-07}
}

@article{kletzing_electron_1994,
  title = {Electron Acceleration by Kinetic {{Alfv\'en}} Waves},
  author = {Kletzing, C. A.},
  year = 1994,
  journal = {Journal of Geophysical Research: Space Physics},
  volume = {99},
  number = {A6},
  pages = {11095--11103},
  issn = {2156-2202},
  doi = {10.1029/94JA00345},
  urldate = {2025-03-05}
}

@article{knudsen_alfven_1992,
  title = {Alfv\'en Waves in the Auroral Ionosphere: {{A}} Numerical Model Compared with Measurements},
  shorttitle = {Alfv\'en Waves in the Auroral Ionosphere},
  author = {Knudsen, D. J. and Kelley, M. C. and Vickrey, J. F.},
  year = 1992,
  journal = {Journal of Geophysical Research: Space Physics},
  volume = {97},
  number = {A1},
  pages = {77--90},
  issn = {2156-2202},
  doi = {10.1029/91JA02300},
  urldate = {2025-12-14},
  copyright = {Copyright 1992 by the American Geophysical Union.}
}

@article{kreiss_problems_1980,
  title = {Problems with Different Time Scales for Partial Differential Equations},
  author = {Kreiss, Heinz-Otto},
  year = 1980,
  journal = {Communications on Pure and Applied Mathematics},
  volume = {33},
  number = {3},
  pages = {399--439},
  issn = {1097-0312},
  doi = {10.1002/cpa.3160330310},
  urldate = {2025-01-29},
  copyright = {Copyright \copyright{} 1980 Wiley Periodicals, Inc., A Wiley Company},
  langid = {english}
}

@article{laundal_global_2025,
  title = {Global Inductive Magnetosphere-Ionosphere- Thermosphere Coupling},
  author = {Laundal, Karl M. and Skeidsvoll, Andreas S. and Popescu Braileanu, Beatrice and Hatch, Spencer M. and Olsen, Nils and Vanham{\"a}ki, Heikki},
  year = 2025,
  month = dec,
  journal = {Annales Geophysicae},
  volume = {43},
  number = {2},
  pages = {803--833},
  publisher = {Copernicus GmbH},
  issn = {0992-7689},
  doi = {10.5194/angeo-43-803-2025},
  urldate = {2026-10-27},
  langid = {english}
}

@book{lemaire_earths_1998,
  title = {The {{Earth}}'s {{Plasmasphere}}},
  author = {Lemaire, J. F. and Gringauz, K. I.},
  year = 1998,
  series = {Cambridge {{Atmospheric}} and {{Space Science Series}}},
  volume = {1},
  publisher = {Cambridge University Press},
  address = {Cambridge},
  doi = {10.1017/CBO9780511600098},
  urldate = {2026-07-30},
  isbn = {978-0-521-43091-3}
}

@article{lynch_multiple-point_1999,
  title = {Multiple-Point Electron Measurements in a Nightside Auroral Arc: {{Auroral}} Turbulence {{II}} Particle Observations},
  shorttitle = {Multiple-Point Electron Measurements in a Nightside Auroral Arc},
  author = {Lynch, K. A. and Pietrowski, D. and Torbert, R. B. and Ivchenko, N. and Marklund, G. and Primdahl, F.},
  year = 1999,
  journal = {Geophysical Research Letters},
  volume = {26},
  number = {22},
  pages = {3361--3364},
  issn = {1944-8007},
  doi = {10.1029/1999GL900599},
  urldate = {2026-03-11},
  copyright = {Copyright 1999 by the American Geophysical Union.},
  langid = {english}
}

@article{lysak_feedback_1991,
  title = {Feedback Instability of the Ionospheric Resonant Cavity},
  author = {Lysak, Robert L.},
  year = 1991,
  journal = {Journal of Geophysical Research: Space Physics},
  volume = {96},
  number = {A2},
  pages = {1553--1568},
  issn = {2156-2202},
  doi = {10.1029/90JA02154},
  urldate = {2026-10-26},
  langid = {english}
}

@incollection{lysak_generalized_1993,
  title = {Generalized {{Model}} of the {{Ionospheric Alfv\'en Resonator}}},
  booktitle = {Auroral {{Plasma Dynamics}}},
  author = {Lysak, Robert L.},
  year = 1993,
  pages = {121--128},
  publisher = {American Geophysical Union (AGU)},
  doi = {10.1029/GM080p0121},
  urldate = {2025-06-04},
  isbn = {978-1-118-66400-1},
  langid = {english}
}

@article{lysak_kinetic_2023,
  title = {Kinetic {{Alfv\'en}} Waves and Auroral Particle Acceleration: A Review},
  shorttitle = {Kinetic {{Alfv\'en}} Waves and Auroral Particle Acceleration},
  author = {Lysak, R. L.},
  year = 2023,
  journal = {Reviews of Modern Plasma Physics},
  volume = {7},
  number = {1},
  pages = {6},
  issn = {2367-3192},
  doi = {10.1007/s41614-022-00111-2},
  urldate = {2025-07-18},
  pmcid = {PMC9825542},
  pmid = {36632354}
}

@article{lysak_magnetosphere-ionosphere_2004,
  title = {Magnetosphere-Ionosphere Coupling by {{Alfv\'en}} Waves at Midlatitudes},
  author = {Lysak, R. L.},
  year = 2004,
  journal = {Journal of Geophysical Research: Space Physics},
  volume = {109},
  number = {A7},
  issn = {2156-2202},
  doi = {10.1029/2004JA010454},
  urldate = {2026-02-03},
  copyright = {Copyright 2004 by the American Geophysical Union.},
  langid = {english}
}

@article{merkin_effects_2010,
  title = {Effects of the Low-Latitude Ionospheric Boundary Condition on the Global Magnetosphere},
  author = {Merkin, V. G. and Lyon, J. G.},
  year = 2010,
  journal = {Journal of Geophysical Research: Space Physics},
  volume = {115},
  number = {A10},
  issn = {2156-2202},
  doi = {10.1029/2010JA015461},
  urldate = {2026-10-16},
  langid = {english}
}

@article{milanese_dynamic_2020,
  title = {Dynamic {{Phase Alignment}} in {{Inertial Alfv\'en Turbulence}}},
  author = {Milanese, Lucio M. and Loureiro, Nuno F. and Daschner, Maximilian and Boldyrev, Stanislav},
  year = 2020,
  month = dec,
  journal = {Physical Review Letters},
  volume = {125},
  number = {26},
  pages = {265101},
  issn = {0031-9007, 1079-7114},
  doi = {10.1103/PhysRevLett.125.265101},
  urldate = {2025-10-08}
}

@article{miles_alfvenic_2018,
  title = {Alfv\'enic {{Dynamics}} and {{Fine Structuring}} of {{Discrete Auroral Arcs}}: {{Swarm}} and e-{{POP Observations}}},
  shorttitle = {Alfv\'enic {{Dynamics}} and {{Fine Structuring}} of {{Discrete Auroral Arcs}}},
  author = {Miles, D. M. and Mann, I. R. and Pakhotin, I. P. and Burchill, J. K. and Howarth, A. D. and Knudsen, D. J. and Lysak, R. L. and Wallis, D. D. and Cogger, L. L. and Yau, A. W.},
  year = 2018,
  journal = {Geophysical Research Letters},
  volume = {45},
  number = {2},
  pages = {545--555},
  issn = {1944-8007},
  doi = {10.1002/2017GL076051},
  urldate = {2026-01-16},
  copyright = {\copyright 2018. American Geophysical Union. All Rights Reserved.},
  langid = {english}
}

@article{mukhopadhyay_global_2022,
  title = {Global {{Driving}} of {{Auroral Precipitation}}: 1. {{Balance}} of {{Sources}}},
  shorttitle = {Global {{Driving}} of {{Auroral Precipitation}}},
  author = {Mukhopadhyay, Agnit and Welling, Daniel and Liemohn, Michael and Ridley, Aaron and Burleigh, Meghan and Wu, Chen and Zou, Shasha and Connor, Hyunju and Vandegriff, Elizabeth and Dredger, Pauline and T{\'o}th, Gabor},
  year = 2022,
  journal = {Journal of Geophysical Research: Space Physics},
  volume = {127},
  number = {7},
  pages = {e2022JA030323},
  issn = {2169-9402},
  doi = {10.1029/2022JA030323},
  urldate = {2026-10-27},
  langid = {english}
}

@inproceedings{okuta_cupy_2017,
  title = {{{CuPy}} : {{A NumPy-Compatible Library}} for {{NVIDIA GPU Calculations}}},
  shorttitle = {{{CuPy}}},
  booktitle = {Proceedings of {{Workshop}} on {{Machine Learning Systems}} ({{LearningSys}}) in {{The Thirty}}-First {{Annual Conference}} on {{Neural Information Processing Systems}} ({{NIPS}}), 2017},
  author = {Okuta, Ryosuke and Unno, Y. and Nishino, Daisuke and Hido, Shohei and {Crissman}},
  year = 2017,
  address = {Long Beach, CA, USA},
  urldate = {2025-08-28}
}

@article{perez_direct_2013,
  title = {Direct {{Numerical Simulations}} of {{Reflection}}-Driven, {{Reduced Magnetohydrodynamic Turbulence}} from the {{Sun}} to the {{Alfv\'en Critical Point}}},
  author = {Perez, Jean Carlos and Chandran, Benjamin D. G.},
  year = 2013,
  month = oct,
  journal = {The Astrophysical Journal},
  volume = {776},
  pages = {124},
  publisher = {IOP},
  issn = {0004-637X},
  doi = {10.1088/0004-637X/776/2/124},
  urldate = {2026-04-27}
}

@article{potapov_morphology_2021,
  title = {Morphology and Diagnostic Potential of the Ionospheric {{Alfv\'en}} Resonator},
  author = {Potapov, Alexander and Polyushkina, Tatyana and Tsegmed, Battuulai},
  year = 2021,
  month = sep,
  journal = {Solar-Terrestrial Physics},
  volume = {7},
  number = {3},
  pages = {36--52},
  doi = {10.12737/stp-73202104},
  urldate = {2026-02-13},
  langid = {english}
}

@article{sakaki_convective_2024,
  title = {Convective {{Growth}} of {{Auroral Arcs Through}} the {{Feedback Instability}} in a {{Dipole Geometry}}},
  author = {Sakaki, T. and Watanabe, T.-H. and Maeyama, S.},
  year = 2024,
  journal = {Journal of Geophysical Research: Space Physics},
  volume = {129},
  number = {12},
  pages = {e2023JA032407},
  issn = {2169-9402},
  doi = {10.1029/2023JA032407},
  urldate = {2026-02-03},
  copyright = {\copyright{} 2024. American Geophysical Union. All Rights Reserved.},
  langid = {english}
}

@article{sakaki_nonlinear_2026,
  title = {Nonlinear {{Evolution}} of {{Auroral Fine Structures Through}} the {{Feedback Instability}} in a {{Dipole Field Configuration}}},
  author = {Sakaki, T. and Watanabe, T.-H.},
  year = 2026,
  journal = {Journal of Geophysical Research: Space Physics},
  volume = {131},
  number = {6},
  pages = {e2025JA034812},
  issn = {2169-9402},
  doi = {10.1029/2025JA034812},
  urldate = {2026-09-17},
  copyright = {\copyright{} 2026. The Author(s).},
  langid = {english}
}

@article{schekochihin_mhd_2022,
  title = {{{MHD}} Turbulence: A Biased Review},
  shorttitle = {{{MHD}} Turbulence},
  author = {Schekochihin, Alexander A.},
  year = 2022,
  month = oct,
  journal = {Journal of Plasma Physics},
  volume = {88},
  number = {5},
  pages = {155880501},
  issn = {0022-3778, 1469-7807},
  doi = {10.1017/S0022377822000721},
  urldate = {2025-10-10},
  langid = {english}
}

@article{schroeder_laboratory_2021,
  title = {Laboratory Measurements of the Physics of Auroral Electron Acceleration by {{Alfv\'en}} Waves},
  author = {Schroeder, J. W. R. and Howes, G. G. and Kletzing, C. A. and Skiff, F. and Carter, T. A. and Vincena, S. and Dorfman, S.},
  year = 2021,
  month = jun,
  journal = {Nature Communications},
  volume = {12},
  number = {1},
  pages = {3103},
  publisher = {Nature Publishing Group},
  issn = {2041-1723},
  doi = {10.1038/s41467-021-23377-5},
  urldate = {2026-02-10},
  copyright = {2021 The Author(s)}
}

@article{seyler_mathematical_1990,
  title = {A Mathematical Model of the Structure and Evolution of Small-Scale Discrete Auroral Arcs},
  author = {Seyler, Charles E.},
  year = 1990,
  journal = {Journal of Geophysical Research: Space Physics},
  volume = {95},
  number = {A10},
  pages = {17199--17215},
  issn = {2156-2202},
  doi = {10.1029/JA095iA10p17199},
  urldate = {2026-07-30},
  copyright = {Copyright 1990 by the American Geophysical Union.},
  langid = {english}
}

@article{sheeley_empirical_2001,
  title = {An Empirical Plasmasphere and Trough Density Model: {{CRRES}} Observations},
  shorttitle = {An Empirical Plasmasphere and Trough Density Model},
  author = {Sheeley, B. W. and Moldwin, M. B. and Rassoul, H. K. and Anderson, R. R.},
  year = 2001,
  journal = {Journal of Geophysical Research: Space Physics},
  volume = {106},
  number = {A11},
  pages = {25631--25641},
  issn = {2156-2202},
  doi = {10.1029/2000JA000286},
  urldate = {2026-07-10},
  copyright = {Copyright 2001 by the American Geophysical Union.},
  langid = {english}
}

@article{southwood_curvature_1985,
  title = {Curvature Coupling of Slow and {{Alfv\'en MHD}} Waves in a Magnetotail Field Configuration},
  author = {Southwood, D. J. and Saunders, M. A.},
  year = 1985,
  month = jan,
  journal = {Planetary and Space Science},
  volume = {33},
  number = {1},
  pages = {127--134},
  issn = {0032-0633},
  doi = {10.1016/0032-0633(85)90149-7},
  urldate = {2025-08-04}
}

@article{spencer_analysis_2007,
  title = {Analysis of the 3--7 {{October}} 2000 and 15--24 {{April}} 2002 Geomagnetic Storms with an Optimized Nonlinear Dynamical Model},
  author = {Spencer, E. and Horton, W. and Mays, M. L. and Doxas, I. and Kozyra, J.},
  year = 2007,
  journal = {Journal of Geophysical Research: Space Physics},
  volume = {112},
  number = {A4},
  issn = {2156-2202},
  doi = {10.1029/2006JA012019},
  urldate = {2026-10-06},
  copyright = {Copyright 2007 by the American Geophysical Union.},
  langid = {english}
}

@misc{squire_transport_2026,
  title = {A {{Transport Theory}} of {{Turbulent Coronal Heating}} in {{General Geometry}}},
  author = {Squire, Jonathan and Chandran, Benjamin D. G. and Adkins, Toby and Clarke, William A. and Meyrand, Romain and Kunz, Matthew W.},
  year = 2026,
  month = jul,
  number = {arXiv:2607.08036},
  eprint = {2607.08036},
  primaryclass = {astro-ph.SR},
  publisher = {arXiv},
  doi = {10.48550/arXiv.2607.08036},
  urldate = {2026-07-13},
  archiveprefix = {arXiv}
}

@article{stasiewicz_identification_2000,
  title = {Identification of Widespread Turbulence of Dispersive {{Alfv\'en}} Waves},
  author = {Stasiewicz, K. and Khotyaintsev, Y. and Berthomier, M. and Wahlund, J -E.},
  year = 2000,
  journal = {Geophysical Research Letters},
  volume = {27},
  number = {2},
  pages = {173--176},
  issn = {1944-8007},
  doi = {10.1029/1999GL010696},
  urldate = {2025-12-03},
  copyright = {Copyright 1999 by the American Geophysical Union.},
  langid = {english}
}

@article{stasiewicz_origin_1984,
  title = {On the Origin of the Auroral Inverted-{{V}} Electron Spectra},
  author = {Stasiewicz, Krzysztof},
  year = 1984,
  month = mar,
  journal = {Planetary and Space Science},
  volume = {32},
  number = {3},
  pages = {379--389},
  issn = {0032-0633},
  doi = {10.1016/0032-0633(84)90172-7},
  urldate = {2026-03-05}
}

@article{stasiewicz_small_2000,
  title = {Small {{Scale Alfv\'enic Structure}} in the {{Aurora}}},
  author = {Stasiewicz, K. and Bellan, P. and Chaston, C. and Kletzing, C. and Lysak, R. and Maggs, J. and Pokhotelov, O. and Seyler, C. and Shukla, P. and Stenflo, L. and Streltsov, A. and Wahlund, J.-E.},
  year = 2000,
  month = may,
  journal = {Space Science Reviews},
  volume = {92},
  number = {3},
  pages = {423--533},
  issn = {1572-9672},
  doi = {10.1023/A:1005207202143},
  urldate = {2025-07-14}
}

@article{strauss_nonlinear_1976,
  title = {Nonlinear, Three-dimensional Magnetohydrodynamics of Noncircular Tokamaks},
  author = {Strauss, H. R.},
  year = 1976,
  month = jan,
  journal = {The Physics of Fluids},
  volume = {19},
  number = {1},
  pages = {134--140},
  issn = {0031-9171},
  doi = {10.1063/1.861310},
  urldate = {2025-01-27}
}

@article{strauss_reduced_1997,
  title = {Reduced {{MHD}} in Nearly Potential Magnetic Fields},
  author = {Strauss, H. R.},
  year = 1997,
  month = jan,
  journal = {Journal of Plasma Physics},
  volume = {57},
  number = {1},
  pages = {83--87},
  issn = {1469-7807, 0022-3778},
  doi = {10.1017/S0022377896005296},
  urldate = {2026-02-03},
  langid = {english}
}

@article{sydorenko_nonlinear_2008,
  title = {Nonlinear Effects in the Ionospheric {{Alfv\'en}} Resonator},
  author = {Sydorenko, D. and Rankin, R. and Kabin, K.},
  year = 2008,
  month = oct,
  journal = {Journal of Geophysical Research: Space Physics},
  volume = {113},
  number = {A10},
  pages = {2008JA013579},
  issn = {0148-0227},
  doi = {10.1029/2008JA013579},
  urldate = {2026-07-20},
  copyright = {http://onlinelibrary.wiley.com/termsAndConditions\#vor},
  langid = {english}
}

@article{thompson_electron_1996,
  title = {Electron Acceleration by Inertial {{Alfv\'en}} Waves},
  author = {Thompson, B. J. and Lysak, R. L.},
  year = 1996,
  journal = {Journal of Geophysical Research: Space Physics},
  volume = {101},
  number = {A3},
  pages = {5359--5369},
  issn = {2156-2202},
  doi = {10.1029/95JA03622},
  urldate = {2025-06-18}
}

@article{tian_evidence_2026,
  title = {Evidence for {{Alfv\'en}} Waves Powering Auroral Arc via a Static Electric Potential Drop},
  author = {Tian, S. and Yao, Z. and Wygant, J. R. and Lysak, R. L. and Bortnik, J. and Lyons, L. R. and Liang, J. and Shi, R. and Ferradas, C. P. and Shen, Y. and Reeves, G. D.},
  year = 2026,
  month = jan,
  journal = {Nature Communications},
  volume = {17},
  number = {1},
  pages = {297},
  publisher = {Nature Publishing Group},
  issn = {2041-1723},
  doi = {10.1038/s41467-025-65819-4},
  urldate = {2026-01-15},
  copyright = {2025 The Author(s)}
}

@article{wang_substorm_2026,
  title = {Substorm Expansion Embedded in a Global Cycle of Field-Aligned Currents and Auroral Electrojets},
  author = {Wang, Tonghui and Dai, Lei and Escoubet, C. Philippe and Gonzalez, Walter and Ren, Yong and Zhu, Minghui and Wang, Shan and Wang, Chi and Wang, Xu and Wang, Kailai and Liu, Jinjuan},
  year = 2026,
  month = feb,
  journal = {Nature Communications},
  volume = {17},
  number = {1},
  pages = {2970},
  publisher = {Nature Publishing Group},
  issn = {2041-1723},
  doi = {10.1038/s41467-026-69753-x},
  urldate = {2026-05-01},
  copyright = {2026 The Author(s)},
  langid = {english}
}

@article{watanabe_feedback_2010,
  title = {Feedback Instability in the Magnetosphere-Ionosphere Coupling System: {{Revisited}}},
  shorttitle = {Feedback Instability in the Magnetosphere-Ionosphere Coupling System},
  author = {Watanabe, T.-H.},
  year = 2010,
  month = feb,
  journal = {Physics of Plasmas},
  volume = {17},
  number = {2},
  pages = {022904},
  issn = {1070-664X},
  doi = {10.1063/1.3304237},
  urldate = {2026-02-03}
}

@article{watanabe_generation_2016,
  title = {Generation of Auroral Turbulence through the Magnetosphere--Ionosphere Coupling},
  author = {Watanabe, Tomo-Hiko and Kurata, Hiroaki and Maeyama, Shinya},
  year = 2016,
  month = dec,
  journal = {New Journal of Physics},
  volume = {18},
  number = {12},
  pages = {125010},
  publisher = {IOP Publishing},
  issn = {1367-2630},
  doi = {10.1088/1367-2630/aa532a},
  urldate = {2026-03-17},
  langid = {english}
}

@article{watt_inertial_2006,
  title = {Inertial {{Alfv\'en}} Waves and Acceleration of Electrons in Nonuniform Magnetic Fields},
  author = {Watt, C. E. J. and Rankin, R. and Rae, I. J. and Wright, D. M.},
  year = 2006,
  journal = {Geophysical Research Letters},
  volume = {33},
  number = {2},
  issn = {1944-8007},
  doi = {10.1029/2005GL024779},
  urldate = {2026-03-12},
  copyright = {Copyright 2006 by the American Geophysical Union.},
  langid = {english}
}

@article{woodroffe_ultra-low_2012,
  title = {Ultra-Low Frequency Wave Coupling in the Ionospheric {{Alfv\'en}} Resonator: {{Characteristics}} and Implications for the Interpretation of Ground Magnetic Fields},
  shorttitle = {Ultra-Low Frequency Wave Coupling in the Ionospheric {{Alfv\'en}} Resonator},
  author = {Woodroffe, J. R. and Lysak, R. L.},
  year = 2012,
  journal = {Journal of Geophysical Research: Space Physics},
  volume = {117},
  number = {A3},
  issn = {2156-2202},
  doi = {10.1029/2011JA017057},
  urldate = {2026-02-17},
  langid = {english}
}

@article{yu_including_2010,
  title = {Including Gap Region Field-aligned Currents and Magnetospheric Currents in the {{MHD}} Calculation of Ground-based Magnetic Field Perturbations},
  author = {Yu, Yiqun and Ridley, Aaron J. and Welling, Dan T. and T{\'o}th, Gabor},
  year = 2010,
  month = aug,
  journal = {Journal of Geophysical Research: Space Physics},
  volume = {115},
  number = {A8},
  pages = {2009JA014869},
  issn = {0148-0227},
  doi = {10.1029/2009JA014869},
  urldate = {2026-07-17},
  copyright = {http://onlinelibrary.wiley.com/termsAndConditions\#vor},
  langid = {english}
}

@article{yu_modeling_2015,
  title = {Modeling Subauroral Polarization Streams during the 17 {{March}} 2013 Storm},
  author = {Yu, Yiqun and Jordanova, Vania and Zou, Shasha and Heelis, Roderick and Ruohoniemi, Mike and Wygant, John},
  year = 2015,
  journal = {Journal of Geophysical Research: Space Physics},
  volume = {120},
  number = {3},
  pages = {1738--1750},
  issn = {2169-9402},
  doi = {10.1002/2014JA020371},
  urldate = {2026-07-28},
  langid = {english}
}

@article{zank_equations_1992,
  title = {The Equations of Reduced Magnetohydrodynamics},
  author = {Zank, G. P. and Matthaeus, W. H.},
  year = 1992,
  month = aug,
  journal = {Journal of Plasma Physics},
  volume = {48},
  number = {1},
  pages = {85--100},
  issn = {1469-7807, 0022-3778},
  doi = {10.1017/S002237780001638X},
  urldate = {2026-02-17},
  langid = {english}
}

@article{zhang_electron_2015,
  title = {Electron Precipitation Models in Global Magnetosphere Simulations},
  author = {Zhang, B. and Lotko, W. and Brambles, O. and Wiltberger, M. and Lyon, J.},
  year = 2015,
  journal = {Journal of Geophysical Research: Space Physics},
  volume = {120},
  number = {2},
  pages = {1035--1056},
  issn = {2169-9402},
  doi = {10.1002/2014JA020615},
  urldate = {2026-10-27},
  copyright = {\copyright 2015. American Geophysical Union. All Rights Reserved.},
  langid = {english}
}

\end{document}